\RequirePackage{ifpdf}
\documentclass[hyper,letterpaper]{JHEP3}
\usepackage{amssymb,amsfonts,bm,amsmath,empheq}
\usepackage{cite}
\usepackage{graphicx}
\usepackage{multirow}
\usepackage{verbatim}
\usepackage{appendix}

\usepackage{url}
\usepackage{float}

\newcommand{\bea}{\begin{eqnarray}}
\newcommand{\eea}{\end{eqnarray}}
\newcommand{\be}{\begin{equation}}
\newcommand{\ee}{\end{equation}}

\def\IN{\mathbb {N}}
\def\IZ{\mathbb {Z}}

\def\IC{\mathbb {C}}
\def\IP{\mathbb {P}}

\renewcommand{\hat}{\widehat}

\title{Super-quantum curves from super-eigenvalue models}

\author{Pawe{\l} Ciosmak$^{1}$, Leszek Hadasz$^{2}$, Masahide Manabe$^{3}$ and Piotr Su{\l}kowski$^{3,4}$
\\
$^1$ Faculty of Mathematics, Informatics and Mechanics, University of Warsaw, ul. Banacha 2, 02-097 Warsaw, Poland  \\
$^2$ M.\ Smoluchowski Institute of Physics, Jagiellonian University, ul. {\L}ojasiewicza 11, 30-348 Krak{\'o}w, Poland  \\
$^3$ Faculty of Physics, University of Warsaw, ul. Pasteura 5, 02-093 Warsaw, Poland  \\
$^4$ Walter Burke Institute for Theoretical Physics, California Institute of Technology, Pasadena, CA 91125, USA}

\abstract{In modern mathematical and theoretical physics various generalizations, in particular supersymmetric or quantum, of Riemann surfaces and complex algebraic curves play a prominent role. We show that such supersymmetric and quantum generalizations can be combined together, and construct supersymmetric quantum curves, or super-quantum curves for short. Our analysis is conducted in the formalism of super-eigenvalue models: we introduce $\beta$-deformed version of those models, and derive differential equations for associated $\alpha/\beta$-deformed super-matrix integrals. We show that for a given model there exists an infinite number of such differential equations, which we identify as super-quantum curves, and which  are in one-to-one correspondence with, and have the structure of, super-Virasoro singular vectors. We discuss potential applications of super-quantum curves and prospects of other generalizations.
\\
\\
\\
\\
\\
\\
\\
\\
\\
\\
\\
{\tt CALT-2016-021}}

\begin{document}


\newpage

\section{Introduction}  \label{sec-intro}

Riemann surfaces and complex algebraic curves play a prominent role in mathematics and admit various generalizations, which often have origin in modern mathematical and theoretical physics. In this paper we consider two such generalizations: supersymmetric and quantum. Supersymmetric Riemann surfaces play an important role in superstring theory \cite{Baranov:1987df,Manin:1987tga,Giddings:1987wn,Voronov:1987xf,Belopolsky:1997bg}, and they have received renewed interest recently \cite{Witten:2012ga,Witten:2012bh,Donagi:2013dua,Penner:2015xla,Ip:2016ojn}. They can also be represented, at least in some specific cases, as supersymmetric algebraic curves \cite{Rabin:1987rg,Rabin:1993bw}.

Another generalization of algebraic curves, also to much extent motivated by physics, is related to quantization and non-commutative geometry, and leads to the notion of quantum curves. Quantum curves can be thought of as differential operators $\widehat{A}(\hat x, \hat y)$ that impose linear, differential, Schroedinger-like equations on certain wave-functions $\Psi(x)$
\be
\widehat{A}(\hat x, \hat y)\Psi(x) = 0,      \label{Ahatxy}
\ee
where $\hat x$ and $\hat y$, represented respectively by multiplication by $x$ and $\hbar\partial_x$, are operators satisfying the commutation relation
\be
[\hat  y, \hat x] = \hbar.
\ee
In the limit $\hbar\to 0$ these operators commute and can be identified with complex numbers, and the quantum curve equation reduces to a ``classical'' algebraic curve
\be
A(x,y) = 0.     \label{Axy}
\ee
Quantum curves arise in various physical contexts, such as intersecting branes in type IIA string theory \cite{DHSV,DHS}, B-branes in (refined) topological string theory \cite{ADKMV,ACDKV}, surface operators \cite{Kozcaz:2010af,ACDKV}, and they are objects of active studies in knot theory \cite{DijkgraafFuji-2,Borot:2012cw,superA} and from other mathematical perspectives \cite{Dunin-Barkowski:2013wca,Schwarz:2014hfa,Norbury-quantum,Marino:2015nla,Dumitrescu:2015mpa,Bouchard:2016obz}. It has been conjectured that in all situations mentioned above quantum curves can be constructed by means of the topological recursion\cite{abmodel}, and in many cases they can also be constructed explicitly in the formalism of matrix models. In this case the classical algebraic curve (\ref{Axy}) is identified with the spectral curve of a matrix model, and the differential equation (\ref{Ahatxy}) is the equation satisfied by a determinant expectation value. It has been shown recently -- in the formalism of matrix models and the topological recursion -- that to a given algebraic curve one can in fact associate an infinite family of quantum curves, which are in one-to-one correspondence with, and have the structure of, Virasoro singular vectors \cite{Manabe:2015kbj}. This statement can be viewed as a consequence
of the familiar relation between matrix models and conformal field theory \cite{Fukuma:1990jw,Dijkgraaf:1990rs}.

In this paper we combine the two above generalizations and construct supersymmetric quantum curves, or super-quantum curves for short. Our construction is based on so-called super-eigenvalue models, introduced in \cite{AlvarezGaume:1991jd} and then analyzed in \cite{Becker:1992rk,McArthur:1993hw,Plefka:1996tt,Semenoff:1996vm,Itoyama:2003mv}, which can be regarded as supersymmetric generalizations of matrix models. In this paper we first generalize super-eigenvalue models to the $\beta$-deformed case, and then formulate a supersymmetric generalization of the construction presented in \cite{Manabe:2015kbj}. Analogously as in \cite{Manabe:2015kbj}, we find that to a given super-eigenvalue model one can associate an infinite family of super-quantum curves, which have the structure of super-Virasoro singular vectors. These super-quantum curves take form of differential equations that annihilate wave-functions, which are constructed as certain $\alpha/\beta$-deformed super-matrix integrals.

Let us stress that our results illustrate more general phenomenon, i.e.\ the existence of infinite families of quantum curves associated to certain underlying symmetries. This phenomenon is manifest in the context of matrix models or their generalizations. In the case of ordinary hermitian matrix models there is an underlying Virasoro symmetry, which implies the existence of quantum curves having the structure of Virasoro singular vectors  \cite{Manabe:2015kbj}. In the case of super-eigenvalue models considered in this paper, there is an underlying super-Virasoro symmetry, and quantum curves take form of super-Virasoro singular vectors. For other types of matrix models, e.g.\ involving other ensembles of matrices, or deformations of the measure or the potential, or multi-matrix models, etc., there is also an underlying symmetry, whose manifestation must be the existence of a family of quantum curves, having the structure of associated singular vectors. Explicit analysis of such generalizations is an interesting task for future work.

It would also be interesting to generalize our results beyond the realm of matrix models. In the case of hermitian matrix models such a generalization can be formulated by means of the topological recursion, so that one can essentially associate a family of quantum curves to a given algebraic curve (irrespective of the existence of a matrix model). Reformulation of other types of matrix models, in particular the super-eigenvalue model, in terms of generalized topological recursions would enable to identify yet more general quantum curves. 

There are many other directions to explore. It would be interesting to relate our super-quantum curve to the representation of super-Virasoro singular vectors in terms of super-Jack polynomials \cite{mimachi1995,Desrosiers:2012zt,Blondeau-Fournier:2016jth}. While our super-quantum curves correspond to singular vectors in the Neveu-Schwarz sector, it is desirable to identify analogous curves for the Ramond sector, possibly following some ideas in \cite{Blondeau-Fournier:2016jth,Rim2016}. It would be interesting to interpret our results in the context of a relation between super-Liouville theory and gauge theories on ALE spaces \cite{Bonelli:2011kv}, or to find links to some putative supersymmetric version of topological string theory \cite{Schwarz:1995ak,Jia:2016jlo}.

As a word of warning, we stress that super-quantum curves introduced in this paper should not be confused with (quantum) super-A-polynomials \cite{superA}. In the latter case the prefix \emph{super} has to do with homological knot invariants (in analogy with superpolynomials), while in this paper \emph{super} refers to a supersymmetric generalization. 

\bigskip

Let us briefly summarize the results of this paper. First, we introduce a $\beta$-deformed super-eigenvalue model as a formal integral
\begin{equation}
Z= \int\prod_{a=1}^Ndz_ad\vartheta_a\Delta(z,\vartheta)^{\beta}e^{-\frac{\sqrt{\beta}}{\hbar}\sum_{a=1}^NV(z_a,\vartheta_a)},  
\label{matrix_def_intro}
\end{equation}
where $\Delta(z,\vartheta)=\prod_{1\le a<b\le N}(z_a-z_b-\vartheta_a\vartheta_b)$ and the potential takes form of a series
\be
V(x,\theta)=V_{B}(x)+V_{F}(x)\theta,\qquad V_{B}(x)=\sum_{n=0}^{\infty}t_nx^n,\qquad
V_{F}(x)=\sum_{n=0}^{\infty}\xi_{n+1/2}x^n,
\ee
where $t_n$ and $\xi_{n+1/2}$ are respectively bosonic and fermionic parameters called times. The partition function (\ref{matrix_def_intro}) is referred to as the super-eigenvalue integral, with bosonic $z_a$ and their fermionic superpartners $\vartheta_a$ interpreted as super-eigenvalues of some putative ensemble of matrices. Although one way of writing (\ref{matrix_def_intro}) at $\beta=1$ as an integral over some ensemble of matrices was proposed in \cite{Takama:1992np} and an explicit form of such an integral was given for small $N$, it is not possible to write it explicitly for general $N$. Nonetheless, it is convenient to call (\ref{matrix_def_intro}) as a super-eigenvalue model, or a super-matrix model, due to the analogy with bosonic matrix models. 

It is known that the bosonic partition function satisfies a set of Virasoro constraints, and one can derive analogous super-Virasoro constraints in case of the super-eigenvalue model (\ref{matrix_def_intro}). In this paper we derive such constraints in the $\beta$-deformed case and show that they take form
\begin{equation}
g_{n+1/2}Z=\ell_{n}Z=0,\qquad \textrm{for}\ n\ge -1,  \label{superVirasoro-Z-intro}
\end{equation}
where $g_{n+1/2}$ and $\ell_{n}$ are differential operators written in terms of times $t_n$ and $\xi_{n+1/2}$, which form a representation of the super-Virasoro algebra. The above equation means that the matrix model partition function encodes the Neveu-Schwarz (NS) vacuum in an auxiliary superconformal field theory associated to the super-eigenvalue model. We also show that in the large $N$ limit a distribution of eigenvalues is encoded in a super-spectral curve, which is described by a pair of equations (\ref{b_f_sp_curve}) 
\be
y_B(x)y_F(x)+G(x)=y_B(x)^2+y_F'(x)y_F(x)+2L(x)=0,  \label{sp-curve-intro}
\ee
where $G(x)$ and $L(x)$ are defined in (\ref{def_GL}). One can also eliminate the differential $y_F'(x)$ of $y_F(x)$ with respect to $x$ from (\ref{sp-curve-intro}) and write this equation in the form of a supersymmetric algebraic curve (\ref{sp_r_curve}). Such a super-spectral curve was identified already in \cite{AlvarezGaume:1991jd}. 

Having discussed properties of the $\beta$-deformed super-eigenvalue model, we associate to it a family of wave-functions parametrized by a parameter $\alpha$; we also refer to these wave-functions as $\alpha/\beta$-deformed super-matrix integrals. These wave-functions are a generalization of a determinant (characteristic polynomial) expectation value in a bosonic matrix model and take form
\be
\widehat{\chi}_{\alpha}(x,\theta) = e^{\frac{\alpha}{\hbar^2}V_B(x)+\frac{\alpha}{\hbar^2}V_F(x)\theta}\left< e^{-\frac{\sqrt{\beta}}{\hbar}\sum_{a=1}^N  \alpha\left(\log(x-z_a)-\frac{\theta}{x-z_a}\vartheta_a\right)} \right> \equiv \widehat{\chi}_{B,\alpha}(x)+\widehat{\chi}_{F,\alpha}(x)\theta.    \label{wave-function-intro}
\ee
In the right hand side we identified bosonic and fermionic components of the wave-function, and denoted them $\widehat{\chi}_{B,\alpha}(x)$ and $\widehat{\chi}_{F,\alpha}(x)$ respectively. We show that the $\alpha/\beta$-deformed integral (\ref{wave-function-intro}) also satisfies super-Virasoro constraints that generalize (\ref{superVirasoro-Z-intro}), and find corresponding super-Virasoro generators $g^{\alpha}_{n+1/2}(x,\theta)$ and $\ell_{n}^{\alpha}(x,\theta)$.

Having defined the wave-function $\widehat{\chi}_{\alpha}(x,\theta)$, we pose the crucial question in this paper: does it satisfy a differential equation in $x$ of a finite order? We find a very interesting answer to this question: $\widehat{\chi}_{\alpha}(x,\theta)$ satisfies such a differential equation only for special values of $\alpha$, labeled by two positive integers $p$ and $q$
\begin{equation}
\alpha=\alpha_{p,q} = \frac{(p-1)\beta^{1/2}-(q-1)\beta^{-1/2}}{2}\hbar,
\qquad \textrm{with $p-q\in 2{\IZ}$.}  \label{s_vir_sing_momenta-intro}
\end{equation}
Remarkably, these are the values of NS degenerate momenta in super-Virasoro algebra, and corresponding wave-functions may be referred to as NS wave-functions. We then show that for the above values of $\alpha$ the wave-function can be identified as an expectation value of a super-Virasoro degenerate primary field in an auxiliary superconformal field theory associated to the super-eigenvalue model, while the differential equation annihilating the wave-function has the structure of the corresponding super-Virasoro singular vector at level $n=pq/2$, when expressed in terms of appropriate generators of the super-Virasoro algebra. We identify such a differential equation as a super-quantum curve and for a given value of $\alpha$ write it as (\ref{Achi0})
\begin{equation}
\widehat{A}_{n}^{\alpha}\widehat{\chi}_{\alpha}(x,\theta)=0.   \label{Achi0-intro}
\end{equation}
Moreover, we show that at the lowest nontrivial level $3/2$ this super-quantum curve reduces to the classical super-spectral curve (\ref{sp-curve-intro}).

In more detail, we find two representations of super-Virasoro algebra, given in (\ref{h_g_chi_rep}) and (\ref{h_g_bf_chi_rep}), which can be used to write a super-quantum curve, and which act respectively on the wave-function $\widehat{\chi}_{\alpha}(x,\theta)$ and its bosonic component $\widehat{\chi}_{B,\alpha}(x)$ defined in (\ref{wave-function-intro}). We show that the bosonic component $\widehat{\chi}_{B,\alpha}(x)$ is annihilated
\be
\widehat{\mathsf{A}}_{n}^{(0)}\widehat{\chi}_{B,\alpha_{p,q}}(x) = 0  \label{A0-chiB-intro}
\ee
by an operator $\widehat{\mathsf{A}}_{n}^{(0)}$, which takes form of the operator encoding a super-Virasoro singular vector, written in terms of generators (\ref{h_g_bf_chi_rep}). This operator can be easily transformed into an operator (super-quantum curve)
\begin{equation}
\widehat{A}_{n}^{\alpha}=
\widehat{A}_{n}^{(0)}-\theta\partial_{\theta}\widehat{A}_{n}^{(0)}
-\theta\widehat{A}_{n}^{(1)},
\label{q_curve_gen_c-intro}
\end{equation}
that annihilates the whole wave-function $\widehat{\chi}_{\alpha}(x,\theta)$. The operator (\ref{q_curve_gen_c-intro}) is expressed in terms of generators (\ref{h_g_chi_rep}), and its component  $\widehat{A}_{n}^{(1)}$ is related to $\widehat{A}_{n}^{(0)}$ simply by the action of $\widehat{\mathsf{G}}_{-1/2}$, as we explain in detail in section \ref{sec-super-quantum}.

A very interesting feature of our result is that at a given level $n$ super-quantum curves are written in a universal form, which depends on the momentum $\alpha$, and substituting a particular value $\alpha=\alpha_{p,q}$ produces a familiar expression encoding the corresponding singular vector. Moreover, one can substitute values of $\alpha$ corresponding to singular vectors at levels lower than $n$, and then our expression (up to a simple factor) reduces to appropriate expressions encoding corresponding singular vectors at lower levels. Therefore at a given level $n$ we find a universal $\alpha$-dependent expression for the super-quantum curve, which encodes all singular vectors at levels equal or lower than $n$. To our knowledge such universal expressions for super-Virasoro singular vectors have not been previously known.

Let us illustrate the above statements with some examples. We find that an operator $\widehat{\mathsf{A}}_{n}^{(0)}$ at level $n=3/2$ takes form
\be
\widehat{\mathsf{A}}_{3/2}^{(0)}=
\widehat{\mathsf{L}}_{-1}\widehat{\mathsf{G}}_{-1/2}-\frac{\alpha^2}{\hbar^2}\widehat{\mathsf{G}}_{-3/2},
\ee
and it indeed encodes familiar expressions for singular vectors at level $3/2$ upon a substitution of $\alpha=\pm \beta^{\pm 1/2} \hbar$, which correspond to degenerate momenta (\ref{s_vir_sing_momenta-intro}) at this level. Then, as explained in section \ref{sec-super-quantum}, from (\ref{q_curve_gen_c-intro}) the full super-quantum curve is reconstructed as
\be
\widehat{A}_{3/2}^{\alpha}=
-\partial_x\partial_{\theta}-\frac{\alpha^2}{\hbar^2}\widehat{G}_{-3/2}
-\theta\Big(\partial_x^2-\frac{2\alpha^2}{\hbar^2}\widehat{L}_{-2}\Big).
\label{q_curve_eq_3_2-intro}
\ee
Similarly, at level $n=2$ we find
\be
\widehat{\mathsf{A}}_{2}^{(0)}=
\widehat{\mathsf{L}}_{-1}^2-\frac{2\alpha^2}{\hbar^2}\widehat{\mathsf{L}}_{-2}
+\frac{2\alpha^2+Q\hbar\alpha-\hbar^2}{\hbar^2}\widehat{\mathsf{G}}_{-3/2}\widehat{\mathsf{G}}_{-1/2}.
\ee
For the degenerate momentum $\alpha=\alpha_{2,2}=(\beta^{1/2}-\beta^{-1/2})\hbar/2$ this expression indeed reduces to a familiar expression encoding a super-Virasoro singular vector at level 2, and for momenta at lower levels it reduces to expressions for corresponding singular vectors at those lower levels. From the above result we can also easily reconstruct the full form of $\widehat{A}_{2}^{\alpha}$, which is given in (\ref{q_curve_eq_2}). Furthermore, at level $n=5/2$, in (\ref{q_curve_eq_c_5_2}) we find analogous universal $\alpha$-dependent expression for $\widehat{\mathsf{A}}_{5/2}^{(0)}$ that encodes singular vectors up to this level, and then the full super-quantum curve $\widehat{A}_{5/2}^{\alpha}$ is given in (\ref{q_curve_eq_5_2}).

Having introduced and explicitly identified super-quantum curves we analyze some of their properties. Among others we discuss two quantum structures encoded in their form, and analyze two corresponding classical limits: the 't Hooft limit and the classical super-Liouville (or equivalently the Nekrasov-Shatashvili \cite{NS}) limit. (As the second word of warning, one needs to be careful in deciphering acronyms when analyzing the NS limit of the NS wave-function.)

Apart from analyzing super-eigenvalue models with general potentials (involving generic times $t_n$ and $\xi_r$), we also consider a specialization of our formalism to models with fixed potentials, such as the super-gaussian and the super-multi-Penner model. In particular we show that familiar objects in super-Liouville theory (the form of correlation functions, super-BPZ equations, etc.) arise from the latter specialization. Furthermore, for completeness and in order to compare classical limit of super-quantum curves with classical super-spectral curves, we conduct an analysis of planar solutions of super-eigenvalue models.

\bigskip

The plan of this paper is as follows. In section \ref{sec-superVirasoro} we review properties of the super-Virasoro algebra and a construction of its singular vectors. In section \ref{sec-beta} we introduce a $\beta$-deformed super-eigenvalue model, show that its partition function satisfies super-Virasoro constraints, and identify its super-spectral curve. In section \ref{sec-alpha-beta} we introduce an $\alpha/\beta$-deformed matrix integral, find a representation of super-Virasoro algebra associated to it, as well as a related representation, which provides building blocks of super-quantum curves. In section \ref{sec-super-quantum} we present a general construction of those super-quantum curves and illustrate it in several examples at levels $1/2$, $3/2$, $2$, and $5/2$. In section \ref{sec-limits} we discuss a double quantum structure encoded in super-quantum curves, and two corresponding classical limits: the 't Hooft limit and the Nekrasov-Shatashvili (or the classical super-Liouville) limit. In section \ref{sec-examples} we specialize our consideration to super-eigenvalue models with the super-gaussian and the super-multi-Penner potentials. Finally, in the appendix we discuss various operator expressions used in calculations throughout the paper, and also present a general form of a planar solution of the super-eigenvalue model.


\section{Super-Virasoro algebra, singular vectors, and free field realization}   \label{sec-superVirasoro}

The N = 1 super-Virasoro algebra is defined by the following (anti)commutation relations
\begin{align}
\begin{split}
\{G_r,G_s\}&=2L_{r+s}+\frac{c}{3}\big(r^2-\frac14\big)\delta_{r+s,0},\\
[L_m,G_r]&=\big(\frac{m}{2}-r\big)G_{m+r},\\
[L_m,L_n]&=(m-n)L_{m+n}+\frac{c}{12}(m^3-m)\delta_{m+n,0}.
\label{s_vir_alg}
\end{split}
\end{align}
In this paper we consider primarily the NS (Neveu-Schwarz) sector of (\ref{s_vir_alg}), in which
the indices of generators $G_r$ take half-integer values, $r\in\mathbb{Z}+\frac{1}{2}.$
Indices $n$ of generators $G_n,$ forming the Virasoro subalgebra of the NS superalgebra, are integers.

Let $\nu_{\Delta}$ be the highest weight state with respect to the NS
algebra (\ref{s_vir_alg})
\begin{equation}
\label{highest}
L_0\nu_{\Delta}=\Delta\nu_{\Delta},
\hskip 5mm
L_m\nu_{\Delta}= G_r\nu_{\Delta} =0,
\hskip 5mm
m, r > 0.
\end{equation}
Denote by $\pi_{\rm\scriptscriptstyle NS}^n(\Delta)$ the free vector space  generated by all vectors
of the form
\begin{equation}
\label{basis}
\nu_{\Delta,MR}
\; = \;
L_{-M}g_{-R}\nu_{\Delta}
\; \equiv \;
L_{-m_j}\ldots L_{-m_1} G_{-r_k}\ldots G_{-r_1}\nu_{\Delta},
\end{equation}
where $R$ and $M$ are multi-indices such that
\[
0 < r_1   < r_2\ldots < r_k ,
\hskip 1cm
0 \leqslant m_1 \leqslant m_2 \ldots \leqslant m_j,
\]
and
\[
|M| + |R| \equiv m_1 + \ldots m_j + r_1 + \ldots r_k = n.
\]

The ${1\over 2}\mathbb{Z}$-graded representation of the NS superconformal algebra
determined on the space
$$
\pi_{\rm\scriptscriptstyle NS}(\Delta)
=
\hspace*{-3mm}
\bigoplus\limits_{{n}\in {1\over 2}\mathbb{N}\cup\{0\}}
\hspace*{-1mm}
\pi^n_{\rm\scriptscriptstyle NS}(\Delta),
\hskip 5mm
\pi^0_{\rm\scriptscriptstyle NS}(\Delta)=\mathbb{C}\, \nu_\Delta,
$$
by the relations
(\ref{s_vir_alg}) and (\ref{highest}) is called the NS supermodule
of the highest weight $\Delta$ and the central charge $c.$
Each $\pi^n_{\rm\scriptscriptstyle NS}(\Delta)$ is an eigenspace of $L_0$ with the eigenvalue
$\Delta +{n}$. The space $\pi_{\rm\scriptscriptstyle NS}(\Delta)$ has also a natural $\mathbb{Z}_2$-grading
\[
\pi_{\rm\scriptscriptstyle NS}(\Delta)
\; = \;
\pi^+_{\rm\scriptscriptstyle NS}(\Delta) \oplus
\pi^-_{\rm\scriptscriptstyle NS}(\Delta),
\hskip 5mm
\pi^+_{\rm\scriptscriptstyle NS}(\Delta)
\; = \hskip -5pt
\bigoplus\limits_{m\in \mathbb{N}\cup \{0\}}
\hskip -3pt
\pi^m_{\rm\scriptscriptstyle NS}(\Delta),
\hskip 5mm
\pi^-_{\rm\scriptscriptstyle NS}(\Delta)
\; = \hskip -5pt
\bigoplus\limits_{m\in \mathbb{N}\cup \{0\}}
\hskip -3pt
\pi^{m+\frac12}_{\rm\scriptscriptstyle NS}(\Delta),
\]
where $\pi^\pm_{\rm\scriptscriptstyle NS}(\Delta)$ are eigenspaces of the
parity operator $(-1)^F= (-1)^{2(L_0-\Delta)}$. NS generators $L_m$ and $G_r$ are,
respectively, even and odd with respect to this grading,
\[
\left[(-1)^F,L_m\right] \; = \; \left\{(-1)^F,G_r\right\} \; = \; 0.
\]

A nonzero element ${\chi}\in \pi_{\rm\scriptscriptstyle NS}(\Delta)$ of degree $n > 0$
is called a singular vector if
it satisfies the highest weight conditions (\ref{highest}) with
$
L_0{\chi} =(\Delta +n){\chi}.
$
It generates its own NS supermodule $\pi_{\rm\scriptscriptstyle NS}(\Delta+n)$, which is a
submodule of $\pi_{\rm\scriptscriptstyle NS}(\Delta).$

There exists on $\pi_{\rm\scriptscriptstyle NS}(\Delta)$ a natural, symmetric bilinear form
$\langle \cdot \,, \cdot \rangle$ uniquely determined  by the relations
$(L_{m})^{\dag}=L_{-m},(G_{r})^{\dag}=G_{-r}$
and normalization condition $\langle\nu_{\Delta},\nu_{\Delta}\rangle =1.$
The singular vector $\chi$  is orthogonal with respect to
$\langle \cdot \,, \cdot\rangle$ to all vectors in $\pi_{\rm\scriptscriptstyle NS}(\Delta)$
including itself,
\begin{equation}
\forall \xi \in \pi_{\rm\scriptscriptstyle NS}(\Delta): \hskip 5pt \langle\,\xi\,|\,\chi\,\rangle = 0.
\end{equation}
Consequently, if we denote by $G^{\,{n}}_{c,\Delta}$ the matrix of $\langle \cdot \,, \cdot \rangle$
on $\pi^n_{\rm\scriptscriptstyle NS}(\Delta)$ calculated
in the basis (\ref{basis})
\begin{equation}
\label{matrix}
\left[ G^{\,{n}}_{c,\Delta}\right]_{MR,NS}
= \;
\langle \nu_{\Delta,MR},\nu_{\Delta,NS}\rangle,
\hskip 1cm |M|+|R| = |N|+|S| = n,
\end{equation}
then it is nonsingular if and only if the supermodule $\pi_{\rm\scriptscriptstyle NS}(\Delta)$ does
not contain singular vectors of degrees ${1\over 2},1,\dots,{n}.$
The determinant of this matrix is given by the Kac theorem \cite{Kac}
\begin{equation}
\label{Kac}
\det G^{\,{n}}_{c,\Delta}
\; = \;
K_{n}\hskip -2mm
\prod\limits_{1\leqslant pq \leqslant 2{n}}
\big(\Delta-\Delta_{p,q}(c)\big)^{P_{\rm\scriptscriptstyle NS}({n}-\frac{pq}{2})}.
\end{equation}
Here $K_{n}$ depends only on the level $n,$ the sum $p+q$ must be even,
the multiplicity of each zero of the Kac determinant is given by
$P_{\rm\scriptscriptstyle NS}({n})= \dim \pi^{n}_{\rm\scriptscriptstyle NS}(\Delta)$
and
\begin{eqnarray}
\label{delta:rs}
\Delta_{p,q}(c)
& = &
-\frac{pq-1}{4} + \frac{p^2-1}{8}\beta + \frac{q^2-1}{8}\frac{1}{\beta},
\end{eqnarray}
where
\begin{eqnarray}
\nonumber
c & = & \frac32 -3\left(\beta^{-1/2}-\beta^{1/2}\right)^2.
\end{eqnarray}
The combination
\begin{equation}
Q=\beta^{-1/2}-\beta^{1/2},
\end{equation}
which we shall frequently use in what follows, corresponds to the background charge in the $\mathcal{N}=1$ super-Liouville field theory.
Finally, if we parametrize $\Delta$ as
\begin{equation}
\label{Delta:through:alpha}
\Delta_{\alpha} = \frac{\alpha}{2\hbar}\left(\frac{\alpha}{\hbar}-Q\right) ,
\end{equation}
then the equation (\ref{delta:rs}) implies
\[
\alpha=\alpha_{p,q} = \frac{(p-1)\beta^{1/2} - (q-1)\beta^{-1/2}}{2}\hbar,
\]
in agreement with (\ref{s_vir_sing_momenta-intro}).

Partial results containing the form of Neveu-Schwarz singular vectors can be found in
\cite{Benoit:1991pa,Benoit:1992fw,Huang:1992zi,Huang:1992nq,Benoit:1994ct}.
For the first few levels one can also compute their
form directly from the definition. Below we give some examples (for more examples see \cite{Zamolodchikov:2003yb}).
\begin{itemize}
\item
For $n = 1/2$ we have a singular vector for $p = q = 1$, which corresponds to $\Delta = 0$ (i.e.\ the singular vector  is
in the NS supermodule of the vacuum) given by
\begin{equation}
\label{singular:vector:level:12}
\chi_{11} = G_{-1/2}\nu_{11}, \hskip 1cm \nu_{pq} \equiv \nu_{\Delta_{p,q}}.
\end{equation}
\item
For $n = 3/2$ we have either $(p,q) = (3,1)$ or $(p,q) = (1,3).$ The former case corresponds to
\begin{equation}
\label{singular:vector:level:32}
\chi_{31} = \left(L_{-1}G_{-1/2} - \beta G_{-3/2}\right)\nu_{31},
\end{equation}
while the latter is obtained by changing $\beta$ to $-\beta^{-1}$ and $\nu_{31}$ to $\nu_{13}.$
\item
For $n = 2$ we have $(p,q) = (2,2)$ and
\begin{equation}
\label{singular:vector:level:2}
\chi_{22} = \left(L_{-1}^2 -\frac12\left(\beta^{-1/2} - \beta^{1/2}\right)^2 L_{-2} - G_{-3/2}G_{-1/2}\right)\nu_{22}.
\end{equation}
\item
Finally, for $n = 5/2$ we have $(p,q) = (5,1)$ or $(p,q) = (1,5).$ In the former case
\begin{equation}
\label{singular:vector:level:52}
\chi_{51} = \left(L_{-1}^2 G_{-1/2} +2\beta(3\beta -1) -3\beta G_{-3/2}L_{-1} -2\beta L_{-2} G_{-1/2}\right)\nu_{51}.
\end{equation}
The singular vector $\chi_{15}$ is obtained by changing $\beta$ to $-\beta^{-1}$ and $\nu_{51}$ to $\nu_{15}.$
\end{itemize}

In what follows we will also take advantage of the free field realization of the super-Virasoro algebra, in terms of the free bosonic and free fermionic operators on a two-sphere $S^2$, such that
\begin{align}
\phi(x_1)\phi(x_2)&=\log(x_1-x_2)+\ldots,\\
\psi(x_1)\psi(x_2)&=\frac{1}{x_1-x_2}+\ldots.
\end{align}
From these operators one can construct the superconformal current
\begin{equation}
S(x)=:\!\psi(x)\partial_x\phi(x)\!:+Q\partial_x\psi(x)
\label{s_current}
\end{equation}
and the energy-momentum tensor
\begin{equation}
T(x)=\frac12:\!\partial_x\phi(x)\partial_x\phi(x)\!:+\frac12:\!\partial_x\psi(x)\psi(x)\!:
+\frac12Q\partial_x^2\phi(x).
\label{e_m_tensor}
\end{equation}
It follows that
\begin{align}
\begin{split}
&
S(x_1)S(x_2)=\frac{2c}{3(x_1-x_2)^3}+\frac{2}{x_1-x_2}T(x_2)+\ldots,\\
&
T(x_1)S(x_2)=\frac{3}{2(x_1-x_2)^2}S(x_2)+\frac{1}{x_1-x_2}S'(x_2)+\ldots,\\
&
T(x_1)T(x_2)=\frac{c}{2(x_1-x_2)^4}+\frac{2}{(x_1-x_2)^2}T(x_2)+\frac{1}{x_1-x_2}T'(x_2)+\ldots,
\label{ope_s_em}
\end{split}
\end{align}
where the central charge is given by
\begin{equation}
c=\frac{3}{2}-3Q^2.
\end{equation}
From the OPE (\ref{ope_s_em}) it follows that the modes $G_r$ and $L_n$ of the superconformal current and the energy-momentum tensor
\begin{equation}
S(x)=\sum_{r\in{\IZ}+1/2}G_rx^{-r-3/2},\ \ \ \
T(x)=\sum_{n\in{\IZ}}L_nx^{-n-2},
\label{mode_exp_st}
\end{equation}
satisfy the commutation relations of the super-Virasoro algebra (\ref{s_vir_alg}).

Note that $G_r$ and $L_n$ discussed above are abstract generators of the super-Virasoro algebra. In the rest of this paper we will find several explicit representations of such super-Virasoro generators, which are expressed in terms of bosonic and fermionic times encoded in the super-eigenvalue model, as well as arguments $x$ and $\theta$ of supersymmetric wave-functions. To avoid confusion we write generators in those different representations using different fonts.


\section{$\beta$-deformed super-eigenvalue models}   \label{sec-beta}

In this section we introduce the $\beta$-deformed super-eigenvalue model. This model is a $\beta$-deformation of the super-eigenvalue model formulated in \cite{AlvarezGaume:1991jd}, which itself is a generalization of an eigenvalue representation of the bosonic, hermitian matrix model to the supersymmetric case. The analysis of expectation values in the $\beta$-deformed super-eigenvalue model will be the main tool in our considerations. The partition function of the $\beta$-deformed super-eigenvalue model can be represented as a formal integral over an even number $N$ of bosonic eigenvalues $z_a$ and fermionic eigenvalues $\vartheta_a$
\begin{equation}
Z= \int\prod_{a=1}^Ndz_ad\vartheta_a\Delta(z,\vartheta)^{\beta}e^{-\frac{\sqrt{\beta}}{\hbar}\sum_{a=1}^NV(z_a,\vartheta_a)},  
\label{matrix_def}
\end{equation}
where
\begin{equation}
\Delta(z,\vartheta)=\prod_{1\le a<b\le N}(z_a-z_b-\vartheta_a\vartheta_b),
\end{equation}
and the potential is given by
\begin{align}
\begin{split}
V(x,\theta)&=V_{B}(x)+V_{F}(x)\theta,\\
V_{B}(x)&=\sum_{n=0}^{\infty}t_nx^n,\ \ \ \
V_{F}(x)=\sum_{n=0}^{\infty}\xi_{n+1/2}x^n.
\end{split}
\end{align}
Here $t_n$ are bosonic parameters and $\xi_{n+1/2}$ are fermionic parameters with $\{\vartheta_a, \xi_{n+1/2}\}=0$, which we call (even and odd, respectively) times.
Instead of parameters $\hbar$ and $\beta$, sometimes it is convenient to use parameters
\begin{equation}
\epsilon_1=-\beta^{1/2}\hbar,\qquad   \epsilon_2=\beta^{-1/2}\hbar.   \label{e1e2}
\end{equation}
For $V_F(x)=0$ and $\beta=1$ the partition function $Z$ is equal to the square of the partition function of the usual bosonic matrix model (which involves integration over eigenvalues $z_a$ only, so that the dependence on fermionic variables is absent in the above formulas) with the same (bosonic) potential $V_B(x)$ \cite{AlvarezGaume:1991jd,Becker:1992rk,Plefka:1996tt}. In what follows by $\left<\cdots\right>$ we mean an unnormalized expectation value, for an operator $\mathcal{O}$ defined as
\begin{equation}
\left<\mathcal{O}\right> = \int\prod_{a=1}^Ndz_ad\vartheta_a\Delta(z,\vartheta)^{\beta}e^{-\frac{\sqrt{\beta}}{\hbar}\sum_{a=1}^NV(z_a,\vartheta_a)}\mathcal{O}.
\label{def_unnorm_exp}
\end{equation}

In case of the usual bosonic matrix model, for some special values of $\beta$, the partition function is indeed a representation of an integral over some ensemble of matrices (hermitian, orthogonal, or symplectic). Moreover, for all values of $\beta$, the  partition function of the bosonic model satisfies Virasoro constraints, i.e.\ it is annihilated by operators $\ell_n$, for $n\geq -1$, which have an explicit representation in terms of times $t_k$ and satisfy the Virasoro algebra; in this sense the partition function is analogous to the vacuum state in a conformal field theory. For the super-eigenvalue model the situation is different -- its partition function cannot be represented as an integral over some (supersymmetric) matrices, for any values of $\beta$. However a remarkable feature of this model, and the original motivation to construct it, is the fact that its partition function satisfies super-Virasoro constraints, i.e.\ it is annihilated by a relevant set of super-Virasoro generators, analogously to the supersymmetric vacuum in a superconformal field theory.

\subsection{Ward identities and super-Virasoro constraints}   \label{ssec-Ward}

In what follows we derive super-Virasoro constraints for the super-eigenvalue model, for arbitrary values of $\beta$. Let us first write down Ward identities, which in the context of matrix models take form of loop equations, for the $\beta$-deformed super-eigenvalue model. Note that the partition function (\ref{matrix_def}) is invariant under a fermionic shift
\begin{equation}
z_a\ \to\ z_a+\frac{\vartheta_a \delta}{x-z_a},\qquad
\vartheta_a\ \to\ \vartheta_a+\frac{\delta}{x-z_a},
\end{equation}
where $\delta$ is a fermionic constant. In consequence one obtains a fermionic Ward identity
\begin{equation}
\int\prod_{a=1}^Ndz_ad\vartheta_a
\sum_{a=1}^N\big(\vartheta_a\partial_{z_a}-\partial_{\vartheta_a}\big)
\bigg[\frac{1}{x-z_a}
\Delta(z,\vartheta)^{\beta}
e^{-\frac{\sqrt{\beta}}{\hbar}\sum_{a=1}^NV(z_a,\vartheta_a)}\bigg]=0.
\label{f_ward_id}
\end{equation}
Using
\begin{equation}
\frac{1}{z_a-z_b-\vartheta_a\vartheta_b}=\frac{z_a-z_b+\vartheta_a\vartheta_b}{(z_a-z_b-\vartheta_a\vartheta_b)(z_a-z_b+\vartheta_a\vartheta_b)}
=\frac{1}{z_a-z_b}+\frac{\vartheta_a\vartheta_b}{(z_a-z_b)^2},
\label{rel_vand}
\end{equation}
the fermionic Ward identity (\ref{f_ward_id}) can be written as
\begin{equation}
\left<S_+(x)\right>=0,
\label{f_ward_id2}
\end{equation}
where
\begin{align}
S_+(x)&=
\beta\sum_{a,b=1}^N\frac{\vartheta_a}{(x-z_a)(x-z_b)}
+(1-\beta)\sum_{a=1}^N\frac{\vartheta_a}{(x-z_a)^2} +
\nonumber\\
&\ \ \
-\frac{\sqrt{\beta}}{\hbar}\sum_{a=1}^N\frac{1}{x-z_a}
\big(V_{B}'(z_a)\vartheta_a+V_{F}(z_a)\big).
\label{s_p_x}
\end{align}
Similarly, the invariance of the partition function (\ref{matrix_def}) under the bosonic infinitesimal transformation
\begin{equation}
z_a\ \to\ z_a+\frac{\varepsilon}{x-z_a},\qquad
\vartheta_a\ \to\ \vartheta_a+\frac{\varepsilon\vartheta_a}{2(x-z_a)^2},
\end{equation}
with an infinitesimal bosonic constant $\varepsilon$, leads to a bosonic Ward identity
\begin{equation}
\int\prod_{a=1}^Ndz_ad\vartheta_a
\sum_{a=1}^N\bigg(\partial_{z_a}-\partial_{\vartheta_a}\frac{\vartheta_a}{2(x-z_a)}\bigg)
\bigg[\frac{1}{x-z_a}
\Delta(z,\vartheta)^{\beta}
e^{-\frac{\sqrt{\beta}}{\hbar}\sum_{a=1}^NV(z_a,\vartheta_a)}\bigg]=0.
\label{b_ward_id}
\end{equation}
Using the relation (\ref{rel_vand}), this Ward identity is written as
\begin{equation}
\left<T_+(x)\right>=0,
\label{b_ward_id2}
\end{equation}
where
\begin{align}
T_+(x)&=
\frac{\beta}{2}\sum_{a,b=1}^N\frac{1}{(x-z_a)(x-z_b)}
+\frac{\beta}{2}\sum_{a,b=1}^N\frac{\vartheta_a\vartheta_b}{(x-z_a)(x-z_b)^2}
+\frac{1}{2}(1-\beta)\sum_{a=1}^N\frac{1}{(x-z_a)^2} +
\nonumber\\
&\qquad
-\frac{\sqrt{\beta}}{\hbar}\sum_{a=1}^N\frac{1}{x-z_a}
\big(V_{B}'(z_a)+V_{F}'(z_a)\vartheta_a\big)
-\frac{\sqrt{\beta}}{2\hbar}\sum_{a=1}^N\frac{V_{F}(z_a)\vartheta_a}{(x-z_a)^2}.
\label{t_p_x}
\end{align}

The quantities $S_+(x)$ and $T_+(x)$, which implement Ward identities given above, can be identified with positive parts of the superconformal current and the energy-momentum tensor of an auxiliary superconformal field theory, which can be associated to the super-eigenvalue model. Note that time derivatives of the partition function (\ref{matrix_def}) can be identified with expectation values of the combinations of eigenvalues, by the following identification
\begin{equation}
\sum_{a=1}^Nz_a^n\ \longleftrightarrow\ -\frac{\hbar}{\sqrt{\beta}}\partial_{t_{n}},\qquad
\sum_{a=1}^Nz_a^n\vartheta_a\ \longleftrightarrow\ -\frac{\hbar}{\sqrt{\beta}}\partial_{\xi_{n+1/2}}.
\label{id_mat_free}
\end{equation}
We can therefore associate to the super-eigenvalue model a free boson and a free fermion fields, which in terms of the above identifications can be written as
\begin{align}
\phi(x)&=\frac{1}{\hbar}V_{B}(x)-\sqrt{\beta}\sum_{a=1}^N\log(x-z_a),
\label{free_boson}
\\
\psi(x)&=\frac{1}{\hbar}V_{F}(x)-\sqrt{\beta}\sum_{a=1}^N\frac{\vartheta_a}{x-z_a}.
\label{free_fermion}
\end{align}
We can then build the superconformal current as in (\ref{s_current}). We find that
\begin{equation}
S(x)=S_-(x)+S_+(x),
\label{s_current_rep}
\end{equation}
where $S_+(x)$ is given by (\ref{s_p_x}) and
\begin{equation}
S_-(x)=\frac{1}{\hbar^2}V_{F}(x)V_{B}'(x)+\frac{Q}{\hbar}V_{F}'(x)-\frac{\sqrt{\beta}}{\hbar}\sum_{a=1}^N\frac{V_{F}(x)-V_{F}(z_a)}{x-z_a}
-\frac{\sqrt{\beta}}{\hbar}\sum_{a=1}^N\frac{\big(V_{B}'(x)-V_{B}'(z_a)\big)\vartheta_a}{x-z_a}.
\end{equation}
Similarly, the energy-momentum tensor (\ref{e_m_tensor}) takes the form
\begin{equation}
T(x)=T_-(x)+T_+(x),
\label{e_m_tensor_rep}
\end{equation}
where $T_+(x)$ is given in (\ref{t_p_x}) and
\begin{align}
\begin{split}
T_-(x)&=\frac{1}{2\hbar^2}\big(V_{B}'(x)^2+V_{F}'(x)V_{F}(x)\big)+\frac{Q}{2\hbar}V_{B}''(x)
-\frac{\sqrt{\beta}}{\hbar}\sum_{a=1}^N\frac{V_{B}'(x)-V_{B}'(z_a)}{x-z_a}\\
&\qquad
-\frac{\sqrt{\beta}}{2\hbar}\sum_{a=1}^N\frac{\big(V_{F}'(x)-V_{F}'(z_a)\big)\vartheta_a}{x-z_a}-\frac{\sqrt{\beta}}{2\hbar}\sum_{a=1}^N\frac{V^{(2)}_{F}(x,z_a)\vartheta_a}{(x-z_a)^2},
\end{split}
\end{align}
where we have defined
\begin{equation}
V^{(2)}_{F}(x,z_a)=V_{F}(x)-V_{F}(z_a)-(x-z_a)V_{F}'(z_a).
\label{v_f_2_def}
\end{equation}
Finally, expanding (\ref{s_current_rep}) and (\ref{e_m_tensor_rep}) in the modes defined as in (\ref{mode_exp_st}), for $n\geq -1$ we find
\be
g_{n+1/2}
=\sum_{k=1}^{\infty}kt_k\partial_{\xi_{k+n+1/2}}+\sum_{k=0}^{\infty}\xi_{k+1/2}\partial_{t_{k+n+1}}+\hbar^2\sum_{k=0}^n\partial_{\xi_{k+1/2}}\partial_{t_{n-k}}  -Q \hbar(n+1)\partial_{\xi_{n+1/2}}
\ee
and
\be
\begin{split}
\ell_{n}&=\sum_{k=1}^{\infty}kt_k\partial_{t_{k+n}}+\frac{\hbar^2}{2}\sum_{k=1}^{n}\partial_{t_k}\partial_{t_{n-k}}+\sum_{k=0}^{\infty}\big(k+\frac{n+1}{2}\big)\xi_{k+1/2}\partial_{\xi_{k+n+1/2}} +   \\
&\qquad \qquad
+\frac{\hbar^2}{2}\sum_{k=1}^{n}k\partial_{\xi_{n-k+1/2}}\partial_{\xi_{k-1/2}}
-\frac12Q\hbar(n+1)\partial_{t_n},
\end{split}
\ee
as well as the modes for $n<-1$ given explicitly in (\ref{ggn}) and (\ref{lln}). In terms of the above generators the fermionic Ward identity (\ref{f_ward_id2}) can be rewritten in the form of super-Virasoro constraints
\begin{equation}
g_{n+1/2}Z=0,\qquad n\ge -1.
\label{s_vir_const}
\end{equation}
Similarly, the bosonic Ward identity (\ref{b_ward_id2}) gives constraints
\begin{equation}
\ell_{n}Z=0,\qquad n\ge -1,
\label{vir_const}
\end{equation}
which can be derived also from (\ref{s_vir_const}) using the commutation relations (\ref{s_vir_alg}). These are the super-Virasoro constraints of the $\beta$-deformed super-eigenvalue model that we have been after.

\subsection{Super-spectral curve}

In the analysis of bosonic matrix models an algebraic curve, called the spectral curve, plays a crucial role. It encodes information about the equilibrium distribution of eigenvalues and can be determined from the large $N$ limit of the loop equation. Upon quantization it is turned into an operator that annihilates a wave-function, defined as a certain expectation value. In this paper we show that an analogous situation arises in super-eigenvalue models -- however in this case the resulting spectral and quantum curves are supersymmetric. In this section we show how a supersymmetric algebraic curve emerges from the large $N$ limit of loop equations (\ref{f_ward_id2}) and (\ref{b_ward_id2}), while corresponding quantum curves are analyzed in the following sections.

By the large $N$ (or classical) limit we understand the limit
\begin{equation}
N\to \infty,\qquad \hbar\to 0,\qquad \textrm{with}\quad \mu=\beta^{1/2} \hbar N=const,
\label{large_N_lim}
\end{equation}
and we also denote
\begin{equation}
\widehat{\hbar}=(\beta^{1/2}-\beta^{-1/2})\hbar.
\end{equation}
In this limit we consider the following expectation values (defined as in (\ref{def_unnorm_exp})) 
\begin{equation}
Y_B(x;\widehat{\hbar})\equiv Y_B(x) = \lim_{\begin{subarray}{c}N\to\infty\\\widehat{\hbar}\,\textrm{fixed}\end{subarray}}\frac{\hbar}{Z}\big<\partial_x\phi(x)\big>,\qquad
Y_F(x;\widehat{\hbar})\equiv Y_F(x) = \lim_{\begin{subarray}{c}N\to\infty\\\widehat{\hbar}\,\textrm{fixed}\end{subarray}}\frac{\hbar}{Z}\big<\psi(x)\big>.
\label{def_y_bf}
\end{equation}
With this notation Ward identities (\ref{f_ward_id2}) and (\ref{b_ward_id2}) yield respectively 
\begin{equation}
Y_B(x)Y_F(x)-V_B'(x)V_F(x)+\widehat{\hbar}\big(V_F'(x)-Y_F'(x)\big)-h^{(0)}(x)=0,
\label{largeN_Ward_b}
\end{equation}
and
\begin{equation}
Y_B(x)^2+Y_F'(x)Y_F(x)-V_B'(x)^2-V_F'(x)V_F(x)+\widehat{\hbar}\big(V_B'(x)-Y_B'(x)\big)-2f^{(0)}(x)=0,
\label{largeN_Ward_f}
\end{equation}
where
\begin{align}
h^{(0)}(x)& =
h^{(0)}(x;\widehat{\hbar}) =
\nonumber\\
& = -\lim_{\begin{subarray}{c}N\to\infty\\\widehat{\hbar}\ \textrm{fixed}\end{subarray}}\frac{\sqrt{\beta}\hbar}{Z}\bigg<
\sum_{a=1}^{N}\bigg(\frac{V_F(x)-V_F(z_a)}{x-z_a}+\frac{\big(V_B'(x)-V_B'(z_a)\big)\vartheta_a}{x-z_a}\bigg)
\bigg>,
\label{h_x_0_cl}
\\
f^{(0)}(x)&=
f^{(0)}(x;\widehat{\hbar}) =
\nonumber\\
&=
-\lim_{\begin{subarray}{c}N\to\infty\\\widehat{\hbar}\ \textrm{fixed}\end{subarray}}\frac{\sqrt{\beta}\hbar}{Z}\bigg<
\sum_{a=1}^{N}\bigg(\frac{V_B'(x)-V_B'(z_a)}{x-z_a}
+\frac{\big(V'_F(x)-V'_F(z_a)\big)\vartheta_a}{2(x-z_a)}
+\frac{V^{(2)}_{F}(x,z_a)\vartheta_a}{2(x-z_a)^2}\bigg)\bigg>,
\label{f_x_0_cl}
\end{align}
where $V^{(2)}_{F}(x,z_a)$ is defined in (\ref{v_f_2_def}). $h^{(0)}(x)$ and $f^{(0)}(x)$ are polynomials in $x$ if the potentials $V_B(x)$ and $V_F(x)$ are polynomials. For $\widehat{\hbar}=0$, or in particular for $\beta=1$, denoting
\be
y_B(x) = Y_B(x;0), \qquad  y_F(x) = Y_F(x;0),   \label{yByF}
\ee
the Ward identities (\ref{largeN_Ward_b}) and (\ref{largeN_Ward_f}) in the large $N$ limit yield  \cite{AlvarezGaume:1991jd}
\begin{align}
\begin{cases}
& A_F(x,y_B|y_F)\equiv y_B(x)y_F(x)+G(x)=0,
\\
& A_B(x,y_B|y_F)\equiv y_B(x)^2+y_F'(x)y_F(x)+2L(x)=0,
\label{b_f_sp_curve}
\end{cases}
\end{align}
where
\begin{align}
\begin{split}
G(x) & = -V_B'(x)V_F(x)-h_{cl}(x),\\
L(x) & = -\frac12 V_B'(x)^2-\frac12 V_F'(x)V_F(x)-f_{cl}(x),\\
h_{cl}(x) & = h^{(0)}(x;0),\qquad   f_{cl}(x) = f^{(0)}(x;0).
\label{def_GL}
\end{split}
\end{align}
Using the first equation in (\ref{b_f_sp_curve}), the differential $y_F'(x)$ in the second equation can be removed by rewriting it as
\begin{align}
\begin{split}
\widetilde{A}_B(x,y_B|y_F)&\equiv y_B(x)A_B(x,y_B|y_F)+y_F(x)\partial_xA_F(x,y_B|y_F)  \\
&
=y_B(x)^3+2L(x)y_B(x)+y_F(x)G'(x)=0.
\end{split}
\end{align}
Then the solution space $\mathcal{C}$ of the equations (\ref{b_f_sp_curve})
can be embedded in a supersymmetric algebraic curve $\widetilde{\mathcal{C}}$
\begin{align}
\begin{split}
\mathcal{C}&=\Big\{(x,y_B|\theta,y_F)\in {\IC}^{2|2}\ \Big|\
A_F(x,y_B|y_F)=\theta A_B(x,y_B|y_F)\Big\}
\\
&
\subset\
\widetilde{\mathcal{C}}=\Big\{(x,y_B|\theta,y_F)\in {\IC}^{2|2}\ \Big|\
y_BA_F(x,y_B|y_F)=\theta \widetilde{A}_B(x,y_B|y_F)\Big\},
\label{sp_r_curve}
\end{split}
\end{align}
where we have introduced a fermionic variable $\theta$ conjugate to $y_F$, $\{\theta, y_F\}=0$.

\section{$\alpha/\beta$-deformed matrix integrals and the wave-function $\widehat{\chi}_{\alpha}(x,\theta)$}   \label{sec-alpha-beta}

The aim of this paper is to construct supersymmetric quantum curves. These objects quantize the solution space $\mathcal{C}$ in (\ref{sp_r_curve}) and by definition impose differential equations for the following wave-function
\begin{equation}
\widehat{\chi}_{\alpha}(x,\theta)=\left<e^{\frac{\alpha}{\hbar}\big(\phi(x) + \psi(x)\theta \big)}\right>,
\label{chi_hat_def}
\end{equation}
where $\theta$ is a fermionic variable with $\{\theta, \vartheta_a\}=\{\theta, \xi_{n+1/2}\}=0$. The expectation value $\left<\cdots\right>$ is understood as in (\ref{def_unnorm_exp}), and we also call the expression (\ref{chi_hat_def}) as the $\alpha/\beta$-deformed matrix integral. The momentum $\alpha$ is a bosonic parameter, and we will see that finite order differential equations for the wave-function arise only for some special values of $\alpha$. Using (\ref{free_boson}) and (\ref{free_fermion}), the wave-function (\ref{chi_hat_def}) can be written as
\begin{equation}
\widehat{\chi}_{\alpha}(x,\theta)=e^{\frac{\alpha}{\hbar^2}V_B(x)+\frac{\alpha}{\hbar^2}V_F(x)\theta}\left<\chi_{\alpha}^{\textrm{ins}}(x,\theta)\right>
\equiv
e^{\frac{\alpha}{\hbar^2}V_B(x)+\frac{\alpha}{\hbar^2}V_F(x)\theta}\chi_{\alpha}(x,\theta),
\label{chi_hat_w}
\end{equation}
where by omitting the proportionality factor we introduced $\chi_{\alpha}(x,\theta)$, and
\begin{equation}
\chi_{\alpha}^{\textrm{ins}}(x,\theta)=e^{-\frac{\sqrt{\beta}}{\hbar}\sum_{a=1}^N  \alpha\left(\log(x-z_a)-\frac{\theta}{x-z_a}\vartheta_a\right)}
=\Big(1+\frac{\sqrt{\beta}}{\hbar}\sum_{a=1}^N\frac{\alpha\theta\vartheta_a}{x-z_a}\Big)\prod_{a=1}^N(x-z_a)^{-\frac{\sqrt{\beta}}{\hbar}\alpha}.
\label{chi_ins_def}
\end{equation}
It is also convenient to decompose the wave-function (\ref{chi_hat_def}) into a bosonic component $\widehat{\chi}_{B,\alpha}(x)$ and its fermionic partner $\widehat{\chi}_{F,\alpha}(x)$
\be
\begin{split}
\widehat{\chi}_{\alpha}(x,\theta) = &
\widehat{\chi}_{B,\alpha}(x)+\widehat{\chi}_{F,\alpha}(x)\theta,  \\
& \widehat{\chi}_{B,\alpha}(x)\equiv
\left<e^{\frac{\alpha}{\hbar}\phi(x)}\right>=\widehat{\chi}_{\alpha}(x,0), \\
& \widehat{\chi}_{F,\alpha}(x)\equiv
\frac{\alpha}{\hbar}\left<\psi(x)e^{\frac{\alpha}{\hbar}\phi(x)}\right>=-\partial_{\theta}\widehat{\chi}_{\alpha}(x,\theta).
\label{wave_bose_fermi}
\end{split}
\ee
In some situations --- in particular in the analysis of the classical limit of super-quantum curves --- we also consider a wave-function normalized by the partition function (\ref{matrix_def}), which we denote
\begin{equation}
\Psi_{\alpha}(x,\theta) = \frac{\widehat{\chi}_{\alpha}(x,\theta)}{Z}.   \label{Psi-def}
\end{equation}
Note that, analogously to the bosonic case \cite{Manabe:2015kbj}, the normalized wave-function can be expressed in terms of connected differentials
\begin{align}
W_{(h,0)}(x_1,\ldots,x_h)&=\beta^{h/2}\bigg<\prod_{i=1}^h\sum_{a=1}^N\frac{dx_i}{x_i-z_a}\bigg>^{(\mathrm{c})},
\label{h_conn_diff}
\\
W_{(h,1)}(x_1,\ldots,x_h|x,\theta)&=\beta^{(h+1)/2}\bigg<\prod_{i=1}^h\sum_{a=1}^N\frac{dx_i}{x_i-z_a}\cdot \sum_{a=1}^N\frac{\vartheta_a\theta}{x-z_a}\bigg>^{(\mathrm{c})},
\label{hf_conn_diff}
\end{align}
where $\left<\cdots \right>^{(\mathrm{c})}$ denotes the connected part of the normalized expectation value $\left<\cdots \right>/Z$. From the definition (\ref{chi_hat_def}) we find the formula
\begin{align}
\begin{split}
\log \Psi_{\alpha}(x,\theta)&=\frac{\alpha}{\hbar^2}V(x,\theta) +
\\
&\quad
+\sum_{h=1}^{\infty}\frac{1}{h!}\Big(-\frac{\alpha}{\hbar}\Big)^h
\int^{x}_{\infty}\cdots \int^{x}_{\infty}\Big(W_{(h,0)}(x_1,\ldots,x_h)+W_{(h,1)}(x_1,\ldots,x_h|x,\theta)\Big).
\label{recon_wave}
\end{split}
\end{align}

Recall that in the bosonic case the differentials (\ref{h_conn_diff}) can be reconstructed by means of the topological recursion \cite{Chekhov:2005rr,Chekhov:2006rq,eyn-or}, with the initial condition given by an algebraic curve. In consequence the wave-function (\ref{recon_wave}) and the quantum curve equation it satisfies can be reconstructed in more general situations, even if a matrix model formulation of a given problem is not known \cite{abmodel}. A supersymmetric version of the topological recursion that would enable to compute both types of differentials (\ref{h_conn_diff}) and (\ref{hf_conn_diff}) in super-eigenvalue models has not been formulated yet. Once such a formulation is established, the expression (\ref{recon_wave}) would enable construction of super-quantum curves beyond the realm of super-eigenvalue models.

In this section we analyze certain properties of the wave-function (\ref{chi_hat_def}) and identify associated representation of the super-Virasoro algebra. This representation will turn out to provide building blocks of super-quantum curves, which will be constructed in the next section.

\subsection{Deformed currents and Ward identities for the wave-function}

Let us first derive Ward identities for the wave-function, analogously as for the original partition function $Z$ in section (\ref{ssec-Ward}). For simplicity, we consider first $\chi_{\alpha}(x,\theta)$ defined in (\ref{chi_hat_w}), with the prefactor removed. Such a wave-function can be regarded as the original super-eigenvalue model, however with deformed potentials
\be
\begin{split}
\widetilde{V}_B(y;x) & =V_B(y)+\alpha\log(x-y),     \\
\widetilde{V}_F(y;x,\theta) & =V_F(y)-\frac{\alpha\theta}{x-y}.    \label{V_BF-deformed}
\end{split}
\ee
Our strategy is to consider bosonic and fermionic fields corresponding to the model with such potentials, and to write Ward identities in terms of the corresponding currents. The deformation of the potentials given above leads to a deformation of bosonic and fermionic fields (\ref{free_boson}) and (\ref{free_fermion}), so that the super-conformal current $S(y)$ in (\ref{s_current_rep}) is replaced by
\begin{equation}
S(y;x,\theta)=S_-(y;x,\theta)+S_+(y;x,\theta),   \label{S-yxtheta}
\end{equation}
where
\be
\begin{split}
S_-(y;x,\theta)&
=\frac{1}{\hbar^2}\widetilde{V}_{F}(y;x,\theta)\partial_y\widetilde{V}_{B}(y;x)
-\frac{\sqrt{\beta}}{\hbar}\sum_{a=1}^N\frac{\widetilde{V}_{F}(y;x,\theta)-\widetilde{V}_{F}(z_a;x,\theta)}{y-z_a}  + \\
&\qquad + \frac{Q}{\hbar}\partial_y\widetilde{V}_{F}(y;x,\theta)
-\frac{\sqrt{\beta}}{\hbar}\sum_{a=1}^N\frac{\big(\partial_y\widetilde{V}_{B}(y;x)-\partial_{z_a}\widetilde{V}_{B}(z_a;x)\big)\vartheta_a}{y-z_a} = \\
\label{s_yx_m}
& = \frac{2\Delta_{\alpha}\theta}{(y-x)^2}+
\frac{1}{y-x}\Big[\theta\Big(\partial_x+\frac{\alpha}{\hbar^2}V_{B}'(y)\Big)-\Big(\partial_{\theta}-\frac{\alpha}{\hbar^2}V_F(y)\Big)\Big] + \\
&\qquad +\frac{1}{\hbar^2}\Big[V_{B}'(y)V_{F}(y)+Q\hbar V_{F}'(y)+\widehat{h}(y)\Big].
\end{split}
\ee
The second expression above is written in terms of the operator $\widehat{h}(y)$ defined in (\ref{h_x_op}), and deriving the action of $S_-(y;x,\theta)$ on $\chi_{\alpha}(x,\theta)$ we took advantage of derivatives $\partial_{\theta}\chi_{\alpha}^{\textrm{ins}}(x,\theta)$ and $\partial_x\chi_{\alpha}^{\textrm{ins}}(x,\theta)$ given in (\ref{d_th1_chi_ins}). Above, as in (\ref{Delta:through:alpha}), we denoted
\begin{equation}
\Delta_{\alpha} = \frac{\alpha}{2\hbar}\Big(\frac{\alpha}{\hbar}-Q\Big).
\end{equation}
We also find
\be
\begin{split}
S_+(y;x,\theta) &=
\beta\sum_{a,b=1}^N\frac{\vartheta_a}{(y-z_a)(y-z_b)}
+(1-\beta)\sum_{a=1}^N\frac{\vartheta_a}{(y-z_a)^2}  + \\
&\qquad
-\frac{\sqrt{\beta}}{\hbar}\sum_{a=1}^N\frac{1}{y-z_a}
\big(\partial_{z_a}\widetilde{V}_{B}(z_a;x)\vartheta_a+\widetilde{V}_{F}(z_a;x,\theta)\big) =       \label{s_yx_p}  \\
& = \frac{\alpha\sqrt{\beta}}{\hbar}\sum_{a=1}^N\frac{\vartheta_a}{(x-z_a)(y-z_a)}
+Q\sqrt{\beta}\sum_{a=1}^N\frac{\vartheta_a}{(y-z_a)^2}
+\beta\sum_{a,b=1}^N\frac{\vartheta_a}{(y-z_a)(y-z_b)}   +   \\
&\qquad
+\frac{\alpha\theta\sqrt{\beta}}{\hbar}\sum_{a=1}^N\frac{1}{(x-z_a)(y-z_a)}
-\frac{\sqrt{\beta}}{\hbar}\sum_{a=1}^N\frac{V_B'(y)\vartheta_a+V_F(y)}{y-z_a}
-\frac{1}{\hbar^2}\widehat{h}(y).
\end{split}
\ee

Similarly the energy-momentum tensor $T(y)$ in (\ref{e_m_tensor_rep}) is replaced by
\begin{equation}
T(y;x,\theta)=T_-(y;x,\theta)+T_+(y;x,\theta),   \label{T-yxtheta}
\end{equation}
where
\be
\begin{split}
T_-(y;x,\theta)&=\frac{1}{2\hbar^2}\big((\partial_y\widetilde{V}_{B}(y;x))^2+(\partial_y\widetilde{V}_{F}(y;x,\theta))\widetilde{V}_{F}(y;x,\theta)\big)+\frac{Q}{2\hbar}\partial_y^2\widetilde{V}_{B}(y;x) + \\
&\qquad
-\frac{\sqrt{\beta}}{\hbar}\sum_{a=1}^N\frac{\partial_y\widetilde{V}_{B}(y;x)-\partial_{z_a}\widetilde{V}_{B}(z_a;x)}{y-z_a}
-\frac{\sqrt{\beta}}{2\hbar}\sum_{a=1}^N\frac{\widetilde{V}^{(2)}_{F}(y,z_a;x,\theta)\vartheta_a}{(y-z_a)^2}   + \\
&\qquad
-\frac{\sqrt{\beta}}{2\hbar}\sum_{a=1}^N\frac{\big(\partial_y\widetilde{V}_{F}(y;x,\theta)-\partial_{z_a}\widetilde{V}_{F}(z_a;x,\theta)\big)\vartheta_a}{y-z_a} =
\label{em_yx_m}  \\  
&=\frac{1}{2(y-x)^2}\Big[2\Delta_{\alpha}+\theta\Big(\partial_{\theta}-\frac{\alpha}{\hbar^2}V_F(y)\Big)\Big]+
\frac{1}{y-x}\Big[\partial_x+\frac{\alpha}{\hbar^2}V_B'(y)-\frac{\alpha\theta}{2\hbar^2}V_F'(y)\Big] + \\
&\qquad
+\frac{1}{\hbar^2}\Big[\frac12V_{B}'(y)^2+\frac12V_F'(y)V_F(y)+\frac12Q\hbar V_{B}''(y)+\widehat{f}(y)\Big],
\end{split}
\ee
where we defined
\begin{equation}
\widetilde{V}^{(2)}_{F}(y,z_a;x,\theta)=\widetilde{V}_{F}(y;x,\theta)-\widetilde{V}_{F}(z_a;x,\theta)-(y-z_a)\partial_{z_a}\widetilde{V}_{F}(z_a;x,\theta).
\end{equation}
The second expression in (\ref{em_yx_m}) is written in terms of the operator $\widehat{f}(y)$ defined in (\ref{f_x_op}), and to find the representation of $T_-(y;x,\theta)$ on $\chi_{\alpha}(x,\theta)$ we took advantage of derivatives (\ref{d_th1_chi_ins}). We also find
\be
\begin{split}
T_+(y;x,\theta) &= \frac{\beta}{2}\sum_{a,b=1}^N\frac{1}{(y-z_a)(y-z_b)}
+\frac{\beta}{2}\sum_{a,b=1}^N\frac{\vartheta_a\vartheta_b}{(y-z_a)(y-z_b)^2}
+\frac{1}{2}(1-\beta)\sum_{a=1}^N\frac{1}{(y-z_a)^2} + \\
&\qquad -\frac{\sqrt{\beta}}{\hbar}\sum_{a=1}^N\frac{1}{y-z_a}
\big(\partial_{z_a}\widetilde{V}_{B}(z_a;x)+\partial_{z_a}\widetilde{V}_{F}(z_a;x,\theta)\vartheta_a\big)
-\frac{\sqrt{\beta}}{2\hbar}\sum_{a=1}^N\frac{\widetilde{V}_{F}(z_a;x,\theta)\vartheta_a}{(y-z_a)^2},
\label{em_yx_p}
\end{split}
\ee
which can be rewritten as
\begin{align}
T_+(y;x,\theta)&=
\frac{\alpha\sqrt{\beta}}{\hbar}\sum_{a=1}^N\frac{1}{(x-z_a)(y-z_a)}
+\frac{Q\sqrt{\beta}}{2}\sum_{a=1}^N\frac{1}{(y-z_a)^2}
+\frac{\beta}{2}\sum_{a,b=1}^N\frac{1}{(y-z_a)(y-z_b)} +
\nonumber\\
&\ \ \
+\frac{\beta}{2}\sum_{a,b=1}^N\frac{\vartheta_a\vartheta_b}{(y-z_a)(y-z_b)^2}
+\frac{\alpha\theta\sqrt{\beta}}{\hbar}\sum_{a=1}^N\frac{\vartheta_a}{(x-z_a)^2(y-z_a)} +
\nonumber\\
&\ \ \
+\frac{\alpha\theta\sqrt{\beta}}{2\hbar}\sum_{a=1}^N\frac{\vartheta_a}{(x-z_a)(y-z_a)^2}
-\frac{\sqrt{\beta}}{2\hbar}\sum_{a=1}^N\frac{2V_B'(y)+V_F'(y)\vartheta_a}{y-z_a}   +
\nonumber\\
&\ \ \
-\frac{\sqrt{\beta}}{2\hbar}\sum_{a=1}^N\frac{V_F(y)\vartheta_a}{(y-z_a)^2}
-\frac{1}{\hbar^2}\widehat{f}(y).
\label{em_yx_p_ex}
\end{align}

In terms of the above currents, the fermionic and bosonic Ward identities for the super-eigenvalue model with deformed potentials (\ref{V_BF-deformed}) take a simple form
\be
\begin{split}
\big<S_+(y;x,\theta)\chi_{\alpha}^{\textrm{ins}}(x,\theta)\big> & = 0, \label{f_ward_s_wave}  \\ 
\big<T_+(y;x,\theta)\chi_{\alpha}^{\textrm{ins}}(x,\theta)\big> & = 0.
\end{split}
\ee

\subsection{Super-Virasoro operators for $\alpha/\beta$-deformed matrix integrals}   \label{ssec-abVirasoro}

By expanding the superconformal current $S(y;x,\theta)$ in (\ref{S-yxtheta}) and the energy-momentum tensor $T(y;x,\theta)$ in (\ref{T-yxtheta}) in powers of $y$, and writing them in the form
\begin{equation}
S(y;x,\theta)=\sum_{r\in{\IZ}+1/2}g^{\alpha}_r(x,\theta) y^{-r-3/2},\qquad
T(y;x,\theta)=\sum_{n\in{\IZ}}\ell_n^{\alpha}(x,\theta) y^{-n-2},
\end{equation}
we find the following representation of the super-Virasoro algebra (as acting on $\chi_{\alpha}(x,\theta)$)
\begin{equation}
g_{n+1/2}^{\alpha}(x,\theta)=
\begin{cases}
g_{n+1/2} - \theta x^{n+1}\partial_x + x^{n+1}\partial_{\theta} + \\
\qquad + \alpha\sum_{k=0}^n x^{n-k} \partial_{\xi_{k+1/2}} +  \alpha\theta\sum_{k=0}^n x^{n-k} \partial_{t_k},
&
\mbox{if}\ n\ge -1
\\
g_{n+1/2} -2\Delta_{\alpha}\theta(n+1)x^n - \theta x^{n+1}\partial_x + x^{n+1}\partial_{\theta} + \\
\qquad -\frac{\alpha}{\hbar^2} \sum_{k=0}^{-n-2} \xi_{k+1/2} x^{k+n+1} - \frac{\alpha}{\hbar^2}\theta \sum_{k=0}^{-n-2} (k+1) t_{k+1} x^{k+n+1},
&
\mbox{if}\ n\le -2
\end{cases}\qquad    \label{Gn}
\end{equation}
where
\begin{equation}
g_{n+1/2} =
\begin{cases}
\sum_{k=1}^{\infty}kt_k\partial_{\xi_{k+n+1/2}}+\sum_{k=0}^{\infty}\xi_{k+1/2}\partial_{t_{k+n+1}} + \\
\qquad +\hbar^2\sum_{k=0}^n\partial_{\xi_{k+1/2}}\partial_{t_{n-k}} -Q\hbar(n+1)\partial_{\xi_{n+1/2}},
&
\mbox{if}\ n\ge -1
\\
\frac{1}{\hbar^2} \sum_{k=0}^{-n-2} (k+1) t_{k+1}\xi_{-n-k-3/2} - \frac{n+1}{\hbar^2}Q\xi_{-n-1/2} + \\
\qquad +\sum_{k=-n-1}^{\infty} \xi_{k+1/2}\partial_{t_{k+n+1}} + \sum_{k=-n}^{\infty} k t_k\partial_{\xi_{k+n+1/2}},
&
\mbox{if}\ n\le -2
\end{cases}\qquad    \label{ggn}
\end{equation}
Furthermore
\begin{equation}
\ell_n^{\alpha}(x,\theta)=
\begin{cases}
\ell_n -x^{n+1}\partial_x -\frac{n+1}{2} x^n \theta\partial_{\theta} + \alpha\sum_{k=0}^n x^k \partial_{t_{n-k}} + \\
\qquad
+ \alpha \theta\sum_{k=0}^{n-1} \big(\frac{n-1}{2}-k\big) x^k\partial_{\xi_{n-k-1/2}},
&
\mbox{if}\ n\ge -1
\\
\ell_n - (n+1)\Delta_{\alpha} x^n -x^{n+1}\partial_x -\frac{n+1}{2} x^n \theta\partial_{\theta} + \\
\qquad -\frac{\alpha}{\hbar^2}\sum_{k=0}^{-n-2} (k+1)t_{k+1} x^{n+k+1} +
\\
\qquad + \frac{\alpha}{2\hbar^2} \theta \sum_{k=0}^{-n-2} \Big( (n+k+1)x^{n+k} \xi_{k+1/2} + (k+1)x^{n+k+1}\xi_{k+3/2} \Big),
&
\mbox{if}\ n\le -2
\end{cases}\qquad    \label{Ln}
\end{equation}
where
\begin{equation}
\ell_n =
\begin{cases}
\sum_{k=1}^{\infty}kt_k\partial_{t_{k+n}}+\frac{\hbar^2}{2}\sum_{k=1}^{n}\partial_{t_k}\partial_{t_{n-k}}+\sum_{k=0}^{\infty}\big(k+\frac{n+1}{2}\big)\xi_{k+1/2}\partial_{\xi_{k+n+1/2}} +
\\
\qquad +\frac{\hbar^2}{2}\sum_{k=1}^{n}k\partial_{\xi_{n-k+1/2}}\partial_{\xi_{k-1/2}}
-\frac{n+1}{2}Q\hbar\partial_{t_n},
&
\mbox{if}\ n\ge -1
\\
\frac{n(n+1)}{2\hbar}Qt_{-n} - \frac{1}{2\hbar^2}\sum_{k=0}^{-n-2}(k+1)(k+n+1) t_{k+1}t_{-n-k-1} + \\
\qquad + \sum_{k=0}^{\infty} (k-n)t_{k-n}\partial_{t_k}  - \frac{1}{2\hbar^2}\sum_{k=0}^{-n-2} (n+k+1)\xi_{-n-k-1/2}\xi_{k+1/2} + \\
\qquad + \sum_{k=-n}^{\infty} \big(k+\frac{n+1}{2}\big)\xi_{k+1/2}\partial_{\xi_{k+n+1/2}},
&
\mbox{if}\ n\le -2
\end{cases}\qquad    \label{lln}
\end{equation}
Note that super-Virasoro generators $g_{-1/2}^{\alpha}(x,\theta)$, $g_{1/2}^{\alpha}(x,\theta)$ together with $\ell_{-1}^{\alpha}(x,\theta)$, $\ell_0^{\alpha}(x,\theta)$ and $\ell_1^{\alpha}(x,\theta)$ form a representation of $\mathfrak{osp}(1|2)$ algebra, and they impose constraints on the wave-function
\begin{equation}
g_{\pm 1/2}^{\alpha}(x,\theta)\chi_{\alpha}(x,\theta)=\ell_{\pm 1, 0}^{\alpha}(x,\theta)\chi_{\alpha}(x,\theta)=0. \label{osp12}
\end{equation}
In terms of super-eigenvalues $z_a$ and $\vartheta_a$, the generators in the above constraints (as acting on $\chi_{\alpha}(x,\theta)$) take form
\begin{align}
\begin{split}
g_{-1/2}^{\alpha}(x,\theta) &=
-D-\frac{\sqrt{\beta}}{\hbar}\sum_{a=1}^N\big(V_B'(z_a)\vartheta_a+V_F(z_a)\big),   \\
g_{1/2}^{\alpha}(x,\theta) &=
-xD-\frac{\alpha\mu\theta}{\hbar^2}-\frac{\sqrt{\beta}}{\hbar}\sum_{a=1}^N\big(V_B'(z_a)\vartheta_a+V_F(z_a)\big)z_a
+(\mu+Q\hbar-\alpha)\frac{\sqrt{\beta}}{\hbar}\sum_{a=1}^N\vartheta_a,   \label{osp12-g}
\end{split}
\end{align}
and
\begin{align}
\begin{split}
\ell_{-1}^{\alpha}(x,\theta) &=
-\partial_x-\frac{\sqrt{\beta}}{\hbar}\sum_{a=1}^N\big(V_B'(z_a)+V_F'(z_a)\vartheta_a\big),\\
\ell_{0}^{\alpha}(x,\theta) & =
-x\partial_x-\frac{\theta}{2}\partial_{\theta}
+\frac{\mu}{2\hbar^2}(\mu+Q\hbar-2\alpha)
-\frac{\sqrt{\beta}}{\hbar}\sum_{a=1}^N\big(V_B'(z_a)+V_F'(z_a)\vartheta_a\big)z_a  + \\
&\qquad
-\frac{\sqrt{\beta}}{2\hbar}\sum_{a=1}^NV_F(z_a)\vartheta_a,\\
\ell_{1}^{\alpha}(x,\theta) &=
-x^2\partial_x-x\theta\partial_{\theta}
-\frac{\alpha\mu}{\hbar^2}x
-\frac{\sqrt{\beta}}{\hbar}\sum_{a=1}^N\big(V_B'(z_a)+V_F'(z_a)\vartheta_a\big)z_a^2  + \\
&\qquad
-\frac{\sqrt{\beta}}{\hbar}\sum_{a=1}^NV_F(z_a)\vartheta_az_a
+\frac{\sqrt{\beta}}{\hbar}(\mu+Q\hbar-\alpha)\sum_{a=1}^Nz_a,  \label{osp12-l}
\end{split}
\end{align}
where
\begin{equation}
D = -\partial_{\theta}+\theta\partial_x.
\end{equation}
In section \ref{sec-examples} we will see that, at least for super-eigenvalue models with some particular potentials, this symmetry --- by means of the constraints (\ref{osp12}) --- can be used to turn time-dependent quantum curves into time-independent ones.

\subsection{Super-Virasoro operators as building blocks of super-quantum curves}   \label{ssec-sVir-blocks}

We can define now the representation of super-Virasoro algebra on $\chi_{\alpha}(x,\theta)$ at $y=x$. The generators $G_{-n+1/2}$ and $L_{-n}$ for $n\ge 1$ which provide building blocks of super-quantum curves
can be determined as
\be
\begin{split}
G_{n+1/2}\chi_{\alpha}(x,\theta)&=
\oint_{y=x}\frac{dy}{2\pi i}(y-x)^{n+1}S(y;x,\theta)\chi_{\alpha}(x,\theta) =   \\
&= \oint_{y=x}\frac{dy}{2\pi i}(y-x)^{n+1}S_-(y;x,\theta)\chi_{\alpha}(x,\theta),
\label{g_chi_def} \\
L_{n}\chi_{\alpha}(x,\theta)&=
\oint_{y=x}\frac{dy}{2\pi i}(y-x)^{n+1}T(y;x,\theta)\chi_{\alpha}(x,\theta)  =  \\
&= \oint_{y=x}\frac{dy}{2\pi i}(y-x)^{n+1}T_-(y;x,\theta)\chi_{\alpha}(x,\theta),
\end{split}
\ee
where in the second line of each expression above we have used the Ward identity (\ref{f_ward_s_wave}). 
Taking advantage of (\ref{s_yx_m}) and (\ref{em_yx_m}) we then find
\be
\begin{split}
G_{-1/2}&=\theta\Big(\partial_x+\frac{\alpha}{\hbar^2}V_{B}'(x)\Big)-\Big(\partial_{\theta}-\frac{\alpha}{\hbar^2}V_F(x)\Big),  \\
G_{-n+1/2}&=\frac{1}{\hbar^2(n-2)!}\Big(
\partial_x^{n-2}\big(V_B'(x)V_F(x)\big)+\Big(Q\hbar+\frac{\alpha}{n-1}\Big)\partial_x^{n-1}V_F(x)   +   \\
& \qquad
+\frac{\alpha\theta}{n-1}\partial_x^nV_B(x)+\partial_x^{n-2}\widehat{h}(x)\Big),\qquad\qquad \textrm{for $n\ge 2$},
\label{g_chi_rep}  \\ 
L_{-1} & = \partial_x+\frac{\alpha}{\hbar^2}V_B'(x)-\frac{\alpha\theta}{2\hbar^2}V_F'(x),\\
L_{-n} & = \frac{1}{\hbar^2(n-2)!}\Big(\frac12\partial_x^{n-2}\big(V_{B}'(x)\big)^2
+\frac12\partial_x^{n-2}\big(V_F'(x)V_F(x)\big)
+\Big(\frac12Q\hbar+\frac{\alpha}{n-1}\Big)\partial_x^nV_{B}(x)   +  \\
&\qquad  -\frac{(n+1)\alpha\theta}{2n(n-1)}\partial_x^nV_F(x)
+\partial_x^{n-2}\widehat{f}(x)\Big),\qquad \qquad \textrm{for $n\ge 2$},
\end{split}
\ee
where $\partial_x^{n}\widehat{h}(x)$ and $\partial_x^{n}\widehat{f}(x)$ are defined in (\ref{dxn-hhat}) and (\ref{dxn-fhat}).

We can also include the prefactor in (\ref{chi_hat_w}) and consider the representation of super-Virasoro algebra on $\widehat{\chi}_{\alpha}(x,\theta)$, defined by
\be
\begin{split}
\widehat{G}_{-n+1/2}\widehat{\chi}_{\alpha}(x,\theta) &=
\oint_{y=x}\frac{dy}{2\pi i}(y-x)^{-n+1}S_-(y;x,\theta)\widehat{\chi}_{\alpha}(x,\theta), \\
\widehat{L}_{-n}\widehat{\chi}_{\alpha}(x,\theta) &=
\oint_{y=x}\frac{dy}{2\pi i}(y-x)^{-n+1}T_-(y;x,\theta)\widehat{\chi}_{\alpha}(x,\theta).
\end{split}
\ee
From the earlier results, and using the commutation relations (\ref{hf_Vbf_g}), from (\ref{g_chi_rep}), we find
\be
\begin{split}
\widehat{G}_{-1/2} &= \theta\partial_x-\partial_{\theta},   \qquad \qquad \qquad \widehat{L}_{-1} = \partial_x,    \\
\widehat{G}_{-n+1/2} &= \frac{1}{\hbar^2(n-2)!}\Big(
\partial_x^{n-2}\big(V_B'(x)V_F(x)\big)+Q\hbar\partial_x^{n-1}V_F(x)
+\partial_x^{n-2}\widehat{h}(x)\Big),\ \ \textrm{for $n\ge 2$},
\label{h_g_chi_rep}   \\ 
\widehat{L}_{-n} & =\frac{1}{\hbar^2(n-2)!}\Big(\frac12\partial_x^{n-2}\big(V_{B}'(x)\big)^2
+\frac12\partial_x^{n-2}\big(V_F'(x)V_F(x)\big)  +  \\
&\qquad \qquad
+\frac12Q\hbar\partial_x^nV_{B}(x)
+\partial_x^{n-2}\widehat{f}(x)\Big),\ \ \textrm{for $n\ge 2$}.
\end{split}
\ee
Using the commutation relations (\ref{hf_Vbf_g}) and (\ref{hf_hh_g}), one can check  that the above generators indeed satisfy the super-Virasoro algebra, in particular
$\big\{\widehat{G}_{-m-1/2},\widehat{G}_{-n-1/2}\big\}=2\widehat{L}_{-m-n-1}$, $\big[\widehat{L}_{-m},\widehat{G}_{-n-1/2}\big]=\big(n-(m-1)/2\big)\widehat{G}_{-m-n-1/2}$ and $\big[\widehat{L}_{-m},\widehat{L}_{-n}\big]=(n-m)\widehat{L}_{-m-n}$ for $n\ge 1$.


From (\ref{h_g_chi_rep}) we also find the representation of the super-Virasoro algebra, as acting on bosonic and fermionic components $\widehat{\chi}_{B,\alpha}(x)$ and $\widehat{\chi}_{F,\alpha}(x)$ introduced in (\ref{wave_bose_fermi});
for $n\ge 2$ we get
\begin{align}
\begin{split}
&
\widehat{\mathsf{G}}_{-1/2}\widehat{\chi}_{B,\alpha}(x)=
\widehat{\chi}_{F,\alpha}(x),\qquad
\widehat{\mathsf{G}}_{-1/2}\widehat{\chi}_{F,\alpha}(x)=
\partial_x\widehat{\chi}_{B,\alpha}(x),\\
&
\widehat{\mathsf{G}}_{-n+1/2}\widehat{\chi}_{B,\alpha}(x)=
\widehat{G}_{-n+1/2}\widehat{\chi}_{B,\alpha}(x),\qquad
\widehat{\mathsf{G}}_{-n+1/2}\widehat{\chi}_{F,\alpha}(x)=
\widehat{G}_{-n+1/2}\widehat{\chi}_{F,\alpha}(x),
\label{h_g_bf_chi_rep} \\ 
&
\widehat{\mathsf{L}}_{-1}\widehat{\chi}_{B,\alpha}(x)=\partial_x\widehat{\chi}_{B,\alpha}(x),\qquad
\widehat{\mathsf{L}}_{-1}\widehat{\chi}_{F,\alpha}(x)=\partial_x\widehat{\chi}_{F,\alpha}(x),\\
&
\widehat{\mathsf{L}}_{-n}\widehat{\chi}_{B,\alpha}(x)=
\widehat{L}_{-n}\widehat{\chi}_{B,\alpha}(x),\qquad
\widehat{\mathsf{L}}_{-n}\widehat{\chi}_{F,\alpha}(x)=
\widehat{L}_{-n}\widehat{\chi}_{F,\alpha}(x).
\end{split}
\end{align}
From this representation, $\widehat{\mathsf{G}}_{-1/2}^2=\widehat{\mathsf{L}}_{-1}$ on $\widehat{\chi}_{B,\alpha}(x)$, and we find commutation relations
$\big\{\widehat{\mathsf{G}}_{-m-1/2},\widehat{\mathsf{G}}_{-n-1/2}\big\}=2\widehat{\mathsf{L}}_{-m-n-1}$, $\big[\widehat{\mathsf{L}}_{-m},\widehat{\mathsf{G}}_{-n-1/2}\big]=\big(n-(m-1)/2\big)\widehat{\mathsf{G}}_{-m-n-1/2}$, as well as $\big[\widehat{\mathsf{L}}_{-m},\widehat{\mathsf{L}}_{-n}\big]=(n-m)\widehat{\mathsf{L}}_{-m-n}$ for $n\ge 1$. Although commutation relations
\begin{equation}
\big\{\widehat{\mathsf{G}}_{-1/2},\widehat{\mathsf{G}}_{-n+1/2}\big\}=2\widehat{\mathsf{L}}_{-n},\qquad
\big[\widehat{\mathsf{G}}_{-1/2},\widehat{\mathsf{L}}_{-n}\big]=\frac{n-1}{2}\widehat{\mathsf{G}}_{-n-1/2},\qquad n\ge 1,
\end{equation}
are not obvious at this stage, they are needed to construct quantum curves in the next section.

Yet another representation of super-Virasoro algebra that we consider is associated to the normalized wave-function (\ref{Psi-def}). In this case the operators $\widehat{G}_{n+1/2}$ and $\widehat{L}_n$ in (\ref{h_g_chi_rep}) are converted to $\widehat{\mathcal{G}}_{-n+1/2}\equiv
Z^{-1}\widehat{G}_{-n+1/2}Z$ and $\widehat{\mathcal{L}}_{-n}\equiv
Z^{-1}\widehat{L}_{-n}Z$, and for $n\geq 2$ they take form
\be
\begin{split}
\widehat{\mathcal{G}}_{-n+1/2} & =
\frac{1}{\hbar^2(n-2)!}\Big(
\partial_x^{n-2}\big(V_B'(x)V_F(x)\big)+Q\hbar\partial_x^{n-1}V_F(x)
+\partial_x^{n-2}\widehat{h}(x) +
\\
&\qquad +\big[\partial_x^{n-2}\widehat{h}(x),\log Z\big]\Big),   \\
\widehat{\mathcal{L}}_{-n} &=
\frac{1}{\hbar^2(n-2)!}\Big(
\frac12\partial_x^{n-2}\big(V_{B}'(x)^2\big)
+\frac12\partial_x^{n-2}\big(V_F'(x)V_F(x)\big)
+\frac12Q\hbar\partial_x^nV_{B}(x)  + \\
&\qquad
+\partial_x^{n-2}\widehat{f}(x)
+\big[\partial_x^{n-2}\widehat{f}(x),\log Z\big]\Big).   \label{Psi-GL}
\end{split}
\ee

\section{Super-quantum curves as super-Virasoro singular vectors}  \label{sec-super-quantum}

In this section we construct supersymmetric quantum curves, or super-quantum curves, denoted $\widehat{A}_{n}^{\alpha}$, which annihilate wave-functions $\widehat{\chi}_{\alpha}(x,\theta)$ introduced in (\ref{chi_hat_def})
\begin{equation}
\widehat{A}_{n}^{\alpha}\widehat{\chi}_{\alpha}(x,\theta)=0.   \label{Achi0}
\end{equation}
Our construction is based on the analysis of Ward identities (\ref{f_ward_s_wave}). We show that these super-quantum curves have the structure of the Neveu-Schwarz (NS) super-Virasoro singular vectors, which have been presented in section \ref{sec-superVirasoro}. As we recalled in that section, NS super-Virasoro singular vectors exist only at levels $n=pq/2$, for integer $p$ and $q$ such that $p-q$ is even, and correspond to momenta of the form
\begin{equation}
\alpha=\alpha_{p,q} = \frac{(p-1)\beta^{1/2}-(q-1)\beta^{-1/2}}{2}\hbar,
\qquad \textrm{with $p-q\in 2{\IZ}$.}  \label{s_vir_sing_momenta}
\end{equation}
We will show that super-quantum curves can be identified also only for those special values of $\alpha$. In our derivation these values arise from the condition that the wave-function $\widehat{\chi}_{\alpha}(x,\theta)$ satisfies a differential equation of a finite order in $x$. Therefore in the notation $\widehat{A}_{n}^{\alpha}$ the superscript $\alpha$ refers to the deformation parameter of the corresponding wave-function, and the subscript $n$ denotes the level of the singular vector encoding the structure of a given super-quantum curve; we also refer to $n$ as the level of a given super-quantum curve. In the next section we will show in particular that the super-quantum curve at level 3/2 reduces to super-spectral curve (\ref{sp_r_curve}) in the classical limit.

The structure of super-Virasoro singular vectors is encoded in operators $\widehat{A}_{n}^{\alpha}$ as follows. First note, that in view of the decomposition (\ref{wave_bose_fermi}) and the representation (\ref{h_g_bf_chi_rep}), we can write the wave-function in the form
\be
\widehat{\chi}_{\alpha}(x,\theta) = (1-\theta  \widehat{\mathsf{G}}_{-1/2} ) \widehat{\chi}_{B,\alpha}(x).
\ee
Therefore the information about the wave-function is essentially encoded in its bosonic component $\widehat{\chi}_{B,\alpha}(x)$. Note also that
\be
\theta \widehat{\chi}_{\alpha}(x,\theta)  = \theta \widehat{\chi}_{B,\alpha}(x), \qquad \widehat{\chi}_{F,\alpha}(x) = -\partial_{\theta} \widehat{\chi}_{\alpha}(x,\theta)  = \widehat{\mathsf{G}}_{-1/2}  \widehat{\chi}_{B,\alpha}(x).     \label{theta-G12}
\ee
In consequence one can introduce an operator $\widehat{\mathsf{A}}_{n}^{(0)}$, expressed in terms of generators (\ref{h_g_bf_chi_rep}), which annihilates the bosonic wave-function
\be
\widehat{\mathsf{A}}_{n}^{(0)}\widehat{\chi}_{B,\alpha_{p,q}}(x) = 0.  \label{A0-chiB}
\ee
As we will show, this is the operator $\widehat{\mathsf{A}}_{n}^{(0)}$ that takes form of a super-Virasoro singular vector, written in terms of generators (\ref{h_g_bf_chi_rep}).

The above equation can be rewritten also in terms of an operator $\widehat{A}_{n}^{(0)}$ -- which arises by replacing $\widehat{\mathsf{G}}_{-1/2}$ by $(-\partial_{\theta})$ in $\widehat{\mathsf{A}}_{n}^{(0)}$ -- acting on $\widehat{\chi}_{\alpha}(x,\theta)$
\begin{equation}
\theta\widehat{A}_{n}^{(0)}\widehat{\chi}_{\alpha_{p,q}}(x,\theta)=
\theta\widehat{\mathsf{A}}_{n}^{(0)}\widehat{\chi}_{B,\alpha_{p,q}}(x)=0.
\label{hq_curve_gen_c}
\end{equation}
Furthermore, the bosonic wave-function is annihilated also by the following operator $\widehat{\mathsf{A}}_{n}^{(1)}$ (expressed in terms of generators (\ref{h_g_bf_chi_rep}))
\begin{equation}
\widehat{\mathsf{A}}_{n}^{(1)}=
\widehat{\mathsf{G}}_{-1/2}\widehat{\mathsf{A}}_{n}^{(0)},
\end{equation}
which can be related to the operator acting on $\widehat{\chi}_{\alpha}(x,\theta)$ as
\begin{equation}
\theta\widehat{A}_{n}^{(1)}\widehat{\chi}_{\alpha_{p,q}}(x,\theta)=
\theta\widehat{\mathsf{A}}_{n}^{(1)}\widehat{\chi}_{B,\alpha_{p,q}}(x)=0.
\end{equation}
Finally, the original super-quantum curve $\widehat{A}_{n}^{\alpha}$ can be reconstructed from the two ingredients $\widehat{A}_{n}^{(0)}$ and $\widehat{A}_{n}^{(1)}$ introduced above, as
\begin{equation}
\widehat{A}_{n}^{\alpha}=
\widehat{A}_{n}^{(0)}-\theta\partial_{\theta}\widehat{A}_{n}^{(0)}
-\theta\widehat{A}_{n}^{(1)},
\label{q_curve_gen_c}
\end{equation}
and this operator is expressed in terms of generators (\ref{h_g_chi_rep}).

One can also transform the operator (\ref{q_curve_gen_c}) into an operator that annihilates the normalized wave-function (\ref{Psi-def}), which takes form
\begin{equation}
\widehat{\mathcal{A}}_n^{\alpha} = Z^{-1}\widehat{A}_n^{\alpha}Z,   \label{A-hat-norm}
\end{equation}
and can be expressed in terms of generators (\ref{Psi-GL}).

In section \ref{ssec-general} we discuss an algorithm that enables to construct (\ref{A0-chiB}) and then super-quantum curves (\ref{q_curve_gen_c}) at arbitrary level, and in following sections we illustrate this algorithm and construct explicitly super-quantum curves at several lowest levels.

\subsection{General construction}    \label{ssec-general}

Our construction of super-quantum curves is based on the analysis of constraint equations (\ref{f_ward_s_wave}) and their generalizations, which we write as
\begin{equation}
\big<S_+^{(k+3/2)}(x,\theta)\chi_{\alpha}^{\textrm{ins}}(x,\theta)\big>=0,
\qquad
\big<T_+^{(k+2)}(x,\theta)\chi_{\alpha}^{\textrm{ins}}(x,\theta)\big>=0,
\label{t_fb_ward_id}
\end{equation}
where
\be
\begin{split}
S_{+}^{(k+3/2)}(x,\theta) &= \frac{(-1)^k}{k!}\frac{\partial^{k}}{\partial y^{k}} S_{+}(y;x,\theta)\Big|_{y=x}, \\
T_{+}^{(k+2)}(x,\theta) &= \frac{(-1)^k}{k!}\frac{\partial^{k}}{\partial y^{k}} T_{+}(y;x,\theta)\Big|_{y=x},
\end{split}
\ee
for $k\geq 0$, denote derivatives of the superconformal current $S_{+}(y;x,\theta)$ given in (\ref{s_yx_p}) and the energy-momentum tensor $T_{+}(y;x,\theta)$ in (\ref{em_yx_p_ex}). We also consider additional constraint equations, expressed in terms of loop insertion operators (\ref{loop-insertion})
\begin{align}
\begin{split}
\Big(\partial_{V_B(x)}^{(1)}\Big)^{k_1}\cdots \Big(\partial_{V_B(x)}^{(p)}\Big)^{k_p}
\Big(\partial_{V_F(x)}^{(1/2)}\Big)^{\ell_1}\cdots \Big(\partial_{V_F(x)}^{(q-1/2)}\Big)^{\ell_q}
\big<S_+^{(k+3/2)}(x,\theta)\chi_{\alpha}^{\textrm{ins}}(x,\theta)\big>&=0,
\\
\Big(\partial_{V_B(x)}^{(1)}\Big)^{k_1}\cdots \Big(\partial_{V_B(x)}^{(p)}\Big)^{k_p}
\Big(\partial_{V_F(x)}^{(1/2)}\Big)^{\ell_1}\cdots \Big(\partial_{V_F(x)}^{(q-1/2)}\Big)^{\ell_q}
\big<T_+^{(k+2)}(x,\theta)\chi_{\alpha}^{\textrm{ins}}(x,\theta)\big>&=0,
\label{t_fb_ward_id_kl}
\end{split}
\end{align}
where $k_1,\ldots,k_p$ are non-negative integers and $\ell_1,\ldots,\ell_q \in \{0,1\}$. Here the actions of the loop insertion operators on $S_+(y;x,\theta)$ and $T_+(y;x,\theta)$ are given by (\ref{loop_op_ward}).

To construct super-quantum curves we consider singular terms in $x$ (for $x\to \infty$), i.e. terms involving $(x-z_k)$ in denominators in the above constraints, and show that they can be expressed in terms of derivatives of $\chi_{\alpha}(x,\theta)$ with respect to $x$ or $\theta$, so that a combination of such derivatives provides a differential equation we are interested in. We argue now that it should always be possible to construct such differential equations, and in following sections we will illustrate such a construction in explicit examples.


Let us discuss first super-quantum curves at half-integer levels $n-\frac{1}{2}$, for $n\in{\IN}$. Consider a linear combination of constraint equations (\ref{t_fb_ward_id_kl}) of the form
\begin{align}
\begin{split}
&
0=\theta\Big<\sum_{p=0}^{n-2}\sum_{|\mu|=p}a_{1;\mu}\vec{\partial}_{V_B(x)}^{(\mu)}
S_{+}^{(n-p-1/2)}(x)\chi^{\textrm{ins}}_{\alpha}(x,\theta)
+\sum_{|\mu|=n-3}a_{2;\mu}\vec{\partial}_{V_B(x)}^{(\mu)}
\partial_{V_F(x)}^{(1/2)}T_{+}^{(2)}(x)\chi^{\textrm{ins}}_{\alpha}(x,\theta)  +  \\
&+
\sum_{|\mu|=n-4}\vec{\partial}_{V_B(x)}^{(\mu)}\Big( a_{3;\mu}\partial_{V_F(x)}^{(3/2)}T_{+}^{(2)}(x)
+a_{4;\mu}\partial_{V_F(x)}^{(1/2)}T_{+}^{(3)}(x)
+a_{5;\mu}\partial_{V_F(x)}^{(1/2)}\partial_{V_F(x)}^{(3/2)}S_{+}^{(3/2)}(x)
\Big)\chi^{\textrm{ins}}_{\alpha}(x,\theta)\Big>,
\label{half_int_const}
\end{split}
\end{align}
where $\mu$ is an integer partition with $|\mu| = \mu_1+\mu_2+\ldots$, and
\begin{equation}
\vec{\partial}_{V_B(x)}^{(\mu)} =
\partial_{V_B(x)}^{(\mu_1)}\partial_{V_B(x)}^{(\mu_2)}\cdots.
\end{equation}
The number of coefficients $a_{i;\mu}$ in (\ref{half_int_const}) is given by
\begin{equation}
a(n) = \sum_{p=0}^{n-2}\mathfrak{p}(p)+\mathfrak{p}(n-3)+3\mathfrak{p}(n-4),
\end{equation}
where $\mathfrak{p}(p)$ denotes the number of partitions of $p$. The most singular terms in the equation (\ref{half_int_const}) are of the form
\begin{equation}
\frac{\theta\vartheta_{a_1}}{(x-z_{a_1})^{\mu_1}(x-z_{a_2})^{\mu_2}\cdots
(x-z_{a_p})^{\mu_p}},
\label{m_sing_half_const}
\end{equation}
with $\mu_1+\mu_2+\ldots+\mu_p=n$. The number of such terms is given by
\begin{equation}
A(n) = \sum_{p=0}^{n-1}\mathfrak{p}(p).
\end{equation}
Now note, that the inequality $\mathfrak{p}(p-1)+\mathfrak{p}(p-2)\ge \mathfrak{p}(p)$ for $p\ge 2$ implies that
\begin{equation}
a(n)+1\ge A(n).
\end{equation}
This means that the number of coefficients in (\ref{half_int_const}), together with an additional freedom to adjust $\alpha$ (taken into account by an additional 1 in the left hand side of the above equation), is sufficient to group the most singular terms of the form (\ref{m_sing_half_const}) (present in (\ref{half_int_const})) into combinations that reproduce the same type of singularities in
\begin{equation}
\theta\partial_x^{n-1}\partial_{\theta}\chi_{\alpha}(x,\theta).
\end{equation}
In an analogous way one can express less singular terms in (\ref{half_int_const}) also in terms of derivatives of the wave-function with respect to $\theta$ and $x$. Therefore a combination of constraint equations (\ref{half_int_const}), with coefficients appropriately adjusted and expressed in terms of derivatives of the wave-function with respect to $\theta$ and $x$, gives rise to a differential equation which we identify as a super-quantum curve.


Similarly one can construct super-quantum curves at an integer level $n\in{\IN}$. To this end we consider now a combination of constraints
\begin{align}
\begin{split}
0&=\theta\Big<\sum_{p=0}^{n-2}\sum_{|\mu|=p}b_{1;\mu}\vec{\partial}_{V_B(x)}^{(\mu)}
T_{+}^{(n-p)}(x)\chi^{\textrm{ins}}_{\alpha}(x,\theta)  +  \\
&\ \ \
+\sum_{p=0}^{n-2}\sum_{q=0}^{n-p-2}
\sum_{|\mu|=p}b_{2;\mu;q}\vec{\partial}_{V_B(x)}^{(\mu)}
\partial_{V_F(x)}^{(q+1/2)}S_{+}^{(n-p-q-1/2)}(x)\chi^{\textrm{ins}}_{\alpha}(x,\theta)\Big>.
\label{int_const}
\end{split}
\end{align}
The number of coefficients $b_{1;\mu}$ and $b_{2;\mu;q}$ is given by
\begin{equation}
b(n) = \sum_{p=0}^{n-2}\mathfrak{p}(p)+\sum_{p=0}^{n-2}(n-p-1)\mathfrak{p}(p),
\end{equation}
while the most singular terms in (\ref{int_const}) take form
\begin{equation}
\frac{\theta}{(x-z_{a_1})^{\mu_1}(x-z_{a_2})^{\mu_2}\cdots
(x-z_{a_p})^{\mu_p}},\qquad
\frac{\theta\vartheta_{a_1}\vartheta_{a_2}}{(x-z_{a_1})^{\mu_1}(x-z_{a_2})^{\mu_2}\cdots
(x-z_{a_p})^{\mu_p}},
\label{m_sing_const}
\end{equation}
where $\mu_1+\mu_2+\ldots+\mu_p=n$. The number of such terms is given by
\begin{equation}
B(n)\equiv \mathfrak{p}(n)+\sum_{p=0}^{n-2}\Big\lfloor \frac{n-p}{2}\Big\rfloor \mathfrak{p}(p),
\end{equation}
where $\lfloor \cdots \rfloor$ is the floor function. From inequalities $\sum_{p=0}^{n-2}\mathfrak{p}(p)+1\ge \mathfrak{p}(n)$ and $p-1\ge \big\lfloor \frac{p}{2}\big\rfloor$ for $p\in{\IN}$, and taking into account the freedom to adjust $\alpha$, it follows that
\begin{equation}
b(n)+1\ge B(n).
\end{equation}
One can therefore adjust coefficients in (\ref{int_const}) so that the most singular terms of the first type in (\ref{m_sing_const}) reproduce the same singular combinations that appear in
\begin{equation}
\theta\partial_x^{n}\chi_{\alpha}(x,\theta),
\end{equation}
while the singularities of the second type in  (\ref{m_sing_const}) are removed (by setting their coefficients to zero). Repeating such operations for less singular terms, one can express constraint equations in terms of derivatives of the wave-function with respect to $\theta$ and $x$, and ultimately interpret them as super-quantum curves.

\subsection{Super-quantum curve at level $1/2$}

The lowest level at which a super-Virasoro singular vector arises is $n=1/2$, which corresponds to values $p=q=1$ in (\ref{s_vir_sing_momenta}), and vanishing momentum $\alpha=0$ in consequence. Therefore the wave-function is identified with the partition function (\ref{matrix_def}) in this case, $\widehat{\chi}_{\alpha=0}(x,\theta)=Z$, which does not depend neither on $x$ nor on $\theta$. In this case the operator $\widehat{A}_{1/2}^{(0)}$ takes a simple form
\begin{equation}
\widehat{A}_{1/2}^{(0)}\widehat{\chi}_{\alpha=0}(x,\theta)=
-\partial_{\theta}\widehat{\chi}_{\alpha=0}(x,\theta)=0,
\end{equation}
and from (\ref{theta-G12}), on the level of the bosonic component $\widehat{\chi}_{B,\alpha=0}(x)$ we get
\begin{equation}
\widehat{\mathsf{A}}_{1/2}^{(0)}=\widehat{\mathsf{G}}_{-1/2},\qquad
\widehat{\mathsf{A}}_{1/2}^{(1)}=
\widehat{\mathsf{G}}_{-1/2}\widehat{\mathsf{A}}_{1/2}^{(0)}=
\widehat{\mathsf{L}}_{-1},
\label{q_curve_eq_c_1_2}
\end{equation}
so that $\widehat{\mathsf{A}}_{1/2}^{(0)}$ indeed has the same form as the operator (\ref{singular:vector:level:12}) encoding a super-Virasoro singular vector at level $1/2$. Finally, from (\ref{q_curve_gen_c}) we find the full super-quantum curve equation at level $1/2$
\begin{equation}
\widehat{A}_{1/2}^{\alpha=0}\widehat{\chi}_{\alpha=0}(x,\theta)=0,\qquad
\widehat{A}_{1/2}^{\alpha=0}=
-\partial_{\theta}-\theta\partial_x.
\end{equation}

\subsection{Super-quantum curves at level $3/2$}

To construct quantum curves at level $3/2$ we consider the first nontrivial constraint equation
\begin{equation}
c\theta\big<S_+^{(3/2)}(x,\theta)\chi_{\alpha}^{\textrm{ins}}(x,\theta)\big>=0,
\label{q_curve_c_3_2}
\end{equation}
where
\be
\begin{split}
\theta S_+^{(3/2)}(x,\theta)&=\frac{(\alpha+Q\hbar)\sqrt{\beta}}{\hbar}\sum_{a=1}^N\frac{\theta\vartheta_a}{(x-z_a)^2}
+\beta\sum_{a,b=1}^N\frac{\theta\vartheta_a}{(x-z_a)(x-z_b)}  + \\
&\qquad
-\frac{\theta\sqrt{\beta}}{\hbar}\sum_{a=1}^N\frac{V_B'(x)\vartheta_a+V_F(x)}{x-z_a}
-\frac{\theta}{\hbar^2}\widehat{h}(x).
\end{split}
\ee
The constraint equation (\ref{q_curve_c_3_2}) can be written in terms of $\theta\partial_x\partial_{\theta}\chi_{\alpha}(x,\theta)$, see (\ref{d_th1x1_chi_ins}), only for
\begin{equation}
c=-\frac{\alpha^2}{\hbar^2},
\end{equation}
with special values of momenta
\begin{equation}
\alpha=0,\quad \beta^{1/2}\hbar,\quad \textrm{or}\quad -\beta^{-1/2}\hbar.
\label{momenta_3_2}
\end{equation}
These values correspond respectively to $(p,q)=(1,1)$, $(p,q)=(3,1)$, and $(p,q)=(1,3)$ in (\ref{s_vir_sing_momenta}), with the first solution corresponding to singular vector at level $1/2$. For the above values of $\alpha$ the equation (\ref{q_curve_c_3_2}) can be written as
\begin{equation}
\theta\Big(\partial_x\partial_{\theta}+\frac{\alpha^2}{\hbar^4}\widehat{h}(x)\Big)\chi_{\alpha}(x,\theta)
+\frac{\theta\alpha^2\sqrt{\beta}}{\hbar^3}
\bigg<\bigg(\sum_{a=1}^N\frac{V_B'(x)\vartheta_a}{x-z_a}
+\sum_{a=1}^N\frac{V_F(x)}{x-z_a}\bigg)
\chi_{\alpha}^{\textrm{ins}}(x,\theta)\bigg>=0,
\label{q_curve_c_3_2_a}
\end{equation}
which using (\ref{d_th1_chi_ins}) can be further rewritten as a differential equation
\begin{equation}
\theta\Big(\partial_x\partial_{\theta}+\frac{\alpha}{\hbar^2}V_B'(x)\partial_{\theta}
-\frac{\alpha}{\hbar^2}V_F(x)\partial_x
+\frac{\alpha^2}{\hbar^4}\widehat{h}(x)\Big)\chi_{\alpha}(x,\theta)=0.
\end{equation}
By including the prefactor in (\ref{chi_hat_w}) and using (\ref{hf_Vbf_g_s}) we then obtain
\begin{equation}
\widehat{A}_{3/2}^{(0)}=-\partial_x\partial_{\theta}
-\frac{\alpha^2}{\hbar^2}\widehat{G}_{-3/2},
\end{equation}
for the momenta $\alpha$ in (\ref{momenta_3_2}). As the operators acting on $\widehat{\chi}_{B,\alpha=0}(x)$ we find
\begin{align}
\begin{split}
&
\widehat{\mathsf{A}}_{3/2}^{(0)}=
\widehat{\mathsf{L}}_{-1}\widehat{\mathsf{G}}_{-1/2}-\frac{\alpha^2}{\hbar^2}\widehat{\mathsf{G}}_{-3/2},
\\
&
\widehat{\mathsf{A}}_{3/2}^{(1)}=\widehat{\mathsf{G}}_{-1/2}\widehat{\mathsf{A}}_{3/2}^{(0)}=
\widehat{\mathsf{L}}_{-1}^2-\frac{2\alpha^2}{\hbar^2}\widehat{\mathsf{L}}_{-2}
+\frac{\alpha^2}{\hbar^2}\widehat{\mathsf{G}}_{-3/2}\widehat{\mathsf{G}}_{-1/2},
\label{q_curve_eq_c_3_2}
\end{split}
\end{align}
so that $\widehat{\mathsf{A}}_{3/2}^{(0)}$ indeed has the form of singular vectors at level $3/2$ given in (\ref{singular:vector:level:32}) for $\alpha=\pm \beta^{\pm 1/2} \hbar$. Moreover, for the remaining value $\alpha=0$ in (\ref{momenta_3_2}), $\widehat{\mathsf{A}}_{3/2}^{(0)}$ essentially reduces to the singular vector at level $1/2$, see (\ref{q_curve_eq_c_1_2}). We also obtain
\begin{equation}
\widehat{A}_{3/2}^{(1)}=
\partial_x^2-\frac{2\alpha^2}{\hbar^2}\widehat{L}_{-2}
-\frac{\alpha^2}{\hbar^2}\widehat{G}_{-3/2}\partial_{\theta},
\end{equation}
and from (\ref{q_curve_gen_c}) we finally find the super-quantum curve equation
\begin{equation}
\widehat{A}_{3/2}^{\alpha}\widehat{\chi}_{\alpha}(x,\theta)=0,\qquad
\widehat{A}_{3/2}^{\alpha}=
-\partial_x\partial_{\theta}-\frac{\alpha^2}{\hbar^2}\widehat{G}_{-3/2}
-\theta\Big(\partial_x^2-\frac{2\alpha^2}{\hbar^2}\widehat{L}_{-2}\Big),
\label{q_curve_eq_3_2}
\end{equation}
for special values of $\alpha$ in (\ref{momenta_3_2}).

\subsection{Super-quantum curve at level $2$}

To identify super-quantum curves at level 2 we consider the constraint
\begin{equation}
\theta\Big<\Big(c_1T_+^{(2)}(x,\theta)
+c_2\partial_{V_F(x)}^{(1/2)}S_+^{(3/2)}(x,\theta)\Big)
\chi_{\alpha}^{\textrm{ins}}(x,\theta)\Big>=0,
\label{q_curve_c_2}
\end{equation}
where
\be
\begin{split}
\theta T_+^{(2)}(x,\theta)&=
\frac{(\alpha+Q\hbar/2)\sqrt{\beta}}{\hbar}\sum_{a=1}^N\frac{\theta}{(x-z_a)^2}
+\frac{\beta}{2}\sum_{a,b=1}^N\frac{\theta}{(x-z_a)(x-z_b)}  +  \\
&\qquad
+\frac{\beta}{2}\sum_{a,b=1}^N\frac{\theta\vartheta_a\vartheta_b}{(x-z_a)(x-z_b)^2}
-\frac{\theta\sqrt{\beta}}{2\hbar}\sum_{a=1}^N\frac{2V_B'(x)+V_F'(x)\vartheta_a}{x-z_a}  +  \\
&\qquad
-\frac{\theta\sqrt{\beta}}{2\hbar}\sum_{a=1}^N\frac{V_F(x)\vartheta_a}{(x-z_a)^2}
-\frac{\theta}{\hbar^2}\widehat{f}(x),
\label{t_p_2_xt}
\end{split}
\ee
and
\be
\begin{split}
\theta\partial_{V_F(x)}^{(1/2)}S_+^{(3/2)}(x,\theta)&=
\sum_{a=1}^N\frac{\theta}{(x-z_a)^2}
+\frac{(\alpha+Q\hbar)\sqrt{\beta}}{\hbar}\sum_{a,b=1}^N\frac{\theta\vartheta_a\vartheta_b}{(x-z_a)(x-z_b)^2}
   + \\
&\qquad
+\frac{\theta\sqrt{\beta}}{\hbar}\sum_{a,b=1}^N\frac{V_F(x)\vartheta_a}{(x-z_a)(x-z_b)}
+\frac{\theta}{\hbar^2}\widehat{h}(x)\sum_{a=1}^N\frac{\vartheta_a}{x-z_a}.
\end{split}
\ee
The constraint (\ref{q_curve_c_2}) can be expressed in terms of  $\theta\partial_x^2\chi_{\alpha}^{\textrm{ins}}(x,\theta)$ given in (\ref{d_th1x1_chi_ins}) if
the coefficients $c_1$ and $c_2$ are fixed as
\begin{equation}
c_1=\frac{2\alpha^2}{\hbar^2},\quad
c_2=-\frac{\beta^{1/2}\alpha(2\alpha^2+Q\hbar\alpha-\hbar^2)}{\hbar^3},
\end{equation}
and momenta take form
\begin{equation}
\alpha=0,\quad \beta^{1/2}\hbar, \quad -\beta^{-1/2}\hbar, \quad -\frac{Q\hbar}{2} =\frac{\beta^{1/2}-\beta^{-1/2}}{2}\hbar.
\label{momenta_2}
\end{equation}
We recognize that values $\alpha=0$ and $\alpha=\pm \beta^{\pm 1/2}\hbar$ correspond respectively to singular vectors at levels $1/2$ and $3/2$, while $\alpha=-Q\hbar/2$ arises for $p=q=2$ in (\ref{s_vir_sing_momenta}) and corresponds to the singular vector at level 2, as we show in what follows. For all values of $\alpha$ in (\ref{momenta_2}), using (\ref{d_th1_chi_ins}), the equation (\ref{q_curve_c_2}) can be rewritten as
\be
\begin{split}
&
\theta\Big(\partial_x^2-\frac{c_1}{2\alpha}V_F'(x)\partial_{\theta}
+\frac{c_1}{\alpha}V_B'(x)\partial_x-\frac{c_1}{\hbar^2}\widehat{f}(x)
+\frac{c_2}{\hbar\sqrt{\beta}\alpha}\widehat{h}(x)\partial_{\theta}\Big)
\chi_{\alpha}(x,\theta)  +  \\
&
-\frac{\theta\sqrt{\beta}}{2\hbar}V_F(x)\bigg<\bigg(
\sum_{a=1}^N\frac{c_1\vartheta_a}{(x-z_a)^2}
-\sum_{a,b=1}^N\frac{2c_2\vartheta_a}{(x-z_a)(x-z_b)}\bigg)
\chi_{\alpha}^{\textrm{ins}}(x,\theta)\bigg>=0.
\label{q_curve_c_2_a}
\end{split}
\ee
Furthermore, adding to the equation (\ref{q_curve_c_2_a}) the following constraint and fixing the value of $c_3$
\begin{equation}
-\frac{\theta\sqrt{\beta}}{2\hbar}V_F(x)\Big<c_3 S_+^{(3/2)}(x,\theta)\chi_{\alpha}^{\textrm{ins}}(x,\theta)\Big>=0,\qquad  c_3=\frac{2\alpha}{\sqrt{\beta}\hbar},
\label{q_curve_c_2_b}
\end{equation}
for the momenta in (\ref{momenta_2}) we obtain a differential equation
\be
\begin{split}
&
\theta\Big(\partial_x^2-\frac{c_1}{2\alpha}V_F'(x)\partial_{\theta}
-\frac{c_1}{\hbar^2}\widehat{f}(x)+\frac{c_1}{\alpha}V_B'(x)\partial_x
+\frac{c_2}{\hbar\sqrt{\beta}\alpha}\widehat{h}(x)\partial_{\theta}
+\frac{2\alpha+Q\hbar}{\hbar^2}V_F(x)
\partial_x\partial_{\theta}   +  \\
&\quad
+\frac{1}{\hbar^2}V_B'(x)V_F(x)\partial_{\theta}
+\frac{\alpha}{\hbar^4}V_F(x)\widehat{h}(x)
\Big)\chi_{\alpha}(x,\theta)=0,
\label{q_curve_c_2_c}
\end{split}
\ee
which now can be expressed in terms of super-Virasoro generators identified in section \ref{ssec-sVir-blocks}. Including the prefactor in (\ref{chi_hat_w}), and using (\ref{hf_Vbf_g_s}), after some algebra we find
\begin{equation}
\widehat{A}_{2}^{(0)}=
\partial_x^2-\frac{2\alpha^2}{\hbar^2}\widehat{L}_{-2}
-\frac{2\alpha^2+Q\hbar\alpha-\hbar^2}{\hbar^2}\widehat{G}_{-3/2}\partial_{\theta},
\end{equation}
for the momenta (\ref{momenta_2}). This operator leads to the following operators acting on $\widehat{\chi}_{B,\alpha}(x)$
\begin{align}
\begin{split}
\widehat{\mathsf{A}}_{2}^{(0)}&=
\widehat{\mathsf{L}}_{-1}^2-\frac{2\alpha^2}{\hbar^2}\widehat{\mathsf{L}}_{-2}
+\frac{2\alpha^2+Q\hbar\alpha-\hbar^2}{\hbar^2}\widehat{\mathsf{G}}_{-3/2}\widehat{\mathsf{G}}_{-1/2} =  \\
&=\widehat{\mathsf{A}}_{3/2}^{(1)}
+\frac{\alpha^2+Q\hbar\alpha-\hbar^2}{\hbar^2}\widehat{\mathsf{G}}_{-3/2}\widehat{\mathsf{G}}_{-1/2},  \\
\widehat{\mathsf{A}}_{2}^{(1)}&=\widehat{\mathsf{G}}_{-1/2}\widehat{\mathsf{A}}_{2}^{(0)}  = \widehat{\mathsf{L}}_{-1}^2\widehat{\mathsf{G}}_{-1/2}
-\frac{2\alpha^2}{\hbar^2}\widehat{\mathsf{L}}_{-2}\widehat{\mathsf{G}}_{-1/2}
-\frac{\alpha^2}{\hbar^2}\widehat{\mathsf{G}}_{-5/2} + \\
&\qquad \qquad \qquad \quad
+\frac{2\alpha^2+Q\hbar\alpha-\hbar^2}{\hbar^2}\big(2\widehat{\mathsf{L}}_{-2}\widehat{\mathsf{G}}_{-1/2}-\widehat{\mathsf{G}}_{-3/2}\widehat{\mathsf{L}}_{-1}\big),
\label{q_curve_eq_c_2}
\end{split}
\end{align}
and for $\alpha=-\frac{Q\hbar}{2}$ the operator $\widehat{\mathsf{A}}_{2}^{(0)}$ takes form of singular vector at level 2 in (\ref{singular:vector:level:2}). Furthermore
\begin{equation}
\widehat{A}_{2}^{(1)}=
-\partial_x^2\partial_{\theta}
+\frac{2\alpha^2}{\hbar^2}\widehat{L}_{-2}\partial_{\theta}
-\frac{\alpha^2}{\hbar^2}\widehat{G}_{-5/2}
-\frac{2\alpha^2+Q\hbar\alpha-\hbar^2}{\hbar^2}\big(2\widehat{L}_{-2}\partial_{\theta}+\widehat{G}_{-3/2}\partial_x\big),
\end{equation}
and finally from (\ref{q_curve_gen_c}) we obtain the super-quantum curve equation at level 2
\begin{align}
\begin{split}
\widehat{A}_{2}^{\alpha}\widehat{\chi}_{\alpha}(x,\theta) &= 0,   \\
\widehat{A}_{2}^{\alpha} &=
\partial_x^2-\frac{2\alpha^2}{\hbar^2}\widehat{L}_{-2}
-\frac{2\alpha^2+Q\hbar\alpha-\hbar^2}{\hbar^2}\widehat{G}_{-3/2}\partial_{\theta}
+\frac{\alpha^2}{\hbar^2}\theta\widehat{G}_{-5/2}  +  \\
&\qquad
+\frac{2\alpha^2+Q\hbar\alpha-\hbar^2}{\hbar^2}\theta
\Big(2\widehat{L}_{-2}\partial_{\theta}+\widehat{G}_{-3/2}\partial_x\Big),
\label{q_curve_eq_2}
\end{split}
\end{align}
for the momenta (\ref{momenta_2}).

Similarly as before we see that these results reduce to results at lower levels, upon substitution of  values of $\alpha$ corresponding to those lower levels (\ref{momenta_3_2}); in particular for those values of $\alpha$ the operator $\widehat{\mathsf{A}}_{2}^{(0)}$, up to a simple factor, coincides with the operator $\widehat{\mathsf{A}}_{3/2}^{(1)}$ in (\ref{q_curve_eq_c_3_2}).

\subsection{Super-quantum curves at level $5/2$}

To construct quantum curves at level $5/2$ we consider
\begin{equation}
\theta\Big<\Big(c_1S_+^{(5/2)}(x,\theta)
+c_2\partial_{V_B(x)}^{(1)}S_+^{(3/2)}(x,\theta)
+c_3\partial_{V_F(x)}^{(1/2)}T_+^{(2)}(x,\theta)\Big)
\chi_{\alpha}^{\textrm{ins}}(x,\theta)\Big>=0,
\label{q_curve_c_5_2}
\end{equation}
where
\be
\begin{split}
\theta S_+^{(5/2)}(x,\theta)&=
\frac{(\alpha+2Q\hbar)\sqrt{\beta}}{\hbar}\sum_{a=1}^N\frac{\theta\vartheta_a}{(x-z_a)^3}+\beta\sum_{a,b=1}^N\frac{\theta\vartheta_a}{(x-z_a)^2(x-z_b)} +  \\
&\qquad
+\beta\sum_{a,b=1}^N\frac{\theta\vartheta_a}{(x-z_a)(x-z_b)^2}
-\frac{\theta\sqrt{\beta}}{\hbar}\sum_{a=1}^N\frac{V_B'(x)\vartheta_a+V_F(x)}{(x-z_a)^2}  +  \\
&\qquad
+\frac{\theta\sqrt{\beta}}{\hbar}\sum_{a=1}^N\frac{V_B''(x)\vartheta_a+V_F'(x)}{x-z_a}
+\frac{\theta}{\hbar^2}\partial_x\widehat{h}(x),
\end{split}
\ee
\be
\begin{split}
\theta\partial_{V_B(x)}^{(1)}S_+^{(3/2)}(x,\theta)&=
\sum_{a=1}^N\frac{\theta\vartheta_a}{(x-z_a)^3}
+\frac{(\alpha+Q\hbar)\sqrt{\beta}}{\hbar}\sum_{a,b=1}^N\frac{\theta\vartheta_a}{(x-z_a)^2(x-z_b)}  + \\
&\qquad
+\beta\sum_{a,b,c=1}^N\frac{\theta\vartheta_a}{(x-z_a)(x-z_b)(x-z_c)}
-\frac{\theta\sqrt{\beta}}{\hbar}\sum_{a,b=1}^N\frac{V_B'(x)\vartheta_a}{(x-z_a)(x-z_b)}   +  \\
&\qquad
-\frac{\theta\sqrt{\beta}}{\hbar}\sum_{a,b=1}^N\frac{V_F(x)}{(x-z_a)(x-z_b)}
-\frac{\theta}{\hbar^2}\widehat{h}(x)\sum_{a=1}^N\frac{1}{x-z_a},
\end{split}
\ee
and
\be
\begin{split}
\theta\partial_{V_F(x)}^{(1/2)}T_+^{(2)}(x,\theta)&=
\frac{3}{2}\sum_{a=1}^N\frac{\theta\vartheta_a}{(x-z_a)^3}
+\frac{(2\alpha+Q\hbar)\sqrt{\beta}}{2\hbar}\sum_{a,b=1}^N\frac{\theta\vartheta_a}{(x-z_a)(x-z_b)^2} + \\
&\qquad
+\frac{\beta}{2}\sum_{a,b,c=1}^N\frac{\theta\vartheta_a}{(x-z_a)(x-z_b)(x-z_c)}
-\frac{\theta\sqrt{\beta}}{\hbar}\sum_{a,b=1}^N\frac{V_B'(x)\vartheta_a}{(x-z_a)(x-z_b)} + \\
&\qquad
-\frac{\theta\sqrt{\beta}}{2\hbar}\sum_{a,b=1}^N\frac{V_F(x)\vartheta_a\vartheta_b}{(x-z_a)^2(x-z_b)}-\frac{\theta}{\hbar^2}\widehat{f}(x)\sum_{a=1}^N\frac{\vartheta_a}{x-z_a}.
\end{split}
\ee
We find that only for momenta
\begin{equation}
\alpha=0,\quad \beta^{1/2}\hbar, \quad -\beta^{-1/2}\hbar, \quad -\frac{Q\hbar}{2},\quad 2\beta^{1/2}\hbar, \quad -2\beta^{-1/2}\hbar,
\label{momenta_5_2}
\end{equation}
the equation (\ref{q_curve_c_5_2}) can be written as
\begin{align}
&
\theta\Big(\partial_x^2\partial_{\theta}
+\frac{c_1}{\alpha}V_B''(x)\partial_{\theta}-\frac{c_1}{\alpha}V_F'(x)\partial_x
+\frac{c_1}{\hbar^2}\partial_x\widehat{h}(x)
+\frac{c_2}{\hbar\sqrt{\beta}\alpha}\widehat{h}(x)\partial_x
-\frac{c_3}{\hbar\sqrt{\beta}\alpha}\widehat{f}(x)\partial_{\theta}\Big)
\chi_{\alpha}(x,\theta)
\nonumber\\
&
-\frac{\theta\sqrt{\beta}}{2\hbar}V_F(x)\bigg<\bigg(
\sum_{a=1}^N\frac{2c_1}{(x-z_a)^2}
+\sum_{a,b=1}^N\frac{2c_2}{(x-z_a)(x-z_b)}
-\sum_{a,b=1}^N\frac{c_3\vartheta_a\vartheta_b}{(x-z_a)(x-z_b)^2}\bigg)
\chi_{\alpha}^{\textrm{ins}}(x,\theta)\bigg>
\nonumber\\
&
-\frac{\theta\sqrt{\beta}}{\hbar}V_B'(x)\bigg<\bigg(
\sum_{a=1}^N\frac{c_1\vartheta_a}{(x-z_a)^2}
+\sum_{a,b=1}^N\frac{(c_2+c_3)\vartheta_a}{(x-z_a)(x-z_b)}\bigg)
\chi_{\alpha}^{\textrm{ins}}(x,\theta)\bigg>=0,
\label{q_curve_c_5_2_a}
\end{align}
where coefficients $c_1, c_2, c_3$ have been fixed as in earlier examples. The momenta $\pm 2\beta^{\pm 1/2}\hbar$ in (\ref{momenta_5_2}) arise respectively for $(p,q)=(5,1)$ and $(p,q)=(1,5)$ in (\ref{s_vir_sing_momenta}), and represent singular vectors at level $5/2$, while other values of $\alpha$ in (\ref{momenta_5_2}) represent singular vectors at all lower levels. By adding to the equation (\ref{q_curve_c_5_2_a}) an additional constraint
\begin{equation}
-\frac{\theta\sqrt{\beta}}{\hbar}\Big<\Big(c_4V_B'(x) S_+^{(3/2)}(x,\theta)
+c_5V_F(x)T_+^{(2)}(x,\theta)
+c_6V_F(x)\partial_{V_F(x)}^{(1/2)}S_+^{(3/2)}(x,\theta)\Big)
\chi_{\alpha}^{\textrm{ins}}(x,\theta)\Big>=0,
\label{q_curve_c_5_2_b}
\end{equation}
and adjusting coefficients $c_4, c_5, c_6$, for the momenta (\ref{momenta_5_2}) we obtain a differential equation
\begin{align}
&
\theta\Big(\partial_x^2\partial_{\theta}
+\frac{c_1}{\alpha}V_B''(x)\partial_{\theta}-\frac{c_1}{\alpha}V_F'(x)\partial_x
+\frac{c_1}{\hbar^2}\partial_x\widehat{h}(x)
+\frac{c_2}{\hbar\sqrt{\beta}\alpha}\widehat{h}(x)\partial_x
-\frac{c_3}{\hbar\sqrt{\beta}\alpha}\widehat{f}(x)\partial_{\theta} +
\nonumber\\
&\quad
+\frac{2\alpha}{\hbar^2}V_B'(x)\partial_x\partial_{\theta}
+\frac{\sqrt{\beta}c_4}{\hbar\alpha}V_B'(x)^2\partial_{\theta}
+\frac{\sqrt{\beta}c_4}{\hbar^3}V_B'(x)\widehat{h}(x)
-\frac{\alpha}{\hbar^2}V_F(x)\partial_x^2 +
\nonumber\\
&\quad
-\frac{\sqrt{\beta}(c_4+c_5)}{\hbar\alpha}V_B'(x)V_F(x)\partial_x
-\frac{\sqrt{\beta}c_5}{2\hbar\alpha}V_F'(x)V_F(x)\partial_{\theta}
+\frac{\sqrt{\beta}c_5}{\hbar^3}V_F(x)\widehat{f}(x)
\Big)\chi_{\alpha}(x,\theta)=0.
\label{q_curve_c_5_2_c}
\end{align}
Including the prefactor in (\ref{chi_hat_w}) this differential equation takes form
\begin{align}
\begin{split}
\widehat{A}_{5/2}^{(0)}&=
-\partial_x^2\partial_{\theta}
+\frac{2\alpha(\alpha^2+Q\hbar\alpha-\hbar^2)}{\hbar^2(3\alpha+2Q\hbar)}\widehat{L}_{-2}\partial_{\theta}
-\frac{\alpha(2\alpha^2+Q\hbar\alpha+\hbar^2)}{\hbar^2(3\alpha+2Q\hbar)}\widehat{G}_{-3/2}\partial_x +
\\
&\ \ \
+\frac{\alpha^2(2\alpha^3+3Q\hbar\alpha^2+(Q^2-5)\hbar^2\alpha-3Q\hbar^3)}{\hbar^4(3\alpha+2Q\hbar)}\widehat{G}_{-5/2},
\end{split}
\end{align}
for the momenta (\ref{momenta_5_2}). Acting on the bosonic component this operator is represented as
\begin{equation}
\widehat{\mathsf{A}}_{5/2}^{(0)}=\widehat{\mathsf{A}}_{2}^{(1)}
+\frac{(\alpha^2+Q\hbar\alpha-\hbar^2)(2\alpha+Q\hbar)}{\hbar^2(3\alpha+2Q\hbar)}
\Big(\frac{\alpha^2}{\hbar^2}\widehat{\mathsf{G}}_{-5/2}
+2\widehat{\mathsf{G}}_{-3/2}\widehat{\mathsf{L}}_{-1}
-4\widehat{\mathsf{L}}_{-2}\widehat{\mathsf{G}}_{-1/2}\Big),
\label{q_curve_eq_c_5_2}
\end{equation}
and, as expected, for $\alpha=\pm 2\beta^{\pm 1/2}\hbar$ it indeed has the same form as a singular vector at level $5/2$ in (\ref{singular:vector:level:52}). Finally, determining $\widehat{\mathsf{A}}_{5/2}^{(1)}$ and then using (\ref{q_curve_gen_c}) we finally obtain the super-quantum curve at level $5/2$
\begin{align}
\begin{split}
&
\widehat{A}_{5/2}^{\alpha}\widehat{\chi}_{\alpha}(x,\theta)=0,
\\
&
\widehat{A}_{5/2}^{\alpha}=
-\partial_x^2\partial_{\theta}
+\frac{2\alpha(\alpha^2+Q\hbar\alpha-\hbar^2)}{\hbar^2(3\alpha+2Q\hbar)}\widehat{L}_{-2}\partial_{\theta}
-\frac{\alpha(2\alpha^2+Q\hbar\alpha+\hbar^2)}{\hbar^2(3\alpha+2Q\hbar)}\widehat{G}_{-3/2}\partial_x +
\\
&\hspace{2.5em}
+\frac{\alpha^2(2\alpha^3+3Q\hbar\alpha^2+(Q^2-5)\hbar^2\alpha-3Q\hbar^3)}{\hbar^4(3\alpha+2Q\hbar)}\widehat{G}_{-5/2}
-\theta\Big(\partial_x^3
-\frac{2\alpha^2}{\hbar^2}\widehat{L}_{-2}\partial_x +
\\
&\hspace{2.5em}
+\frac{\alpha(\alpha^2+Q\hbar\alpha-\hbar^2)}{\hbar^2(3\alpha+2Q\hbar)}\widehat{G}_{-5/2}\partial_{\theta}
+\frac{2\alpha^2(2\alpha^3+3Q\hbar\alpha^2+(Q^2-5)\hbar^2\alpha-3Q\hbar^3)}{\hbar^4(3\alpha+2Q\hbar)}\widehat{L}_{-3}\Big),
\label{q_curve_eq_5_2}
\end{split}
\end{align}
for the momenta (\ref{momenta_5_2}).

Similarly as before, for the momenta (\ref{momenta_2}) at level 2 the operator $\widehat{\mathsf{A}}_{5/2}^{(0)}$ reduces, up to a simple factor, to the operator $\widehat{\mathsf{A}}_{2}^{(1)}$ in (\ref{q_curve_eq_c_2}), and encodes singular vectors up to level 2.

\section{Double quantum structure and special limits}     \label{sec-limits}

In this section we analyze various limits of super-quantum curves. These limits are interesting, as they are related to two quantum structures, which are encoded in super-quantum curves. First, super-quantum curves are quantum in the 't Hooft sense, and their classical limit  can be identified as the large $N$ limit (\ref{large_N_lim}). As we discuss below, for $\beta=1$ the super-quantum spectral curve at level $3/2$ reduces to the (classical) spectral curve of the super-eigenvalue model.

The second interesting limit can be interpreted as the Nekrasov-Shatashvili limit \cite{NS}, or equivalently as the classical limit in super-Liouville theory
\cite{Hadasz:2007nt}.
In terms of parameters (\ref{e1e2}) this limit arises when $\epsilon_1 \to 0$ with $\epsilon_2$ fixed. The dependence of $\beta$ is therefore crucial in this limit. As we will show below, in this limit super-quantum curves reduce to differential equations for certain fields in classical super-Liouville theory.

To avoid singularities in matrix model expectation values that arise in the above limits, one should consider the normalized partition function (\ref{Psi-def}) and the corresponding super-quantum curves (\ref{A-hat-norm}).

\subsection{Classical limit}

We consider first the classical 't Hooft limit. Note that derivatives of the normalized wave-function $\Psi_{\alpha}(x,\theta)$ introduced in (\ref{Psi-def}) with respect to $\theta$ and $x$ take form
\begin{align}
\begin{split}
&
\hbar\partial_{\theta}\Psi_{\alpha}(x,\theta)=
-\frac{\alpha}{Z}\left<\psi(x)e^{\frac{\alpha}{\hbar}\phi(x)+\frac{\alpha}{\hbar}\psi(x)\theta}\right>,\\
&
\hbar\partial_x\Psi_{\alpha}(x,\theta)=
\frac{\alpha}{Z}\left<\big(\partial_x\phi(x)+\partial_x\psi(x)\theta\big)e^{\frac{\alpha}{\hbar}\phi(x)+\frac{\alpha}{\hbar}\psi(x)\theta}\right>.
\end{split}
\end{align}
Therefore in the classical limit (\ref{large_N_lim}) with $\beta=1$
\begin{align}
\begin{split}
&
\hbar\partial_{\theta}\Psi_{\alpha}(x,\theta)\
\stackrel{N\to\infty, \ \beta\to 1}{\longrightarrow}\
-\frac{\alpha}{\hbar}y_F(x)\Psi_{\alpha}(x,\theta),
\\
&
\hbar\partial_x\Psi_{\alpha}(x,\theta)\
\stackrel{N\to\infty, \ \beta\to 1}{\longrightarrow}\
\frac{\alpha}{\hbar}\big(y_B(x)+y_F'(x)\theta\big)
\Psi_{\alpha}(x,\theta),
\label{par_h_classic}
\end{split}
\end{align}
where $y_B(x)$ and $y_F(x)$ are defined in (\ref{def_y_bf}) and (\ref{yByF}).

In particular, consider the super-quantum curve at level $3/2$ in (\ref{q_curve_eq_3_2})
\begin{equation}
-\hbar^2\widehat{\mathcal{A}}_{3/2}^{\alpha}=\hbar^2\partial_x\partial_{\theta}
+\alpha^2\widehat{\mathcal{G}}_{-3/2}
+\hbar^2\theta \partial_x^2
-2\theta\alpha^2\widehat{\mathcal{L}}_{-2},
\end{equation}
with momenta $\alpha=\pm \beta^{\pm 1/2}\hbar$. In the limit (\ref{par_h_classic}) this quantum curve reduces to the super-spectral curve, which defines the solution space $\mathcal{C}$ in (\ref{sp_r_curve})
\begin{equation}
\widehat{\mathcal{A}}_{3/2}^{\alpha}\Psi_{\alpha}(x,\theta)=0\
\stackrel{N\to\infty, \ \beta\to 1}{\longrightarrow}\
A_F(x,y_B|y_F)=\theta A_B(x,y_B|y_F),
\end{equation}
where
\begin{equation}
G(x)=-\lim_{\begin{subarray}{c}\hbar\to 0\\\beta=1\end{subarray}}
\hbar^2\widehat{\mathcal{G}}_{-3/2},\qquad
L(x)=-\lim_{\begin{subarray}{c}\hbar\to 0\\\beta=1\end{subarray}}
\hbar^2\widehat{\mathcal{L}}_{-2}.   \label{GL-classical}
\end{equation}
In this sense the super-quantum curve at level $3/2$ can be regarded as a quantization of the super-spectral curve. We will present how higher level super-quantum curves behave in the classical limit in section \ref{ssec-classical-2}, after discussing Nekrasov-Shatashvili limit.

\subsection{Nekrasov-Shatashvili --- classical super-Liouville limit}

Let us discuss the limit, which in terms of parameters (\ref{e1e2}) arises for $\epsilon_1\to 0$ with $\epsilon_2$ fixed, and which can be identified with the Nekrasov-Shatashvili limit, or a classical limit in the super-Liouville theory. We will consider normalized wave-functions $\Psi_{\alpha_{2p+1,1}}(x,\theta)$ for the momenta
\begin{equation}
\alpha=\alpha_{2p+1,1}=-p\epsilon_1,   \label{alpha-2p1}
\end{equation}
which in the above limit factorize as
\begin{equation}
\Psi_{-p\epsilon_1}^{\textrm{NS}}(x,\theta)\equiv
\lim_{\epsilon_1\to 0}\Psi_{-p\epsilon_1}(x,\theta)=
\left(\Psi_{-\epsilon_1}^{\textrm{NS}}(x,\theta)\right)^p.
\label{wave_ns_factor}
\end{equation}
The corresponding super-quantum curves arise at level $p+1/2$, and it is convenient to rescale them as follows
\begin{equation}
\widehat{\mathcal{A}}_{p+1/2}^{\textrm{NS}}\Psi_{-p\epsilon_1}^{\textrm{NS}}(x,\theta)=0,\qquad
\widehat{\mathcal{A}}_{p+1/2}^{\textrm{NS}} =
-\epsilon_2^{p+1}\lim_{\epsilon_1\to 0}
\widehat{\mathcal{A}}_{p+1/2}^{-p\epsilon_1},
\end{equation}
and express in terms of operators
\be
\begin{split}
\widehat{\mathcal{G}}_{-n+1/2}^{\textrm{NS}} & = \lim_{\epsilon_1\to 0}\epsilon_1\epsilon_2\widehat{\mathcal{G}}_{-n+1/2} = \\
&=-\frac{1}{(n-2)!}\Big(
\partial_x^{n-2}\big(V_B'(x)V_F(x)\big)+\epsilon_2\partial_x^{n-1}V_F(x)
+\partial_x^{n-2}F_{F}^{(0)}(x,\epsilon_2)\Big),    \\
\widehat{\mathcal{L}}_{-n}^{\textrm{NS}} & = \lim_{\epsilon_1\to 0}\epsilon_1\epsilon_2\widehat{\mathcal{L}}_{-n}= \\
& =-\frac{1}{(n-2)!}\Big(
\frac12\partial_x^{n-2}\big(V_{B}'(x)^2\big)
+\frac12\partial_x^{n-2}\big(V_F'(x)V_F(x)\big)
+\frac12\epsilon_2\partial_x^nV_{B}(x)
+\partial_x^{n-2}F_{B}^{(0)}(x,\epsilon_2)\Big),
\end{split}
\ee
where
\begin{equation}
F_{F}^{(0)}(x,\epsilon_2) =
-\frac{1}{\epsilon_1\epsilon_2}\big[\widehat{h}(x), F^{(0)}(\epsilon_2)\big],
\qquad
F_{B}^{(0)}(x,\epsilon_2) =
-\frac{1}{\epsilon_1\epsilon_2}\big[\widehat{f}(x), F^{(0)}(\epsilon_2)\big],
\end{equation}
are defined in terms of the deformed prepotential
\begin{equation}
F^{(0)}(\epsilon_2) =
-\lim_{\epsilon_1\to 0}\epsilon_1\epsilon_2\log Z.
\end{equation}
Note that $\widehat{\mathcal{G}}_{-n+1/2}^{\textrm{NS}}$ and $\widehat{\mathcal{L}}_{-n}^{\textrm{NS}}$ are simply fermionic and bosonic functions of $x$ that satisfy
\be
\partial_x^n\widehat{\mathcal{G}}_{-3/2}^{\textrm{NS}}=n!\widehat{\mathcal{G}}_{-n-3/2}^{\textrm{NS}}, \qquad \partial_x^n\widehat{\mathcal{L}}_{-2}^{\textrm{NS}}=n!\widehat{\mathcal{L}}_{-n-2}^{\textrm{NS}}.   \label{GL-NS-relations}
\ee

Using the above notation, the super-quantum curve equation (\ref{q_curve_eq_3_2}) at level $3/2$ in the limit $\epsilon_1\to 0$ takes form
\begin{equation}
\widehat{\mathcal{A}}_{3/2}^{\textrm{NS}}\Psi_{-\epsilon_1}^{\textrm{NS}}(x,\theta)=
\left(\epsilon_2^2\partial_x\partial_{\theta}-\widehat{\mathcal{G}}_{-3/2}^{\textrm{NS}}
+\theta\epsilon_2^2\partial_x^2
+2\theta\widehat{\mathcal{L}}_{-2}^{\textrm{NS}}\right)
\Psi_{-\epsilon_1}^{\textrm{NS}}(x,\theta)=0,
\label{q_curve_ns_3_2a}
\end{equation}
and can be written equivalently as
\begin{equation}
\theta\left(\epsilon_2^2\partial_x\partial_{\theta}-\widehat{\mathcal{G}}_{-3/2}^{\textrm{NS}}\right)
\Psi_{-\epsilon_1}^{\textrm{NS}}(x,\theta)=
\theta\left(\epsilon_2^2\partial_x^2+\widehat{\mathcal{G}}_{-3/2}^{\textrm{NS}}\partial_{\theta}
+2\widehat{\mathcal{L}}_{-2}^{\textrm{NS}}\right)
\Psi_{-\epsilon_1}^{\textrm{NS}}(x,\theta)=0.
\label{q_curve_ns_3_2}
\end{equation}

The Nekrasov-Shatashvili limit of super-quantum curves at higher levels, corresponding to momenta (\ref{alpha-2p1}), can be determined recursively and written in the form
\begin{equation}
\widehat{\mathcal{A}}_{p+1/2}^{\textrm{NS}}\Psi_{-p\epsilon_1}^{\textrm{NS}}(x,\theta)=
\left(\widehat{f}_{p+1}^{p+1}
+\theta \widehat{f}_{p+1}^{p+1} \partial_{\theta}
+\theta \widehat{b}_{p+1}^{p+1}\right)
\Psi_{-p\epsilon_1}^{\textrm{NS}}(x,\theta)=0,
\end{equation}
or equivalently
\begin{equation}
\theta \widehat{f}_{p+1}^{p+1}\Psi_{-p\epsilon_1}^{\textrm{NS}}(x,\theta)=
\theta \widehat{b}_{p+1}^{p+1}\Psi_{-p\epsilon_1}^{\textrm{NS}}(x,\theta)=0.
\label{q_curve_ns_p}
\end{equation}
Here the operators $\widehat{f}_{p+1}^{p+1}$ and $\widehat{b}_{p+1}^{p+1}$ are determined by the following sequence for $\widehat{f}_{q}^{p+1}$ and $\widehat{b}_{q}^{p+1}$, $q=0,\ldots,p+1$:
\begin{align}
\begin{split}
&
\widehat{f}_{0}^{p+1}=0,\quad \widehat{f}_{1}^{p+1}=\epsilon_2\partial_{\theta},\quad
\widehat{b}_{0}^{p+1}=1,\quad
\widehat{b}_{1}^{p+1}=\epsilon_2\partial_x,
\\
&
\widehat{f}_{q+1}^{p+1}=\epsilon_2\partial_x\widehat{f}_{q}^{p+1}+
2(q-1)(p-q+1)\widehat{\mathcal{L}}_{-2}^{\textrm{NS}}\widehat{f}_{q-1}^{p+1}
-(p-q+1)\widehat{\mathcal{G}}_{-3/2}^{\textrm{NS}}\widehat{b}_{q-1}^{p+1},
\\
&
\widehat{b}_{q+1}^{p+1}=\epsilon_2\partial_x\widehat{b}_{q}^{p+1}+
q\epsilon_2^{-1}\widehat{\mathcal{G}}_{-3/2}^{\textrm{NS}}\widehat{f}_{q}^{p+1}
+2q(p-q+1)\widehat{\mathcal{L}}_{-2}^{\textrm{NS}}\widehat{b}_{q-1}^{p+1}.
\label{ns_rec_fb}
\end{split}
\end{align}
Using the factorization (\ref{wave_ns_factor}) and the super-quantum curve equation at level $3/2$ in (\ref{q_curve_ns_3_2}), by induction we find
\begin{align}
&
\theta \widehat{f}_{q+1}^{p+1}\Psi_{-p\epsilon_1}^{\textrm{NS}}(x,\theta)=
p(p-1)\cdots (p-q)\theta \left(\Psi_{-\epsilon_1}^{\textrm{NS}}(x,\theta)\right)^{p-q-1}
\left(\epsilon_2\partial_{\theta}\Psi_{-\epsilon_1}^{\textrm{NS}}(x,\theta)\right)\left(\epsilon_2\partial_x\Psi_{-\epsilon_1}^{\textrm{NS}}(x,\theta)\right)^q,
\nonumber\\
&
\theta \widehat{b}_{q+1}^{p+1}\Psi_{-p\epsilon_1}^{\textrm{NS}}(x,\theta)=
p(p-1)\cdots (p-q)\theta \left(\Psi_{-\epsilon_1}^{\textrm{NS}}(x,\theta)\right)^{p-q-1}
\left(\epsilon_2\partial_x\Psi_{-\epsilon_1}^{\textrm{NS}}(x,\theta)\right)^{q+1},
\nonumber
\end{align}
and therefore the quantum curve equation (\ref{q_curve_ns_p}) is derived.

Super-quantum curves at several lowest levels obtained from (\ref{ns_rec_fb}) take form
\begin{align}
\begin{split}
\widehat{\mathcal{A}}_{5/2}^{\textrm{NS}}&=
\epsilon_2^3\partial_x^2\partial_{\theta}
+2\epsilon_2\widehat{\mathcal{L}}_{-2}^{\textrm{NS}}\partial_{\theta}
-3\epsilon_2\widehat{\mathcal{G}}_{-3/2}^{\textrm{NS}}\partial_x
-2\epsilon_2\widehat{\mathcal{G}}_{-5/2}^{\textrm{NS}} +
\\
&\quad
+\theta\Big(\epsilon_2^3\partial_x^3
-\epsilon_2\widehat{\mathcal{G}}_{-5/2}^{\textrm{NS}}\partial_{\theta}
+8\epsilon_2\widehat{\mathcal{L}}_{-2}^{\textrm{NS}}\partial_x
+4\epsilon_2\widehat{\mathcal{L}}_{-3}^{\textrm{NS}}\Big),
\\
\widehat{\mathcal{A}}_{7/2}^{\textrm{NS}}&=
\epsilon_2^4\partial_x^3\partial_{\theta}
+8\epsilon_2^2\widehat{\mathcal{L}}_{-2}^{\textrm{NS}}\partial_x\partial_{\theta}
+4\epsilon_2^2\widehat{\mathcal{L}}_{-3}^{\textrm{NS}}\partial_{\theta}
-6\epsilon_2^2\widehat{\mathcal{G}}_{-3/2}^{\textrm{NS}}\partial_x^2
-8\epsilon_2^2\widehat{\mathcal{G}}_{-5/2}^{\textrm{NS}}\partial_x
-6\epsilon_2^2\widehat{\mathcal{G}}_{-7/2}^{\textrm{NS}} +
\\
&\quad
-18\widehat{\mathcal{G}}_{-3/2}^{\textrm{NS}}\widehat{\mathcal{L}}_{-2}^{\textrm{NS}}
+\theta\Big(\epsilon_2^4\partial_x^4
-4\epsilon_2^2\widehat{\mathcal{G}}_{-5/2}^{\textrm{NS}}\partial_x\partial_{\theta}
-4\epsilon_2^2\widehat{\mathcal{G}}_{-7/2}^{\textrm{NS}}\partial_{\theta}
+20\epsilon_2^2\widehat{\mathcal{L}}_{-2}^{\textrm{NS}}\partial_x^2 +
\\
&\quad
+20\epsilon_2^2\widehat{\mathcal{L}}_{-3}^{\textrm{NS}}\partial_x
+12\epsilon_2^2\widehat{\mathcal{L}}_{-4}^{\textrm{NS}}
-9\widehat{\mathcal{G}}_{-3/2}^{\textrm{NS}}\widehat{\mathcal{G}}_{-5/2}^{\textrm{NS}}
+36\big(\widehat{\mathcal{L}}_{-2}^{\textrm{NS}}\big)^2\Big),
\\
\widehat{\mathcal{A}}_{9/2}^{\textrm{NS}}&=
\epsilon_2^5\partial_x^4\partial_{\theta}
+20\epsilon_2^3\widehat{\mathcal{L}}_{-2}^{\textrm{NS}}\partial_x^2\partial_{\theta}
+20\epsilon_2^3\widehat{\mathcal{L}}_{-3}^{\textrm{NS}}\partial_x\partial_{\theta}
+12\epsilon_2^3\widehat{\mathcal{L}}_{-4}^{\textrm{NS}}\partial_{\theta}
-10\epsilon_2^3\widehat{\mathcal{G}}_{-3/2}^{\textrm{NS}}\partial_x^3 +
\\
&\quad
-20\epsilon_2^3\widehat{\mathcal{G}}_{-5/2}^{\textrm{NS}}\partial_x^2
-30\epsilon_2^3\widehat{\mathcal{G}}_{-7/2}^{\textrm{NS}}\partial_x
-24\epsilon_2^3\widehat{\mathcal{G}}_{-9/2}^{\textrm{NS}}
-\epsilon_2\widehat{\mathcal{G}}_{-3/2}^{\textrm{NS}}\widehat{\mathcal{G}}_{-5/2}^{\textrm{NS}}\partial_{\theta} +
\\
&\quad
+36\epsilon_2\big(\widehat{\mathcal{L}}_{-2}^{\textrm{NS}}\big)^2\partial_{\theta}
-110\epsilon_2\widehat{\mathcal{G}}_{-3/2}^{\textrm{NS}}\widehat{\mathcal{L}}_{-2}^{\textrm{NS}}\partial_x
-56\epsilon_2\widehat{\mathcal{G}}_{-3/2}^{\textrm{NS}}\widehat{\mathcal{L}}_{-3}^{\textrm{NS}}
-72\epsilon_2\widehat{\mathcal{G}}_{-5/2}^{\textrm{NS}}\widehat{\mathcal{L}}_{-2}^{\textrm{NS}} +
\\
&\quad
+\theta\Big(\epsilon_2^5\partial_x^5
-10\epsilon_2^3\widehat{\mathcal{G}}_{-5/2}^{\textrm{NS}}\partial_x^2\partial_{\theta}
-20\epsilon_2^3\widehat{\mathcal{G}}_{-7/2}^{\textrm{NS}}\partial_x\partial_{\theta}
-18\epsilon_2^3\widehat{\mathcal{G}}_{-9/2}^{\textrm{NS}}\partial_{\theta}
+40\epsilon_2^3\widehat{\mathcal{L}}_{-2}^{\textrm{NS}}\partial_x^3 +
\\
&\quad
+60\epsilon_2^3\widehat{\mathcal{L}}_{-3}^{\textrm{NS}}\partial_x^2
+72\epsilon_2^3\widehat{\mathcal{L}}_{-4}^{\textrm{NS}}\partial_x
+48\epsilon_2^3\widehat{\mathcal{L}}_{-5}^{\textrm{NS}}
-34\epsilon_2\widehat{\mathcal{G}}_{-5/2}^{\textrm{NS}}\widehat{\mathcal{L}}_{-2}^{\textrm{NS}}\partial_{\theta}
-2\epsilon_2\widehat{\mathcal{G}}_{-3/2}^{\textrm{NS}}\widehat{\mathcal{L}}_{-3}^{\textrm{NS}}\partial_{\theta} +
\\
&\quad
-56\epsilon_2\widehat{\mathcal{G}}_{-3/2}^{\textrm{NS}}\widehat{\mathcal{G}}_{-5/2}^{\textrm{NS}}\partial_x
-56\epsilon_2\widehat{\mathcal{G}}_{-3/2}^{\textrm{NS}}\widehat{\mathcal{G}}_{-7/2}^{\textrm{NS}}
+256\epsilon_2\big(\widehat{\mathcal{L}}_{-2}^{\textrm{NS}}\big)^2\partial_x
+256\epsilon_2\widehat{\mathcal{L}}_{-2}^{\textrm{NS}}\widehat{\mathcal{L}}_{-3}^{\textrm{NS}}\Big),
\label{high_ns_q_ex}
\end{split}
\end{align}
where we have used the relations (\ref{GL-NS-relations}). As expected, the above operators coincide with operators that implement classical equations of motion for certain fields in the classical super-Liouville theory \cite{Belavin:2006pv}.

\subsection{Classical limit of higher level super-quantum curves}   \label{ssec-classical-2}

Having determined the form of super-quantum curves in the Nekrasov-Shatashvili limit at higher levels, we can simply write down super-spectral curves in the 't Hooft classical limit $\widehat{\mathcal{A}}_{p+1/2}^{\textrm{cl}}$ by considering $\epsilon_2\to 0$ limit
\begin{equation}
\widehat{\mathcal{A}}_{p+1/2}^{\textrm{cl}} =
\lim_{\epsilon_2\to 0}\widehat{\mathcal{A}}_{p+1/2}^{\textrm{NS}}.
\end{equation}
As in (\ref{par_h_classic}), in the classical limit $\epsilon_2\to 0$ of $\Psi_{-p\epsilon_1}^{\textrm{NS}}(x,\theta)$ we find
\begin{align}
\begin{split}
&
\epsilon_2\partial_{\theta}\Psi_{-p\epsilon_1}^{\textrm{NS}}(x,\theta)\
\stackrel{\epsilon_2\to 0}{\longrightarrow}\
-p y_F(x)\Psi_{-p\epsilon_1}^{\textrm{NS}}(x,\theta),
\\
&
\epsilon_2\partial_x\Psi_{-p \epsilon_1}^{\textrm{NS}}(x,\theta)\
\stackrel{\epsilon_2\to 0}{\longrightarrow}\
p \big(y_B(x)+y_F'(x)\theta\big)\Psi_{-p \epsilon_1}^{\textrm{NS}}(x,\theta).
\label{par_ns_classic}
\end{split}
\end{align}
For example, taking advantage of the results in (\ref{high_ns_q_ex}) and denoting $G(x)=\lim_{\epsilon_2\to 0}\widehat{\mathcal{G}}_{-3/2}^{\textrm{NS}}$ and $L(x)=\lim_{\epsilon_2\to 0}\widehat{\mathcal{L}}_{-2}^{\textrm{NS}}$ as in (\ref{GL-classical}), we find higher-level classical curves
\begin{align}
\begin{split}
\widehat{\mathcal{A}}_{5/2}^{\textrm{cl}}&=
-2\Big(3y_B(x)A_F(x,y_B|y_F)+2y_F(x)A_B(x,y_B|y_F)\Big) +
\\
&\ \
+2\theta\Big(3y_F'(x)A_F(x,y_B|y_F)+\partial_xA_F(x,y_B|y_F)y_F(x)
+8y_B(x)A_B(x,y_B|y_F)\Big),
\\
\widehat{\mathcal{A}}_{7/2}^{\textrm{cl}}&=
-18\Big(\big(3y_B(x)^2+L(x)\big)A_F(x,y_B|y_F)+3y_F(x)y_B(x)A_B(x,y_B|y_F)\Big) +
\\
&\ \
+9\theta\Big(2\big(9y_B(x)^2+3y_F'(x)y_F(x)+2L(x)\big)A_B(x,y_B|y_F) +
\\
&\ \
+\big(15y_F'(x)y_B(x)+3y_F(x)y_B'(x)+4\partial_xG(x)\big)A_F(x,y_B|y_F) +
\\
&\ \
+3G(x)\partial_xA_F(x,y_B|y_F)\Big).
\end{split}
\end{align}

\section{Examples: super-gaussian and super-multi-Penner models} \label{sec-examples}

In this section we illustrate how our general considerations specialize when the matrix model potential is fixed to some particular form. We consider the gaussian model with quadratic bosonic potential $V_B(x)$, as well as a supersymmetric version of the multi-Penner model. These models are of particular interest: the gaussian is the simplest model and its analysis nicely illustrates various general features that we discussed in earlier sections, and the multi-Penner model turns out to encode many features familiar in super-Liouville theory. We discuss both super-quantum curves in those models, as well as their planar solutions.

\subsection{Super-gaussian model} \label{ssec-gaussian}

As the first example we discuss a supersymmetric version of the gaussian model. To start with, we consider a potential with a fixed quadratic bosonic term, and with bosonic and fermionic linear terms depending on bosonic and fermionic times, $t$ and $\xi$ respectively
\begin{equation}
V_t(x,\theta)=V_{B,t}(x)+V_F(x)\theta,\qquad
V_{B,t}(x)=tx+\frac{1}{2}x^2,\qquad
V_F(x)=\xi x.
\label{s_d_gauss}
\end{equation}
Below we determine a super-quantum curve for this model. We show that it does not involve derivatives with respect to the bosonic time $t$ (so that one eliminate dependence on $t$ e.g. by setting $t=1$), however it involves a derivative with respect to the fermionic time $\xi$. We also analyze its planar solution and show that it agrees with the classical limit of the super-quantum curve.

\subsubsection*{Super-quantum curves}    \label{ssec-quantum-gauss}

For the super-gaussian model with the deformed potential (\ref{s_d_gauss}) we find that super-quantum curve equations (\ref{q_curve_eq_3_2}) at level 3/2 take form
\begin{equation}
\widehat{A}_{3/2}\widehat{\chi}_{\alpha_{\pm}}(x,\theta)=0,\qquad
\widehat{A}_{3/2}=
-\partial_x\partial_{\theta}-\beta^{\pm 1}\widehat{G}_{-3/2}
-\theta\Big(\partial_x^2-2\beta^{\pm 1}\widehat{L}_{-2}\Big),
\label{g_s_curve}
\end{equation}
where $\alpha_{\pm} = \pm\beta^{\pm 1/2}\hbar$. Here $\widehat{G}_{-3/2}$ and $\widehat{L}_{-2}$ in (\ref{h_g_chi_rep}), as operators acting on $\widehat{\chi}_{\alpha}(x,\theta)$, are expressed as
\begin{align}
\begin{split}
\hbar^2\widehat{G}_{-3/2}&=
\xi(x+t)x+Q\hbar\xi-\xi\mu+\alpha(\xi+\theta)-\sqrt{\beta}\hbar
\Big<\sum_{a=1}^N\vartheta_a\widehat{\chi}_{\alpha}^{\textrm{ins}}(x,\theta)\Big>/\widehat{\chi}_{\alpha}(x,\theta),\\
\hbar^2\widehat{L}_{-2}&=\frac{1}{2}(x+t)^2+\frac{1}{2}Q\hbar-\mu+\alpha,
\label{g_op_gl}
\end{split}
\end{align}
where this time we denote
\begin{equation}
\widehat\chi_{\alpha}^{\textrm{ins}}(x,\theta)=
e^{\frac{\alpha}{\hbar}\left(\phi(x)+\psi(x)\theta\right)}.
\end{equation}
Furthermore, from the constraints (\ref{Gn}) and (\ref{Ln}) we obtain
\begin{equation}
\left(\widehat{g}_{-1/2}^{\alpha}-
\xi\widehat{\ell}_{-1}^{\alpha}\right)\widehat{\chi}_{\alpha}(x,\theta)=0,   \label{gaussian-constraints-gl}
\end{equation}
where
\begin{equation}
\widehat{g}_{-1/2}^{\alpha}-\xi\widehat{\ell}_{-1}^{\alpha}=
\partial_{\theta}+\partial_{\xi}
-(\theta-\xi)\Big(\partial_x-\frac{\alpha}{\hbar^2}t\Big)
+t\frac{\xi\mu}{\hbar^2}
-t\frac{\sqrt{\beta}}{\hbar}\Big<\sum_{a=1}^N\vartheta_a\widehat{\chi}_{\alpha}^{\textrm{ins}}(x,\theta)\Big>/\widehat{\chi}_{\alpha}(x,\theta).
\end{equation}
It follows that $\widehat{G}_{-3/2}$ in (\ref{g_op_gl}) can be expressed as a differential operator
\begin{equation}
\hbar^2\widehat{G}_{-3/2}\widehat{\chi}_{\alpha}(x,\theta)=
\Big(\xi(x+t)x+Q\hbar\xi-2\xi(\mu-\alpha)
-\frac{\hbar^2}{t}\big(\partial_{\xi}+\partial_{\theta}+(\xi-\theta)\partial_x\big)\Big)\widehat{\chi}_{\alpha}(x,\theta),
\end{equation}
which can be used to write down super-quantum curve equations (\ref{g_s_curve}). Since the potential (\ref{s_d_gauss}) has only one fermionic parameter $\xi$, it follows that $\partial_{\xi}Z=0$, and so in the classical limit (\ref{par_h_classic}) the above super-quantum curve equations reproduce the super-spectral curve  for $\xi_n=\xi \delta_{n,1}$, derived independently in (\ref{sp_c_gauss}) from the analysis of the planar limit.

\subsubsection*{Planar one-cut solution}

Let us consider now the planar one-cut solution in the gaussian model without a linear bosonic term, however with general fermionic potential
\begin{equation}
V(x,\theta)=V_B(x)+V_F(x)\theta,\qquad
V_B(x)=\frac{1}{2}x^2,\qquad
V_F(x)=\sum_{n=1}^K\xi_nx^n,
\label{s_gauss}
\end{equation}
where $\xi_n$ are fermionic parameters. In this model the first term in the bosonic resolvent (\ref{b_res_sol}) yields the well known resolvent for the standard bosonic gaussian model
\begin{equation}
\omega_B^{(0)}(x)=\oint_C\frac{dz}{2\pi i}\frac{z}{x-z}\sqrt{\frac{(x-q_1)(x-q_2)}{(z-q_1)(z-q_2)}}
=x-\sqrt{x^2-2\mu},
\end{equation}
where the branch points $q_i$ have been determined by the asymptotic condition (\ref{asy_cond_0}). We also obtain the fermionic resolvent (\ref{f_res_sol})
\begin{equation}
\omega_F(x)=\oint_C\frac{dz}{2\pi i}\frac{V_F(z)}{x-z}\sqrt{\frac{z^2-2\mu}{x^2-2\mu}}
+\frac{\mu_f}{\sqrt{x^2-2\mu}}
=V_F(x)-\frac{\Xi(x)-\mu_f}{\sqrt{x^2-2\mu}},
\label{gau_rel_f}
\end{equation}
where we have defined a fermionic moment function
\begin{align}
\begin{split}
\Xi(x)&\equiv
\sum_{n=1}^K\xi_n
\oint_{z=0}\frac{dz}{2\pi i}\frac{\sqrt{1-2\mu z^2}}{z^{n+2}(1-xz)} =
\\
&=\xi_1(x^2-\mu)+\xi_2x(x^2-\mu)+\frac{1}{2}\xi_3(2x^4-2\mu x^2-\mu^2)+\ldots,
\end{split}
\end{align}
with the values $\Xi(q_i)=\Xi_i^{(1)}$ and $\Xi'(q_i)=\Xi_i^{(2)}$ at the branch points. If all even fermionic parameters are turned off
\begin{equation}
\xi_{2}=\xi_{4}=\xi_{6}=\ldots=0,
\end{equation}
we see that $\Xi(x)=\Xi(-x)$ and $\Xi_1^{(1)}=\Xi_2^{(1)}$. In this case, by assuming the solution (\ref{one_cut_muf_c}) we obtain
\begin{equation}
\mu_f=\Xi\big(\sqrt{2\mu}\big)=\xi_1 \mu+\frac{3}{2}\xi_3\mu^2+\ldots,
\label{gauss_mu}
\end{equation}
and
\begin{equation}
\Xi(x)-\mu_f=(x^2-2\mu)\big(\xi_1+\xi_3(x^2+\mu)+\ldots\big).
\end{equation}
We then obtain the bosonic resolvent $\omega_B(x)=\omega_B^{(0)}(x)$ by $\Xi(q_i)-\mu_f=0$ and the fermionic resolvent (\ref{gau_rel_f}).
By (\ref{spect_res}) we find
\begin{equation}
y_B(x)=\sqrt{x^2-2\mu},\qquad
y_F(x)=\sqrt{x^2-2\mu}\big(\xi_1+\xi_3(x^2+\mu)+\ldots\big),
\label{sol_gauss}
\end{equation}
and the super-spectral curve takes form
\begin{align}
\begin{split}
y_By_F &= (x^2-2\mu)\big(\xi_1+\xi_3(x^2+\mu)+\ldots\big),
\\
y_B^2+y_F'(x)y_F(x) &= (x^2-2\mu)(1-2\xi_1\xi_3x+\ldots).
\label{sp_c_gauss}
\end{split}
\end{align}
For $\xi_n=\xi \delta_{n,1}$ this spectral curve agrees with the classical limit of the super-quantum curve derived in (\ref{g_s_curve}).

To complete the planar analysis we note, that the solution (\ref{gauss_mu}) can be confirmed in yet another way. Recall that the matrix model partition function with a general potential satisfies super-Virasoro constraints (\ref{s_vir_const}) and (\ref{vir_const}). These constraints include
\begin{equation}
g_{-1/2}Z=\ell_{-1}Z=0,
\end{equation}
which for the gaussian model with the potential (\ref{s_d_gauss}) take form
\begin{align}
\begin{split}
g_{-1/2}&=-t\frac{\sqrt{\beta}}{\hbar}\frac{1}{Z}\Big<\sum_{a=1}^N\vartheta_a\Big>
+\partial_{\xi}+\xi\partial_{t},\\
\ell_{-1}&=-t\frac{\mu}{\hbar^2}+\partial_{t}
-\xi\frac{\sqrt{\beta}}{\hbar}\frac{1}{Z}\Big<\sum_{a=1}^N\vartheta_a\Big>.
\end{split}
\end{align}
These generators can also be obtained as $\alpha=0$ limit of (\ref{gaussian-constraints-gl}). Since the potential (\ref{s_d_gauss}) has only one fermionic parameter $\xi$, and so $\partial_{\xi}Z=0$, the constraint $(g_{-1/2}-\xi \ell_{-1})Z=0$ yields
\begin{equation}
\frac{\mu}{N}\frac{1}{Z}\Big<\sum_{a=1}^N\vartheta_a\Big>=\xi\mu,
\label{gauss_mu_f_mu}
\end{equation}
so that the left hand side does not depend on the deformation parameter $t$ and does not have quantum corrections. It follows that (\ref{fer_coup}) takes form
\begin{equation}
\mu_f=\xi\mu,
\end{equation}
which is consistent with the solution (\ref{gauss_mu}) and therefore with (\ref{sol_gauss}) for $\xi_n=\xi \delta_{n,1}$.

\subsection{Super-multi-Penner model and super-Liouville theory} \label{ssec-Penner}

Let us discuss now the super-multi-Penner super-eigenvalue model with the potential
\begin{align}
\begin{split}
V(x,\theta) & =\sum_{i=1}^M \alpha_i \log (x-x_i+\theta_i\theta)
=V_B(x)+V_F(x)\theta,    \\
& V_B(x)=\sum_{i=1}^M\alpha_i\log(x-x_i),\quad
V_F(x)=\sum_{i=1}^M\frac{\alpha_i\theta_i}{x-x_i}.
\label{s_m_penner}
\end{split}
\end{align}
In this model it is convenient to rescale the wave-function in (\ref{chi_hat_def}) and introduce the following normalization
\begin{equation}
\widetilde{\chi}_{\alpha}(x,\theta)=\widehat{\chi}_{\alpha}(x,\theta)
\prod_{i\neq j}(x_i-x_j-\theta_i\theta_j)^{\frac{\alpha_i\alpha_j}{2\hbar^2}}.
\label{chi_til_def}
\end{equation}

Note that in this super-multi-Penner model the potential term
\be
e^{-\frac{\sqrt{\beta}}{\hbar}\sum_{a=1}^NV(z_a,\vartheta_a)}
\ee
takes an analogous form as $M$ insertions of $\chi_{\alpha}^{\textrm{ins}}(x,\theta)$ in (\ref{chi_ins_def}). We already showed that, in the context of conformal field theory, an insertion of $\chi_{\alpha}^{\textrm{ins}}(x,\theta)$ can be interpreted as an insertion of a primary field (\ref{chi_hat_def}), with degenerate momentum $\alpha=\alpha_{p,q}$. Therefore the wave-function (\ref{chi_til_def}) represents a correlation function in the super-Liouville theory, which involves $(M+2)$ Neveu-Schwarz primary fields inserted on ${\IP}^1$: $M$ of those fields are encoded in the potential, one field is represented by the insertion of $\chi_{\alpha}^{\textrm{ins}}(x,\theta)$ itself, and one additional field is inserted at $x=\infty \in {\IP}^1$ and has the momentum $\alpha_{\infty}$ given in (\ref{a_inf_def}). Note that if the condition $\alpha_{\infty}=0$ is imposed, the primary field at $x=\infty$ is removed and the model with the above potential describes super-Liouville theory with $(M+1)$ primary fields. It follows that various objects familiar in super-Liouville theory arise upon the specialization of our formalism to the super-multi-Penner potential. In particular super-quantum curve equations can be identified as supersymmetric BPZ equations for correlation functions of several primary fields, Nekrasov-Shatashvili limit of quantum curves coincides with certain classical equations of motion in super-Liouville theory, etc. We discuss these relations in more detail below.

\subsubsection*{$\mathfrak{osp}(1|2)$ invariance of the wave-function}  \label{ssec-constraints}

Recall that a subset of super-Virasoro generators determined in section \ref{ssec-abVirasoro} forms the $\mathfrak{osp}(1|2)$ subalgebra and imposes constraints (\ref{osp12}). Specializing the potential to the super-multi-Penner model case (\ref{s_m_penner}) we find that generators (\ref{osp12-g}) and (\ref{osp12-l}), as acting on the normalized wave-function $\widetilde{\chi}_{\alpha}(x,\theta)$ in (\ref{chi_til_def}), take form
\begin{align}
\begin{split}
\widetilde{g}_{-1/2}^{\alpha}(x,\theta) & = -D-\sum_{i=1}^MD_i,    \\
\widetilde{g}_{1/2}^{\alpha}(x,\theta) &=
-xD-\sum_{i=1}^Mx_iD_i-2\Delta_{\alpha}\theta-2\sum_{i=1}^M\Delta_{\alpha_i}\theta_i-\frac{\alpha_{\infty}}{\hbar^2}\Big(\alpha\theta+\sum_{i=1}^M\alpha_i\theta_i
-\sqrt{\beta}\hbar\sum_{a=1}^N\vartheta_a\Big),   \label{g-tilde}
\end{split}
\end{align}
and
\begin{align}
\begin{split}
\widetilde{\ell}_{-1}^{\alpha}(x,\theta) &=
-\partial_x-\sum_{i=1}^M\partial_{x_i}, \\
\widetilde{\ell}_{0}^{\alpha}(x,\theta) &=
-x\partial_x-\frac{\theta}{2}\partial_{\theta}
-\sum_{i=1}^M\big(x_i\partial_{x_i}+\frac{\theta_i}{2}\partial_{\theta_i}\big)
-\Delta_{\alpha}-\sum_{i=1}^M\Delta_{\alpha_i}
+\Delta_{\alpha_{\infty}},\\
\widetilde{\ell}_{1}^{\alpha}(x,\theta) &=
-x^2\partial_x-x\theta\partial_{\theta}
-\sum_{i=1}^M(x_i^2\partial_{x_i}+x_i\theta_i\partial_{\theta_i}\big)
-2\Delta_{\alpha}x-2\sum_{i=1}^M\Delta_{\alpha_i}x_i   \\
&\qquad  -\frac{\alpha_{\infty}}{\hbar^2}\Big(\alpha x+\sum_{i=1}^M\alpha_ix_i-\sqrt{\beta}\hbar\sum_{a=1}^Nz_a\Big),  \label{l-tilde}
\end{split}
\end{align}
where
\begin{align}
D_i &= -\partial_{\theta_i}+\theta_i\partial_{x_i},
\label{di_def}
\\
\alpha_{\infty} & =
\mu+Q\hbar-\alpha-\sum_{i=1}^M\alpha_i.
\label{a_inf_def}
\end{align}

When $\alpha_{\infty}\neq 0$, a primary field with momentum $\alpha_{\infty}$ has to be located at $x=\infty$ and we can use generators $\widetilde{\ell}_{0}^{\alpha}(x,\theta),\widetilde{\ell}_{1}^{\alpha}(x,\theta)$ and $\widetilde{g}_{1/2}^{\alpha}(x,\theta)$ only under the super-matrix integral, while the eigenvalue-independent representation exists only for two remaining generators, which can be used to impose constraints on the (integrated) wave-function (\ref{chi_til_def})
\begin{equation}
\widetilde{g}_{-1/2}^{\alpha}(x,\theta)\widetilde{\chi}_{\alpha}(x,\theta)=
\widetilde{\ell}_{-1, 0}^{\alpha}(x,\theta)\widetilde{\chi}_{\alpha}(x,\theta)=0.
\end{equation}
We can use these constraints to remove two bosonic derivatives and one fermionic derivative from operators (\ref{G-Penner}) and (\ref{L-Penner}) introduced below, and then from the super-quantum curve equation.

On the other hand, once the condition $\alpha_{\infty}=0$ is imposed, all generators in (\ref{g-tilde}) and (\ref{l-tilde}) do not depend on super-eigenvalues $z_a$ and $\vartheta_a$, and they can be used to fix the full $\mathfrak{osp}(1|2)$ invariance of the wave-function (\ref{chi_til_def})
\begin{equation}
\widetilde{g}_{\pm 1/2}^{\alpha}(x,\theta)\widetilde{\chi}_{\alpha}(x,\theta)=
\widetilde{\ell}_{\pm 1, 0}^{\alpha}(x,\theta)\widetilde{\chi}_{\alpha}(x,\theta)=0.  \label{gl-Penner-a0}
\end{equation}
Using these constraints we can eliminate three bosonic derivatives and two fermionic derivatives from operators (\ref{G-Penner}) and (\ref{L-Penner}) introduced below, and then from the super-quantum curve equation.

In particular for the super-multi-Penner model (\ref{s_m_penner}) with $M=2$ and for $\alpha_{\infty}=0$ the constraints (\ref{gl-Penner-a0}) respectively take form
\begin{align}
\begin{split}
\big((x_1-x_2)D_1+(x-x_2)D+2\Delta_{\alpha}\theta+2\Delta_{\alpha_1}\theta_1+2\Delta_{\alpha_2}\theta_2\big)\widetilde{\chi}_{\alpha}(x,\theta) &=0,\\
\big((x_2-x_1)D_2+(x-x_1)D+2\Delta_{\alpha}\theta+2\Delta_{\alpha_1}\theta_1+2\Delta_{\alpha_2}\theta_2\big)\widetilde{\chi}_{\alpha}(x,\theta) &=0,
\end{split}
\end{align}
and
\begin{align}
\begin{split}
\big(2(x_1-x_2)\partial_{x_1}+2(x-x_2)\partial_x+\theta\partial_{\theta}
+\theta_1\partial_{\theta_1}+\theta_2\partial_{\theta_2}
+2\Delta_{\alpha}+2\Delta_{\alpha_1}+2\Delta_{\alpha_2}\big)\widetilde{\chi}_{\alpha}(x,\theta) &=0,\\
\big(2(x_2-x_1)\partial_{x_2}+2(x-x_1)\partial_x+\theta\partial_{\theta}
+\theta_1\partial_{\theta_1}+\theta_2\partial_{\theta_2}
+2\Delta_{\alpha}+2\Delta_{\alpha_1}+2\Delta_{\alpha_2}\big)\widetilde{\chi}_{\alpha}(x,\theta) &=0,
\end{split}
\end{align}
and they can be used to remove all time derivatives from operators (\ref{G-Penner}) and (\ref{L-Penner}), and so from super-quantum curves in consequence.

\subsubsection*{Super-quantum curves}

We construct now super-quantum curves for the super-multi-Penner model with the potential (\ref{s_m_penner}). First, the fermionic operator $\widehat{h}(x)$ in (\ref{h_x_op}) and the bosonic operator $\widehat{f}(x)$ in (\ref{f_x_op}) take form
\begin{align}
\begin{split}
\widehat{h}(x) &=
\hbar^2\sum_{i=1}^M\frac{1}{x-x_i}D_i,\\
\widehat{f}(x) &=
\hbar^2\sum_{i=1}^M\Big(\frac{1}{x-x_i}\partial_{x_i}
+\frac{\theta_i}{2(x-x_i)^2}\partial_{\theta_i}\Big),
\end{split}
\end{align}
where $D_i=-\partial_{\theta_i}+\theta_i\partial_{x_i}$. Second, we introduce the super-Virasoro generators represented on the wave-function $\widetilde{\chi}_{\alpha}(x,\theta)$ in (\ref{chi_til_def}); taking advantage of (\ref{h_g_chi_rep}), for $n\geq 2$ they take form
\begin{align}
\begin{split}
\widetilde{G}_{-n+1/2} & =
\Big(\prod_{i\neq j}(x_i-x_j-\theta_i\theta_j)^{\frac{\alpha_i\alpha_j}{2\hbar^2}}\Big)
\widehat{G}_{-n+1/2}
\Big(\prod_{i\neq j}(x_i-x_j-\theta_i\theta_j)^{-\frac{\alpha_i\alpha_j}{2\hbar^2}}\Big) =\\
& = \sum_{i=1}^M\bigg(\frac{2(n-1)\Delta_{\alpha_i}\theta_i}{(x_i-x)^n}
-\frac{1}{(x_i-x)^{n-1}}D_i\bigg),   \label{G-Penner}
\end{split}
\end{align}
and
\begin{align}
\begin{split}
\widetilde{L}_{-n} & =
\Big(\prod_{i\neq j}(x_i-x_j-\theta_i\theta_j)^{\frac{\alpha_i\alpha_j}{2\hbar^2}}\Big)
\widehat{L}_{-n}
\Big(\prod_{i\neq j}(x_i-x_j-\theta_i\theta_j)^{-\frac{\alpha_i\alpha_j}{2\hbar^2}}\Big) = \\
& = \sum_{i=1}^M\bigg(\frac{(n-1)\Delta_{\alpha_i}}{(x_i-x)^n}
-\frac{1}{(x_i-x)^{n-1}}\partial_{x_i}
+\frac{(n-1)\theta_i}{2(x_i-x)^n}\partial_{\theta_i}\bigg).   \label{L-Penner}
\end{split}
\end{align}

Super-quantum curves for the super-multi-Penner model can now be constructed using the above representation of super-Virasoro generators in expressions for super-Virasoro singular vectors. For example,  super-quantum curve equations (\ref{q_curve_eq_3_2}) at level $3/2$ take form
\begin{equation}
\widetilde{A}_{3/2}\widetilde{\chi}_{\alpha = \pm \beta^{\pm 1/2}\hbar}(x,\theta)=0,\qquad
\widetilde{A}_{3/2}=
-\partial_x\partial_{\theta}-\beta^{\pm 1}\widetilde{G}_{-3/2}
-\theta\Big(\partial_x^2-2\beta^{\pm 1}\widetilde{L}_{-2}\Big).
\end{equation}

Furthermore, we can use constraints discussed in section \ref{ssec-constraints} to remove time-dependence from operators $\widetilde{G}_{-n+1/2}$ and $\widetilde{L}_{-n}$. In particular, considering the super-multi-Penner model with $M=2$ and fixing $a_{\infty}=0$, we can completely remove time derivatives from these operators, so that their action on $\widetilde{\chi}_{\alpha}(x,\theta)$ takes for example the following form
\begin{equation}
\widetilde{G}_{-3/2}\widetilde{\chi}_{\alpha}(x,\theta)=
\bigg[-\sum_{i=1,2}\frac{1}{x-x_i}D+\sum_{i=1,2}\frac{2\Delta_{\alpha_i}\theta_i}{(x-x_i)^2}-2\frac{\Delta_{\alpha}\theta+\Delta_{\alpha_1}\theta_1+\Delta_{\alpha_2}\theta_2}{(x-x_1)(x-x_2)}\bigg]\widetilde{\chi}_{\alpha}(x,\theta),
\end{equation}
and
\begin{align}
\widetilde{L}_{-2}\widetilde{\chi}_{\alpha}(x,\theta)&=
\bigg[-\sum_{i=1,2}\frac{1}{x-x_i}\partial_x+\sum_{i=1,2}\frac{\theta_iD+2\Delta_{\alpha_i}}{2(x-x_i)^2}
-\frac{\theta\partial_{\theta}+2\Delta_{\alpha}+2\Delta_{\alpha_1}+2\Delta_{\alpha_2}}{2(x-x_1)(x-x_2)} + \\
&
+\frac{1}{(x-x_1)(x-x_2)}\sum_{i=1,2}\frac{\Delta_{\alpha}\theta_i\theta}{x-x_i}
-\frac{\big((x-x_1)\Delta_{\alpha_1}-(x-x_2)\Delta_{\alpha_2}\big)\theta_1\theta_2}{(x-x_1)^2(x-x_2)^2}\bigg]\widetilde{\chi}_{\alpha}(x,\theta).  \nonumber
\end{align}
It follows that super-quantum curve equations in this case are time-independent, and take form of ordinary super-differential equations. These equations are directly related to the ordinary differential equation of the type considered by Dotsenko and Fateev
\cite{Dotsenko:1984nm,Dotsenko:1984ad}. It was analysed in  \cite{Belavin:2006zr} and some properties of its solutions,
which can be expressed in term of certain two-fold contour integrals, were discussed in \cite{Belavin:2007gz}.

For completeness, let us also consider the classical limit (\ref{par_h_classic}) in the above example. We find that the super-quantum curves at level $3/2$ reduces in this limit to a system of equations
\begin{align}
\begin{split}
y_B(x)y_F(x) & = \sum_{i=1,2}\frac{\alpha_i^2\theta_i}{(x-x_i)^2}
-\frac{\alpha_1^2\theta_1+\alpha_2^2\theta_2}{(x-x_1)(x-x_2)},    \\
y_B(x)^2+y_F'(x)y_F(x) & = \sum_{i=1,2}\frac{\alpha_i^2}{(x-x_i)^2}
-\frac{\alpha_1^2+\alpha_2^2}{(x-x_1)(x-x_2)} + \\
&\qquad
-\frac{\big((x-x_1)\alpha_1^2-(x-x_2)\alpha_2^2\big)\theta_1\theta_2}{(x-x_1)^2(x-x_2)^2}.
\label{s_curve_m2_l}
\end{split}
\end{align}
This agrees with the classical super-spectral curve, determined from the analysis of the planar solution of super-multi-Penner model in (\ref{Penner-planar-curve}).

\subsubsection*{Planar one-cut solution}

For completeness, let us also consider a one-cut solution of the super-multi-Penner with the potential (\ref{s_m_penner}).
Under the one-cut ansatz
\begin{equation}
\sigma(x)=(x-q_1)(x-q_2),
\end{equation}
the first term in the bosonic resolvent (\ref{b_res_sol}) is given by
\begin{equation}
\omega_B^{(0)}(x)=\oint_C\frac{dz}{2\pi i}\frac{1}{x-z}
\sum_{i=1}^M\frac{\alpha_i}{z-x_i}\sqrt{\frac{\sigma(x)}{\sigma(z)}}
=V_B'(x)-\sum_{i=1}^M\frac{\alpha_i\sqrt{\sigma(x)}}{(x-x_i)\sqrt{\sigma(x_i)}}.
\end{equation}
The branch points $q_1$ and $q_2$ are determined by the asymptotic condition (\ref{asy_cond_0})
\begin{equation}
\sum_{i=1}^M\frac{\alpha_i}{\sqrt{\sigma(x_i)}}=0,\qquad
\sum_{i=1}^M\frac{\alpha_ix_i}{\sqrt{\sigma(x_i)}}=-\alpha_{\infty},
\label{asy_cond_l}
\end{equation}
where we have defined
\begin{equation}
\alpha_{\infty} = \mu-\sum_{i=1}^M\alpha_i,
\end{equation}
which describes the momentum of the primary field at $x=\infty$. From (\ref{one_cut_muf_c}) we obtain
\begin{equation}
\mu_f=\frac{1}{2}\oint_C\frac{dz}{2\pi i}
\sum_{i=1}^M\frac{\alpha_i\theta_i}{z-x_i}\frac{\sigma'(z)}{\sqrt{\sigma(z)}}
+c(q_1,q_2)\left(\Xi_1^{(1)}-\Xi_2^{(1)}\right)
=\sum_{i=1}^M\alpha_i\theta_i\left(1-\frac{C_P(x_i)}{\sqrt{\sigma(x_i)}}\right),
\end{equation}
where
\begin{equation}
C_P(x)=\frac{1}{2}\sigma'(x)+(q_1-q_2)c(q_1,q_2),\qquad
c(q_1,q_2)=-c(q_2,q_1),
\end{equation}
is an undetermined function. Here the fermionic and bosonic moments (\ref{fer_moment}) and (\ref{bos_moment}) are also obtained as
\begin{align}
\Xi_I^{(1)}-\mu_f &=
\sum_{i=1}^M\alpha_i\theta_i
\left(\frac{\sqrt{\sigma(x_i)}}{q_I-x_i}+\frac{\sigma'(x_i)}{2\sqrt{\sigma(x_i)}}\right)
=\Xi(q_I),\\
\Xi_I^{(2)} &=-\sum_{i=1}^M\frac{\alpha_i\theta_i\sqrt{\sigma(x_i)}}{(q_I-x_i)^2}
=\Xi'(q_I),\\
M_I &=\sum_{i=1}^M\frac{\alpha_i}{(q_I-x_i)\sqrt{\sigma(x_i)}}
=M(q_I),
\end{align}
where we have introduced fermionic and bosonic moment functions as
\begin{align}
\Xi(x) & = \sum_{i=1}^M\alpha_i\theta_i
\left(\frac{\sqrt{\sigma(x_i)}}{x-x_i}+\frac{C_P(x_i)}{\sqrt{\sigma(x_i)}}\right),
\label{f_moment_l}
\\
M(x) & = \sum_{i=1}^M\frac{\alpha_i}{(x-x_i)\sqrt{\sigma(x_i)}}.
\label{b_moment_l}
\end{align}
Then the bosonic resolvent (\ref{b_res_sol}) is obtained as
\begin{equation}
\omega_B(x)=
V_B'(x)-M(x)\sqrt{\sigma(x)}
-\frac{1}{2}\sum_{I=1,2}\frac{\Xi(q_I)\Xi'(q_I)}{M_I\sigma'(q_I)(x-q_I)\sqrt{\sigma(x)}},
\end{equation}
and the fermionic resolvent (\ref{f_res_sol}) takes form
\begin{equation}
\omega_F(x)=V_F(x)-\frac{\Xi(x)}{\sqrt{\sigma(x)}}.
\end{equation}
By (\ref{spect_res}) we equivalently have
\begin{align}
&
y_B(x)=M(x)\sqrt{\sigma(x)}-
\frac{\Xi'(x)\Xi(x)}{2M(x)\sigma(x)^{3/2}}+
\frac{1}{2\sqrt{\sigma(x)}}\sum_{i=1}^M
\mathop{\textrm{Res}}_{z=x_i}\frac{\Xi'(z)\Xi(z)}{M(z)\sigma(z)(x-z)},
\\
&
y_F(x)=\frac{\Xi(x)}{\sqrt{\sigma(x)}}.
\end{align}

Let us consider the case of $M=2$ and explicitly express the super-spectral curve. In this case the bosonic moment function (\ref{b_moment_l}) is given by
\begin{equation}
M(x)=-\frac{\alpha_{\infty}}{(x-x_1)(x-x_2)},
\end{equation}
where we have used the asymptotic condition (\ref{asy_cond_l}), which furthermore implies
\begin{equation}
q_1+q_2=x_1+x_2-\frac{\alpha_1^2-\alpha_2^2}{\alpha_{\infty}^2}(x_1-x_2),\qquad
q_1q_2=x_1x_2-\frac{\alpha_1^2x_2-\alpha_2^2x_1}{\alpha_{\infty}^2}(x_1-x_2),
\end{equation}
so that we determine the branch points
\begin{equation}
q_{1,2}=\frac{\alpha_{\infty}^2(x_1+x_2)-(\alpha_1^2-\alpha_2^2)(x_1-x_2)\pm
(x_1-x_2)\sqrt{\big(\alpha_{\infty}^2-(\alpha_1+\alpha_2)^2\big)
\big(\alpha_{\infty}^2-(\alpha_1-\alpha_2)^2\big)}}{2\alpha_{\infty}^2}.
\end{equation}
The super-spectral curve is then expressed as
\begin{align}
\begin{split}
y_B(x)y_F(x) &= M(x)\Xi(x),\\
y_B(x)^2+y_F'(x)y_F(x) &= M(x)^2\sigma(x)+
\sum_{i=1,2}\mathop{\textrm{Res}}_{z=x_i}\frac{M(x)\Xi'(z)\Xi(z)}{M(z)\sigma(z)(x-z)},
\end{split}
\end{align}
where the second term on the right hand side in the second equation can be written as
\begin{equation}
\sum_{i=1,2}\mathop{\textrm{Res}}_{z=x_i}\frac{M(x)\Xi'(z)\Xi(z)}{M(z)\sigma(z)(x-z)}=
\frac{\alpha_{\infty}\theta_1\theta_2}{(x-x_1)^2(x-x_2)^2}
\sum_{i=1,2}\alpha_i\left(\frac{\sqrt{\sigma(x_i)}}{x_1-x_2}
+\frac{(-1)^iC_P(x_i)}{\sqrt{\sigma(x_i)}}\right)(x-x_i).
\end{equation}
In particular in the limit $\alpha_{\infty} \to 0$, assuming
\begin{equation}
\alpha_{\infty}^2C_P(x_i) \to 0,
\end{equation}
we find the super-spectral curve
\begin{align}
\begin{split}
y_B(x)y_F(x) &= \frac{-(x_1-x_2)}{(x-x_1)(x-x_2)}
\sum_{i=1,2}\frac{(-1)^i\alpha_i^2\theta_i}{x-x_i},
\\
y_B(x)^2+y_F'(x)y_F(x) &= \frac{-(x_1-x_2)}{(x-x_1)(x-x_2)}
\sum_{i=1,2}\frac{(-1)^i\alpha_i^2}{x-x_i} + \\
& \qquad +\frac{\theta_1\theta_2}{(x-x_1)^2(x-x_2)^2}\sum_{i=1,2}(-1)^i\alpha_i^2(x-x_i). \label{Penner-planar-curve}
\end{split}
\end{align}
This curve coincides with the classical limit of the super-quantum curve at level $3/2$ found in (\ref{s_curve_m2_l}).


\acknowledgments{We thank Vincent Bouchard, Zbigniew Jask\'{o}lski, Motohico Mulase and Chaiho Rim for discussions and comments on the manuscript. We thank Simons Center for Geometry and Physics and C. N. Yang Institute for Theoretical Physics, Stony Brook, NY, for hospitality and support. This work is supported by the ERC Starting Grant no. 335739 \emph{``Quantum fields and knot homologies''} funded by the European Research Council under the European Union's Seventh Framework Programme, and the Ministry of Science and Higher Education in Poland.}

\newpage

\appendix

\section{Operators $\widehat{h}(x)$ and $\widehat{f}(x)$}     \label{app-hf}

In this appendix we introduce and discuss properties of operators $\widehat{h}(x)$ and $\widehat{f}(x)$, which are used in various computations in the paper. The fermionic operator $\widehat{h}(x)$ takes form
\be
\begin{split}
\widehat{h}(x) = & \widehat{h}_t(x)+\widehat{h}_{\xi}(x), \label{h_x_op} \\ 
& \widehat{h}_t(x)\equiv\hbar^2\sum_{n=0}^{\infty}x^n\sum_{k=n+1}^{\infty}\xi_{k+1/2}\partial_{t_{k-n-1}},\\
& \widehat{h}_{\xi}(x)\equiv\hbar^2\sum_{n=0}^{\infty}x^n\sum_{k=n+2}^{\infty}kt_{k}\partial_{\xi_{k-n-3/2}},
\end{split}
\ee
and we denote
\begin{equation}
\partial_x^{n}\widehat{h}(x) \equiv \left[\partial_x, \partial_x^{n-1}\widehat{h}(x)\right]. \label{dxn-hhat}
\end{equation}
From the identification (\ref{id_mat_free}) we find that the action of $\widehat{h}(y)$ on the wave-function $\chi_{\alpha}(x,\theta)$, defined in (\ref{chi_hat_w}), can be rewritten as the expectation value involving the function $h(y)$
\be
\begin{split}
\widehat{h}(y)\chi_{\alpha}(x,\theta)=& \left<h(y)\chi^{\textrm{ins}}_{\alpha}(x,\theta)\right>, \qquad h(y) = h_t(y)+h_{\xi}(y),  \label{h_x_chi_op} \\ 
& h_t(y)\equiv -\sqrt{\beta}\hbar\sum_{a=1}^N\frac{V_F(y)-V_F(z_a)}{y-z_a}, \\
& h_{\xi}(y)\equiv -\sqrt{\beta}\hbar\sum_{a=1}^N\frac{\big(V'_B(y)-V'_B(z_a)\big)\vartheta_a}{y-z_a}.
\end{split}
\ee
In the large $N$ limit (\ref{large_N_lim}) the expectation value of $h(x)$ reproduces $h^{(0)}(x)$ in (\ref{h_x_0_cl})
\begin{equation}
\lim_{\begin{subarray}{c}N\to\infty\\\widehat{\hbar}\ \textrm{fixed}\end{subarray}}\frac{1}{Z}\left<h(x)\right>=h^{(0)}(x).
\end{equation}

Similarly we introduce a bosonic operator $\widehat{f}(x)$ defined by
\be
\begin{split}
\widehat{f}(x)=&\widehat{f}_t(x)+\widehat{f}_{\xi}(x),
\label{f_x_op}  \\  
& \widehat{f}_t(x)\equiv\hbar^2\sum_{n=0}^{\infty}x^n\sum_{k=n+2}^{\infty}kt_{k}\partial_{t_{k-n-2}}, \\
& \widehat{f}_{\xi}(x)\equiv\hbar^2\sum_{n=0}^{\infty}x^n\sum_{k=n+2}^{\infty}\Big(k-\frac{n+1}{2}\Big)\xi_{k+1/2}\partial_{\xi_{k-n-3/2}},
\end{split}
\ee
and we denote
\begin{equation}
\partial_x^{n}\widehat{f}(x) \equiv
\left[\partial_x, \partial_x^{n-1}\widehat{f}(x)\right].  \label{dxn-fhat}
\end{equation}
From identifications (\ref{id_mat_free}), the action of $\widehat{f}(y)$ on the wave-function $\chi_{\alpha}(x,\theta)$ in (\ref{chi_hat_w}) can be rewritten as the expectation value involving the function $f(y)$
\be
\begin{split}
\widehat{f}(y)\chi_{\alpha}(x,\theta)=& \left<f(y)\chi^{\textrm{ins}}_{\alpha}(x,\theta)\right>,\qquad f(y) = f_t(y)+f_{\xi}(y),  \label{f_x_chi_op} \\  
& f_t(y)\equiv
-\sqrt{\beta}\hbar\sum_{a=1}^N\frac{V_B'(y)-V_B'(z_a)}{y-z_a},
\nonumber\\
& f_{\xi}(y)\equiv
-\sqrt{\beta}\hbar\sum_{a=1}^N\bigg(\frac{\big(V'_F(y)-V'_F(z_a)\big)\vartheta_a}{2(y-z_a)}+\frac{V^{(2)}_{F}(y,z_a)\vartheta_a}{2(y-z_a)^2}\bigg),
\nonumber
\end{split}
\ee
where $V^{(2)}_{F}(y,z_a)$ is defined in (\ref{v_f_2_def}).
In the large $N$ limit (\ref{large_N_lim}) the expectation value of $f(x)$ reproduces $f^{(0)}(x)$ introduced in (\ref{f_x_0_cl})
\begin{equation}
\lim_{\begin{subarray}{c}N\to\infty\\\widehat{\hbar}\ \textrm{fixed}\end{subarray}}\frac{1}{Z}\left<f(x)\right>=f^{(0)}(x).
\end{equation}

Operators $\widehat{h}(x)$ in (\ref{h_x_op}) and  $\widehat{f}(x)$ in (\ref{f_x_op}) satisfy certain commutation relations, which we take advantage of in various computations. These commutation relations follow from the formula
\begin{equation}
\sum_{p=1}^k\frac{(p+n-2)!}{(p-1)!}=\frac{(k+n-1)!}{n(k-1)!},\qquad
k,n \in {\IN},
\end{equation}
and take form
\begin{align}
\begin{split}
\left[\widehat{h}(x), \partial_x^nV_B(x)\right] &=\frac{1}{n+1}\hbar^2\partial_x^{n+1}V_F(x),
\\
\left\{\widehat{h}(x), \partial_x^nV_F(x)\right\} &=
\left[\widehat{f}(x), \partial_x^nV_B(x)\right]=
\frac{1}{n+1}\hbar^2\partial_x^{n+2}V_B(x),
\\
\left[\widehat{f}(x), \partial_x^nV_F(x)\right] &=\frac{2n+3}{2(n+1)(n+2)}\hbar^2\partial_x^{n+2}V_F(x),
\label{hf_Vbf_g_s}
\end{split}
\end{align}
and
\begin{align}
\begin{split}
\left\{\widehat{h}(x), \partial_x^n\widehat{h}(x)\right\}&=\frac{2}{n+1}\hbar^2\partial_x^{n+1}\widehat{f}(x),
\\
\left[\widehat{h}(x), \partial_x^n\widehat{f}(x)\right]&=\frac{n-1}{2(n+1)(n+2)}\hbar^2\partial_x^{n+2}\widehat{h}(x),
\\
\left[\widehat{f}(x), \partial_x^n\widehat{f}(x)\right]&=\frac{n}{(n+1)(n+2)}\hbar^2\partial_x^{n+2}\widehat{f}(x).
\end{split}
\end{align}
Differentiating the above relations with respect to $x$, we inductively obtain
\begin{align}
\begin{split}
\left[\partial_x^m\widehat{h}(x), \partial_x^nV_B(x)\right]&=\frac{m!n!}{(m+n+1)!}\hbar^2\partial_x^{m+n+1}V_F(x),
\\
\left\{\partial_x^m\widehat{h}(x), \partial_x^nV_F(x)\right\}&=
\left[\partial_x^m\widehat{f}(x), \partial_x^nV_B(x)\right]=
\frac{m!n!}{(m+n+1)!}\hbar^2\partial_x^{m+n+2}V_B(x),
\\
\left[\partial_x^m\widehat{f}(x), \partial_x^nV_F(x)\right]&=\frac{(m+2n+3)m!n!}{2(m+n+2)!}\hbar^2\partial_x^{m+n+2}V_F(x),
\label{hf_Vbf_g}
\end{split}
\end{align}
and
\begin{align}
\begin{split}
\left\{\partial_x^m\widehat{h}(x), \partial_x^n\widehat{h}(x)\right\}&=\frac{2m!n!}{(m+n+1)!}\hbar^2\partial_x^{m+n+1}\widehat{f}(x),
\\
\left[\partial_x^m\widehat{h}(x), \partial_x^n\widehat{f}(x)\right]&=\frac{(n-2m-1)m!n!}{2(m+n+2)!}\hbar^2\partial_x^{m+n+2}\widehat{h}(x),
\\
\left[\partial_x^m\widehat{f}(x), \partial_x^n\widehat{f}(x)\right]&=\frac{(n-m)m!n!}{(m+n+2)!}\hbar^2\partial_x^{m+n+2}\widehat{f}(x).
\label{hf_hh_g}
\end{split}
\end{align}

\section{Derivatives of $\chi_{\alpha}^{\textrm{ins}}(x,\theta)$}

In this appendix we summarize derivatives with respect to $\theta$ and $x$ of
\begin{equation}
\chi_{\alpha}^{\textrm{ins}}(x,\theta)=\Big(1+\frac{\sqrt{\beta}}{\hbar}\sum_{a=1}^N\frac{\alpha\theta\vartheta_a}{x-z_a}\Big)\prod_{a=1}^N(x-z_a)^{-\frac{\sqrt{\beta}}{\hbar}\alpha}
\end{equation}
defined in (\ref{chi_ins_def}). First order derivatives are equal
\be
\begin{split}
&
\partial_{\theta}\chi_{\alpha}^{\textrm{ins}}(x,\theta)=
\frac{\alpha\sqrt{\beta}}{\hbar}\sum_{a=1}^N\frac{\vartheta_a\chi_{\alpha}^{\textrm{ins}}(x,0)}{x-z_a}
=
\frac{\alpha\sqrt{\beta}}{\hbar}\sum_{a=1}^N\frac{\vartheta_a\chi_{\alpha}^{\textrm{ins}}(x,\theta)}{x-z_a},
\label{d_th1_chi_ins}
\\
&
\partial_x\chi_{\alpha}^{\textrm{ins}}(x,\theta)=
-\frac{\alpha\sqrt{\beta}}{\hbar}\sum_{a=1}^N\frac{\chi_{\alpha}^{\textrm{ins}}(x,\theta)}{x-z_a}-\frac{\alpha\sqrt{\beta}\theta}{\hbar}\sum_{a=1}^N\frac{\vartheta_a\chi_{\alpha}^{\textrm{ins}}(x,\theta)}{(x-z_a)^2},
\end{split}
\ee
and second order derivatives take form
\be
\begin{split}
\partial_x\partial_{\theta}\chi_{\alpha}^{\textrm{ins}}(x,\theta) = &
-\frac{\alpha\sqrt{\beta}}{\hbar}\sum_{a=1}^N\frac{\vartheta_a\chi_{\alpha}^{\textrm{ins}}(x,\theta)}{(x-z_a)^2}
-\frac{\alpha^2\beta}{\hbar^2}\sum_{a,b=1}^N\frac{\vartheta_a\chi_{\alpha}^{\textrm{ins}}(x,\theta)}{(x-z_a)(x-z_b)}  +  \\
&
+\frac{\alpha^2\beta\theta}{\hbar^2}\sum_{a,b=1}^N\frac{\vartheta_a\vartheta_b\chi_{\alpha}^{\textrm{ins}}(x,\theta)}{(x-z_a)(x-z_b)^2},
\label{d_th1x1_chi_ins}  \\  
\partial_x^2\chi_{\alpha}^{\textrm{ins}}(x,\theta) = &
\frac{\alpha\sqrt{\beta}}{\hbar}\sum_{a=1}^N\frac{\chi_{\alpha}^{\textrm{ins}}(x,\theta)}{(x-z_a)^2}
+\frac{\alpha^2\beta}{\hbar^2}\sum_{a,b=1}^N\frac{\chi_{\alpha}^{\textrm{ins}}(x,\theta)}{(x-z_a)(x-z_b)}   +  \\
&
+\frac{2\alpha\sqrt{\beta}\theta}{\hbar}\sum_{a=1}^N\frac{\vartheta_a\chi_{\alpha}^{\textrm{ins}}(x,\theta)}{(x-z_a)^3}
+\frac{2\alpha^2\beta\theta}{\hbar^2}\sum_{a,b=1}^N\frac{\vartheta_a\chi_{\alpha}^{\textrm{ins}}(x,\theta)}{(x-z_a)^2(x-z_b)}.
\end{split}
\ee
Note that in effect the right hand side of $\partial_x\partial_{\theta}\chi_{\alpha}^{\textrm{ins}}(x,\theta)$ does not depend on $\theta$.

\section{Loop insertion operators}

We define fermionic and bosonic loop insertion operators as follows
\begin{align}
\begin{split}
\partial_{V_F(x)}^{(1/2)} &=
-\frac{\hbar}{\sqrt{\beta}}\sum_{n=0}^{\infty}\frac{1}{x^{n+1}}\partial_{\xi_{n+1/2}},   \label{loop-insertion} \\
\partial_{V_B(x)}^{(1)} &=
-\frac{\hbar}{\sqrt{\beta}}\sum_{n=0}^{\infty}\frac{1}{x^{n+1}}\partial_{t_n}.
\end{split}
\end{align}
We also consider their higher order generalizations
\begin{align}
\begin{split}
\partial_{V_F(x)}^{(k-1/2)} &=
\frac{(-1)^{k-1}}{(k-1)!}\underbrace{\big[\partial_x, \big[\partial_x, \cdots \big[\partial_x}_{k-1}, \partial_{V_F(x)}^{(1/2)}\big]\cdots\big]\big],\\
\partial_{V_B(x)}^{(k)} &=
\frac{(-1)^{k-1}}{(k-1)!}\underbrace{\big[\partial_x, \big[\partial_x, \cdots \big[\partial_x}_{k-1}, \partial_{V_B(x)}^{(1)}\big]\cdots\big]\big].
\end{split}
\end{align}
These operators act on the partition function $Z$ or the wave-function $\chi_{\alpha}(x,\theta)$ as
\begin{align}
\begin{split}
\partial^{(k-1/2)}_{V_F(x)}  \Big\langle \cdots \Big\rangle &=
\Big\langle \sum_{a=1}^N\frac{\vartheta_a}{(x-z_a)^k} \cdots \Big\rangle,\\
\partial^{(k)}_{V_B(x)}  \Big\langle \cdots \Big\rangle &=
\Big\langle \sum_{a=1}^N\frac{1}{(x-z_a)^k} \cdots \Big\rangle.
\end{split}
\end{align}
The commutation relations
\begin{align}
\begin{split}
\left\{\partial_{V_F(x)}^{(k-1/2)}, V_F(z)\right\}&=-\frac{\hbar}{\sqrt{\beta}}\frac{1}{(x-z)^{k}},\\
\left[\partial_{V_B(x)}^{(k)}, V_B'(z)\right]&=
\left\{\partial_{V_F(x)}^{(k-1/2)}, V_F'(z)\right\}=-\frac{\hbar}{\sqrt{\beta}}\frac{k}{(x-z)^{k+1}},
\end{split}
\end{align}
lead to
\begin{align}
\begin{split}
\left\{\partial_{V_F(x)}^{(k-1/2)}, S_+(y;x,\theta)\right\}&=
\sum_{a=1}^N\frac{1}{(x-z_a)^k(y-z_a)},\\
\left[\partial_{V_B(x)}^{(k)}, S_+(y;x,\theta)\right]&=
\sum_{a=1}^N\frac{k\vartheta_a}{(x-z_a)^{k+1}(y-z_a)},\\
\left[\partial_{V_F(x)}^{(k-1/2)}, T_+(y;x,\theta)\right]&=
\sum_{a=1}^N\frac{k\vartheta_a}{(x-z_a)^{k+1}(y-z_a)}+\sum_{a=1}^N\frac{\vartheta_a}{2(x-z_a)^k(y-z_a)^2},\\
\left[\partial_{V_B(x)}^{(k)}, T_+(y;x,\theta)\right]&=
\sum_{a=1}^N\frac{k}{(x-z_a)^{k+1}(y-z_a)}.
\label{loop_op_ward}
\end{split}
\end{align}

\section{Planar analysis}   \label{sec-planar}

In this appendix we discuss planar solutions in the super-eigenvalue model.

\subsection{Planar free energy}

Let us introduce bosonic and fermionic planar resolvents
\begin{equation}
\omega_B(x)\equiv
\lim_{\begin{subarray}{c}N\to\infty\\\beta=1\end{subarray}}
\frac{\mu}{N}\frac{1}{Z}\Big<\sum_{a=1}^N\frac{1}{x-z_a}\Big>,\qquad
\omega_F(x)\equiv
\lim_{\begin{subarray}{c}N\to\infty\\\beta=1\end{subarray}}
\frac{\mu}{N}\frac{1}{Z}\Big<\sum_{a=1}^N\frac{\vartheta_a}{x-z_a}\Big>,
\label{planar_res}
\end{equation}
where $\mu=\sqrt{\beta}\hbar N$ is the 't Hooft parameter, and $\left<\cdots\right>$ denotes the unnormalized expectation value (\ref{def_unnorm_exp}). Note that the spectral functions $y_B(x)$, $y_F(x)$ defined in (\ref{def_y_bf}), (\ref{yByF}) are written in terms of these resolvents as
\begin{equation}
y_B(x)=V_B'(x)-\omega_B(x),\qquad
y_F(x)=V_F(x)-\omega_F(x).
\label{spect_res}
\end{equation}
In the super-eigenvalue model (\ref{matrix_def}) at $\beta=1$, let us consider the effective potential
\begin{equation}
V_{eff}=\frac{1}{\hbar}\sum_{a=1}^NV(z_a,\vartheta_a)
-\sum_{1\le a<b\le N}\log|z_a-z_b-\vartheta_a\vartheta_b|.
\label{eff_pot}
\end{equation}
Using (\ref{rel_vand}) we then obtain equations of motion for eigenvalues
\begin{equation}
V_B'(z_a)=\hbar\sum_{b\ne a}\frac{1}{z_a-z_b},\qquad
V_F(z_a)=\hbar\sum_{b\ne a}\frac{\vartheta_b}{z_a-z_b}.
\label{eom_eig}
\end{equation}
The eigenvalue distribution is described by bosonic and fermionic density operators
\begin{equation}
\widehat{\rho}_B(z)=\frac{1}{N}\sum_{a=1}^N\delta(z-z_a),\qquad
\widehat{\rho}_F(z)=\frac{1}{N}\sum_{a=1}^N\vartheta_a\delta(z-z_a),
\end{equation}
and the eigenvalue densities at large $N$ can be expressed in terms of the planar resolvents (\ref{planar_res}) as
\begin{align}
\begin{split}
\rho_B(z) &\equiv
\lim_{\begin{subarray}{c}N\to\infty\\\beta=1\end{subarray}}
\frac{1}{Z}\big<\widehat{\rho}_B(z)\big>=
\frac{1}{2\pi i\mu}
\lim_{\epsilon\to 0}\big(\omega_B(z-i\epsilon)-\omega_B(z+i\epsilon)\big),\\
\rho_F(z) &\equiv
\lim_{\begin{subarray}{c}N\to\infty\\\beta=1\end{subarray}}
\frac{1}{Z}\big<\widehat{\rho}_F(z)\big>=
\frac{1}{2\pi i\mu}
\lim_{\epsilon\to 0}\big(\omega_F(z-i\epsilon)-\omega_F(z+i\epsilon)\big).
\label{eig_dens}
\end{split}
\end{align}
From (\ref{eff_pot}) in the large $N$ limit, we then find that the planar free energy can be written in terms of these eigenvalue densities
\begin{align}
S_{eff}(\rho_B, \rho_F)&
=-\mu\int_{D}dz\rho_B(z)V_B(z)
+\frac{\mu^2}{2}\int_{D\times D}dzdz'\rho_B(z)\rho_B(z')\log|z-z'| +
\nonumber\\
&\ \ \
+\mu\int_{D}dz\rho_F(z)V_F(z)
-\frac{\mu^2}{2}\int_{D\times D}dzdz'\frac{\rho_F(z)\rho_F(z')}{z-z'},
\label{pla_free_en}
\end{align}
where $D$ is the support of $\rho_B(z)$ and $\rho_F(z)$, and we have used
\begin{equation}
\log(z_a-z_b-\vartheta_a\vartheta_b)=\log(z_a-z_b)-\frac{\vartheta_a\vartheta_b}{z_a-z_b}.
\end{equation}

\subsection{Planar resolvents}

To compute the resolvents (\ref{planar_res}) one can use planar equations (\ref{b_f_sp_curve}) that, using (\ref{spect_res}), can be written as
\begin{align}
&
\omega_B(x)\omega_F(x)-\oint_C\frac{dz}{2\pi i}\frac{V_F(z)}{x-z}\omega_B(z)
-\oint_C\frac{dz}{2\pi i}\frac{V_B'(z)}{x-z}\omega_F(z)=0,
\label{res_sp_eqs_f}
\\
&
\frac12\omega_B(x)^2+\frac12\omega_F'(x)\omega_F(x)
-\oint_C\frac{dz}{2\pi i}\frac{V_B'(z)}{x-z}\omega_B(z) +
\nonumber
\\
&\hspace{3.5em}
-\oint_C\frac{dz}{2\pi i}\frac{V_F'(z)}{x-z}\omega_F(z)
-\frac12\oint_C\frac{dz}{2\pi i}\frac{V_F(z)}{(x-z)^2}\omega_F(z)=0,
\label{res_sp_eqs_b}
\end{align}
where $C$ is the contour around the support $D$. Around $x=\infty$ the resolvents behave asymptotically as
\begin{equation}
\omega_B(x)=\frac{\mu}{x}+\mathcal{O}(x^{-2}),\qquad
\omega_F(x)=\frac{\mu_{f}}{x}+\mathcal{O}(x^{-2}),
\label{res_asympt}
\end{equation}
where
\begin{equation}
\mu_{f}\equiv
\lim_{\begin{subarray}{c}N\to\infty\\\beta=1\end{subarray}}
\frac{\mu}{N}\frac{1}{Z}\Big<\sum_{a=1}^N\vartheta_a\Big>=\mu \int_Ddz\rho_F(z)
=\oint_C\frac{dz}{2\pi i}\omega_F(z).     \label{fer_coup}
\end{equation}
As shown in \cite{Becker:1992rk, McArthur:1993hw, Plefka:1996tt}, the resolvents have at most second-order fermionic couplings of $\xi_{n+1/2}$
\begin{equation}
\omega_B(x)=\omega_B^{(0)}(x)+\omega_B^{(2)}(x),\qquad
\omega_F(x)=\omega_F^{(1)}(x),
\end{equation}
where additional superscripts denote orders of fermionic couplings. Therefore equations (\ref{res_sp_eqs_f}) and (\ref{res_sp_eqs_b}) are filtered by the orders of fermionic couplings, as in \cite{Plefka:1996tt}
\begin{align}
&
\textrm{order 0}:\qquad
\frac12\omega_B^{(0)}(x)^2=\oint_C\frac{dz}{2\pi i}\frac{V_B'(z)}{x-z}\omega_B^{(0)}(z),
\label{res_pl_0}
\\
&
\textrm{order 1}:\qquad
\widehat{\mathcal{V}}_B'*\omega_F^{(1)}(x)=
-\oint_C\frac{dz}{2\pi i}\frac{V_F(z)}{x-z}\omega_B^{(0)}(z),
\label{res_pl_1}
\\
&
\textrm{order 2}:\qquad
\widehat{\mathcal{V}}_B'*\omega_B^{(2)}(x)=
-\frac12\omega_F^{(1)}(x)\partial_x\omega_F^{(1)}(x)
-\oint_C\frac{dz}{2\pi i}\frac{V_F'(z)}{x-z}\omega_F^{(1)}(z) +
\nonumber\\
&\hspace{12em}
-\frac12\oint_C\frac{dz}{2\pi i}\frac{V_F(z)}{(x-z)^2}\omega_F^{(1)}(z),
\label{res_pl_2}
\\
&
\textrm{order 3}:\qquad
\oint_C\frac{dz}{2\pi i}\frac{V_F(z)}{x-z}\omega_B^{(2)}(z)
-\omega_F^{(1)}(x)\omega_B^{(2)}(x)=0,
\label{res_pl_3}
\end{align}
where we have defined a linear operator $\widehat{\mathcal{V}}_B'$ acting on a function $f(x)$ of $x$ by
\begin{equation}
\widehat{\mathcal{V}}_B'*f(x)\equiv
\oint_C\frac{dz}{2\pi i}\frac{V_B'(z)}{x-z}f(z)
-\omega_B^{(0)}(x)f(x).
\end{equation}

Assuming the $s$-cut ansatz \cite{Migdal:1984gj, Ambjorn:1992gw, Akemann:1996zr}, from the equation (\ref{res_pl_0}) at order 0 we find a solution 
\begin{equation}
\omega_B^{(0)}(x)=\oint_C\frac{dz}{2\pi i}\frac{V_B'(z)}{x-z}\sqrt{\frac{\sigma(x)}{\sigma(z)}},\qquad
\sigma(x)\equiv \prod_{i=1}^{2s}(x-q_i),
\label{res_sol_b0}
\end{equation}
where by the asymptotic condition (\ref{res_asympt}) for $\omega_B(x)$ one finds
the following conditions for $q_i$:
\begin{equation}
\oint_C\frac{dz}{2\pi i}\frac{z^kV_B'(z)}{\sqrt{\sigma(z)}}=\mu\delta_{k,s},\qquad
k=0,1,\ldots,s.
\label{asy_cond_0}
\end{equation}
These conditions imply that the kernel of the operator $\widehat{\mathcal{V}}_B'$ is given by
\begin{equation}
\textrm{ker}(\widehat{\mathcal{V}}_B') \supset
\bigg\{\frac{x^k}{\sqrt{\sigma(x)}},\ k=0,1,\ldots,s\ \bigg|\ \textrm{$q_i$'s satisfy the conditions (\ref{asy_cond_0})}
\bigg\}.
\label{ker_v_b}
\end{equation}
Using the solution (\ref{res_sol_b0}) we then consider the equation (\ref{res_pl_1}) at order 1, and find a solution
\begin{equation}
\omega_F^{(1)}(x)=\oint_C\frac{dz}{2\pi i}\frac{V_F(z)}{x-z}\sqrt{\frac{\sigma(z)}{\sigma(x)}}
+\frac{\Theta(x)}{\sqrt{\sigma(x)}},
\label{res_sol_f1}
\end{equation}
where
\begin{equation}
\Theta(x)\equiv
\sum_{k=0}^{s-1}\mu_{f}^{(k)}x^k,\qquad
\mu_{f}^{(s-1)}\equiv \mu_{f}
\label{fer_undet_fun}
\end{equation}
is an undetermined function with first-order fermionic couplings $\mu_{f}^{(k)}$. Actually, by an exchange of contours
\begin{equation}
\oint_{C_1}\frac{dz_1}{2\pi i}\oint_{C_2}\frac{dz_2}{2\pi i}\frac{V_F(z_1)V_B'(z_2)}{(x-z_1)(z_1-z_2)}\sqrt{\frac{\sigma(z_1)}{\sigma(z_2)}}
=
\oint_{C_1}\frac{dz_1}{2\pi i}\oint_{C_2}\frac{dz_2}{2\pi i}\frac{V_B'(z_1)V_F(z_2)}{(x-z_2)(z_2-z_1)}\sqrt{\frac{\sigma(z_2)}{\sigma(z_1)}},
\end{equation}
we see that the first term of (\ref{res_sol_f1}) satisfies the equation (\ref{res_pl_1}) (the notation $\oint_{C_1}dz_1\oint_{C_2}dz_2$ above means that the contour $C_1$ contains the contour $C_2$; the contribution coming from the pole at $z_1=z_2$ vanishes by the exchange of contours). The second term is in the kernel (\ref{ker_v_b}) of the operator $\widehat{\mathcal{V}}_B'$, and it is provided by the asymptotic condition (\ref{res_asympt}) for $\omega_F(x)$.

By plugging the solution (\ref{res_sol_f1}) into the right hand side of the equation (\ref{res_pl_2}) at order 2 we obtain
\begin{align}
&
\widehat{\mathcal{V}}_B'*\omega_B^{(2)}(x) =
\nonumber\\
&
=
\frac12\oint_{C_1}\frac{dz_1}{2\pi i}\oint_{C_2}\frac{dz_2}{2\pi i}
\frac{V_F(z_1)V_F(z_2)\sqrt{\sigma(z_1)\sigma(z_2)}}{(x-z_1)(x-z_2)^2\sigma(x)}
-\frac12\oint_{C_1}\frac{dz_1}{2\pi i}\oint_{C_2}\frac{dz_2}{2\pi i}
\frac{V_F(z_1)V_F(z_2)\sqrt{\sigma(z_2)}}{(x-z_1)^2(z_1-z_2)\sqrt{\sigma(z_1)}}
\nonumber\\
&\quad
+\frac{\Theta(x)}{2\sigma(x)}\oint_{C_1}\frac{dz_1}{2\pi i}
\frac{V_F(z_1)\sqrt{\sigma(z_1)}}{(x-z_1)^2}
+\frac{\Theta'(x)}{2\sigma(x)}\oint_{C_1}\frac{dz_1}{2\pi i}
\frac{V_F(z_1)\sqrt{\sigma(z_1)}}{x-z_1}
\nonumber\\
&\quad
-\frac12\oint_{C_1}\frac{dz_1}{2\pi i}
\frac{V_F(z_1)\Theta(z_1)}{(x-z_1)^2\sqrt{\sigma(z_1)}}
-\frac{\Theta(x)\Theta'(x)}{2\sigma(x)}
-\oint_{C}\frac{dz}{2\pi i}
\frac{V_F'(z)}{x-z}\omega_F^{(1)}(z),
\label{res_pl_2_r1}
\end{align}
which can be rewritten using the Cauchy's integral formula  as
\begin{align}
&
\widehat{\mathcal{V}}_B'*\omega_B^{(2)}(x) =
\nonumber\\
&
=
\frac12\oint_C\frac{dz}{2\pi i}\oint_{C_1}\frac{dz_1}{2\pi i}\oint_{C_2}\frac{dz_2}{2\pi i}
\frac{V_F(z_1)V_F(z_2)\sqrt{\sigma(z_1)\sigma(z_2)}}{(x-z)(z-z_1)(z-z_2)^2\sigma(z)}
\nonumber\\
&\quad
-\frac12\oint_C\frac{dz}{2\pi i}\oint_{C_1}\frac{dz_1}{2\pi i}\oint_{C_2}\frac{dz_2}{2\pi i}
\frac{V_F(z_1)V_F(z_2)\sqrt{\sigma(z_2)}}{(x-z)(z-z_1)^2(z_1-z_2)\sqrt{\sigma(z_1)}}
\nonumber\\
&\quad
+\frac12\oint_C\frac{dz}{2\pi i}\oint_{C_1}\frac{dz_1}{2\pi i}
\frac{\Theta(z)V_F(z_1)\sqrt{\sigma(z_1)}}{(x-z)(z-z_1)^2\sigma(z)}
+\frac12\oint_C\frac{dz}{2\pi i}\oint_{C_1}\frac{dz_1}{2\pi i}
\frac{\Theta'(z)V_F(z_1)\sqrt{\sigma(z_1)}}{(x-z)(z-z_1)\sigma(z)}
\nonumber\\
&\quad
-\frac12\oint_C\frac{dz}{2\pi i}\oint_{C_1}\frac{dz_1}{2\pi i}
\frac{V_F(z_1)\Theta(z_1)}{(x-z)(z-z_1)^2\sqrt{\sigma(z_1)}}
-\frac{\Theta(x)\Theta'(x)}{2\sigma(x)}
-\oint_{C}\frac{dz}{2\pi i}
\frac{V_F'(z)}{x-z}\omega_F^{(1)}(z).
\label{res_pl_2_r2}
\end{align}
In this expression, by exchanging the contour $C$ with contours $C_1$ and $C_2$, the second, the fifth and the last term cancel out and we obtain
\begin{equation}
\widehat{\mathcal{V}}_B'*\omega_B^{(2)}(x)=
-\frac12\sum_{i=1}^{2s}\frac{1}{(x-q_i)\sigma'(q_i)}
\left(\Xi_i^{(1)}-\Theta(q_i)\right)\left(\Xi_i^{(2)}-\Theta'(q_i)\right),
\label{res_pl_2_r3}
\end{equation}
where we have introduced fermionic moments
\begin{equation}
\Xi_i^{(n)}\equiv
\oint_C\frac{dz}{2\pi i}\frac{V_F(z)}{(z-q_i)^n}\sqrt{\sigma(z)}.
\label{fer_moment}
\end{equation}
By taking into account the kernel (\ref{ker_v_b}) of the operator $\widehat{\mathcal{V}}_B'$ and the asymptotic condition (\ref{res_asympt}) for $\omega_B(x)$, we find a solution of (\ref{res_pl_2_r3}) in the form
\begin{equation}
\omega_B^{(2)}(x)=
-\frac12\sum_{i=1}^{2s}\frac{\big(\Xi_i^{(1)}-\Theta(q_i)\big)\big(\Xi_i^{(2)}-\Theta'(q_i)\big)}{M_i\sigma'(q_i)(x-q_i)\sqrt{\sigma(x)}}
+\frac{\Upsilon(x)}{\sqrt{\sigma(x)}},
\label{res_sol_b2}
\end{equation}
where
\begin{equation}
\Upsilon(x)\equiv \sum_{k=0}^{s-2}\mu_{ff}^{(k)}x^k
\label{bos_undet_fun}
\end{equation}
is an undetermined function with second-order fermionic couplings $\mu_{ff}^{(k)}$, and $M_i$ are the first bosonic momenta
\begin{equation}
M_i\equiv
\oint_C\frac{dz}{2\pi i}\frac{V_B'(z)}{z-q_i}\frac{1}{\sqrt{\sigma(z)}}.
\label{bos_moment}
\end{equation}

Let us finally consider the last equation (\ref{res_pl_3}) at order 3, which gives constraints for undetermined functions $\Theta(x)$ in (\ref{fer_undet_fun}) and $\Upsilon(x)$ in (\ref{bos_undet_fun}). Using the Cauchy's integral formula the equation (\ref{res_pl_3}) is written as
\begin{equation}
\oint_C\frac{dz}{2\pi i}\frac{V_F(z)}{x-z}\omega_B^{(2)}(z)
-\oint_C\frac{dz}{2\pi i}\frac{1}{x-z}\omega_F^{(1)}(z)\omega_B^{(2)}(z)=0.
\end{equation}
Plugging the solutions (\ref{res_sol_f1}) and (\ref{res_sol_b2}) into this equation, we see that the first term is canceled out by a term coming from the exchange of contours in the second summand. We then obtain
\begin{equation}
\sum_{i=1}^{2s}\frac{\Xi_i^{(1)}-\Theta(q_i)}{(x-q_i)\sigma'(q_i)}
\bigg[
-\frac12\sum_{j\ne i}^{2s}\frac{1}{(q_i-q_j)M_j\sigma'(q_j)}
\left(\Xi_j^{(1)}-\Theta(q_j)\right)\left(\Xi_j^{(2)}-\Theta'(q_j)\right)+\Upsilon(q_i)\bigg]=0.
\end{equation}
The residues at $x=q_i, i=1,\ldots,2s$, give $2s$ constraints for $\Theta(x)$ and $\Upsilon(x)$
\begin{equation}
\left(\Xi_i^{(1)}-\Theta(q_i)\right)\bigg[
-\frac12\sum_{j\ne i}^{2s}\frac{1}{(q_i-q_j)M_j\sigma'(q_j)}
\left(\Xi_j^{(1)}-\Theta(q_j)\right)\left(\Xi_j^{(2)}-\Theta'(q_j)\right)+\Upsilon(q_i)\bigg]=0.
\label{res_sol_const}
\end{equation}
These equations give constrains for $s$ indeterminates $\mu_{f}^{(k)}$ in $\Theta(x)$ and $s-1$ indeterminates $\mu_{ff}^{(k)}$ in $\Upsilon(x)$.
In particular in the 1-cut ($s=1$) case $\Upsilon(x)=0$, so that these constraints give
\begin{equation}
\left(\Xi_1^{(1)}-\mu_{f}\right)\left(\Xi_2^{(1)}-\mu_{f}\right)=0, \label{one_const_eq}
\end{equation}
and one finds that one solution can be written as \cite{Plefka:1996tt}
\begin{equation}
\widetilde{\mu}_{f}=\frac{\Xi_1^{(1)}+\Xi_2^{(1)}}{2}
=\frac{1}{2}\oint_C\frac{dz}{2\pi i}\frac{V_F(z)\sigma'(z)}{\sqrt{\sigma(z)}},\qquad
\sigma(z)=(z-q_1)(z-q_2).
\label{one_cut_muf}
\end{equation}
In fact this solution has an ambiguity
\begin{equation}
\mu_{f}=\widetilde{\mu}_{f}+c(q_1,q_2)\left(\Xi_1^{(1)}-\Xi_2^{(1)}\right),
\label{one_cut_muf_c}
\end{equation}
where $c(q_1,q_2)=-c(q_2,q_1)$ is a bosonic function of $q_1$ and $q_2$; if $\Xi_1^{(1)}=\Xi_2^{(1)}$, the equation (\ref{one_const_eq}) is trivially satisfied and does not determine $\mu_f$.

Summarizing the above results, the $s$-cut solution of the planar resolvents is given by
\begin{align}
&
\omega_B(x)=\oint_C\frac{dz}{2\pi i}\frac{V_B'(z)}{x-z}\sqrt{\frac{\sigma(x)}{\sigma(z)}}
-\frac12\sum_{i=1}^{2s}\frac{\big(\Xi_i^{(1)}-\Theta(q_i)\big)\big(\Xi_i^{(2)}-\Theta'(q_i)\big)}{M_i\sigma'(q_i)(x-q_i)\sqrt{\sigma(x)}}
+\frac{\Upsilon(x)}{\sqrt{\sigma(x)}},
\label{b_res_sol}
\\
&
\omega_F(x)=\oint_C\frac{dz}{2\pi i}\frac{V_F(z)}{x-z}\sqrt{\frac{\sigma(z)}{\sigma(x)}}
+\frac{\Theta(x)}{\sqrt{\sigma(x)}},
\label{f_res_sol}
\end{align}
with the constraint equations (\ref{res_sol_const}), where $C$ is the contour around all the branch cuts of $\sqrt{\sigma(x)}$. This formula is the multi-cut version of the one-cut solution derived in \cite{Plefka:1996tt}. By (\ref{spect_res}) we then also obtain the spectral functions $y_B(x)$ and $y_F(x)$. To determine the branch points $q_i$ one can use the $s+1$ constraint equations (\ref{asy_cond_0}). For $s\ge 2$, as in the usual bosonic model, in addition to this asymptotic condition we also need to consider $s-1$ constraint equations representing stability conditions among cuts, or the filling fractions along each cut (see e.g. \cite{Akemann:1996zr, Marino:2004eq}).

We consider examples of planar one-cut solutions in the super-gaussian and the super-multi-Penner models in sections \ref{ssec-gaussian} and \ref{ssec-Penner}.

\newpage

\bibliographystyle{JHEP}
\bibliography{abmodel}

\end{document}